%% file: JHEP.tex
\documentclass{cernyrep} 
\usepackage{graphics}
\usepackage{texnames}
\usepackage[T1]{fontenc}
\usepackage{rotating}

\usepackage{latexsym}
\usepackage{amsfonts,amsmath,amssymb}
\usepackage{url}
\usepackage[utf8]{inputenc}
\usepackage{fancyref}
\usepackage[normalem]{ulem}
\usepackage{color}

\newcommand{\cms}{{\rm cm}^{-2}{\rm s}^{-1}}
\newcommand{\epem}{{\rm e}^{+}{\rm e}^{-}}
\newcommand{\mpmm}{\mu^+\mu^-}
\newcommand{\epemto}{{\rm e}^{+}{\rm e}^{-}\to}

\newcommand{\ttbar}{{\rm t}\bar{\rm t}}
\newcommand{\bbbar}{{\rm b}\bar{\rm b}}
\newcommand{\nnbar}{\nu\bar\nu}
\newcommand{\infb}{{\rm fb}^{-1}}
\newcommand{\inab}{{\rm ab}^{-1}}
\newcommand{\ALR}{A_{\rm LR}}
\newcommand{\sintw}{\sin^2\theta_{\rm W}^{\rm eff}}

\begin{document}

\title{First Look at the Physics Case of TLEP}

\author{The TLEP Design Study Working Group}
\institute{{\small (See next pages for the list of authors)}\vskip 3cm}

\maketitle 

\bibliographystyle{naturemag}
 
\begin{abstract}
The discovery by the ATLAS and CMS experiments of a
new boson with mass around 125 GeV and with measured properties
compatible with those of a Standard-Model Higgs boson, coupled with the
absence of discoveries of phenomena beyond the Standard Model at the TeV
scale, has triggered interest in ideas for future Higgs factories. 
A new circular $\epem$ collider hosted in a 80~to~100~km tunnel, 
TLEP, is among the most attractive solutions proposed so far.
It has a clean experimental environment, produces high
luminosity for top-quark, Higgs boson, W and Z studies, accommodates
multiple detectors, and can reach energies up to the $\ttbar$ threshold
and beyond. It will enable measurements of the Higgs boson properties
and of Electroweak Symmetry-Breaking (EWSB) parameters with unequalled
precision, offering exploration of physics beyond the Standard Model in
the multi-TeV range. Moreover, being the natural precursor of the
VHE-LHC, a 100~TeV hadron machine in the same tunnel, it builds up a
long-term vision for particle physics. Altogether, the combination of
TLEP and the VHE-LHC offers, for a great cost effectiveness, the best
precision and the best search reach of all options presently on the
market. This paper presents a first appraisal of the salient features of
the TLEP physics potential, to serve as a baseline for a more extensive
design study.
\end{abstract}
\vfill\eject

\input{authors.tex}
\vfill\eject

\section{Introduction}
\label{sec:intro}

\begin{figure}[htbp]
\begin{center}
\includegraphics[width=0.65\columnwidth]{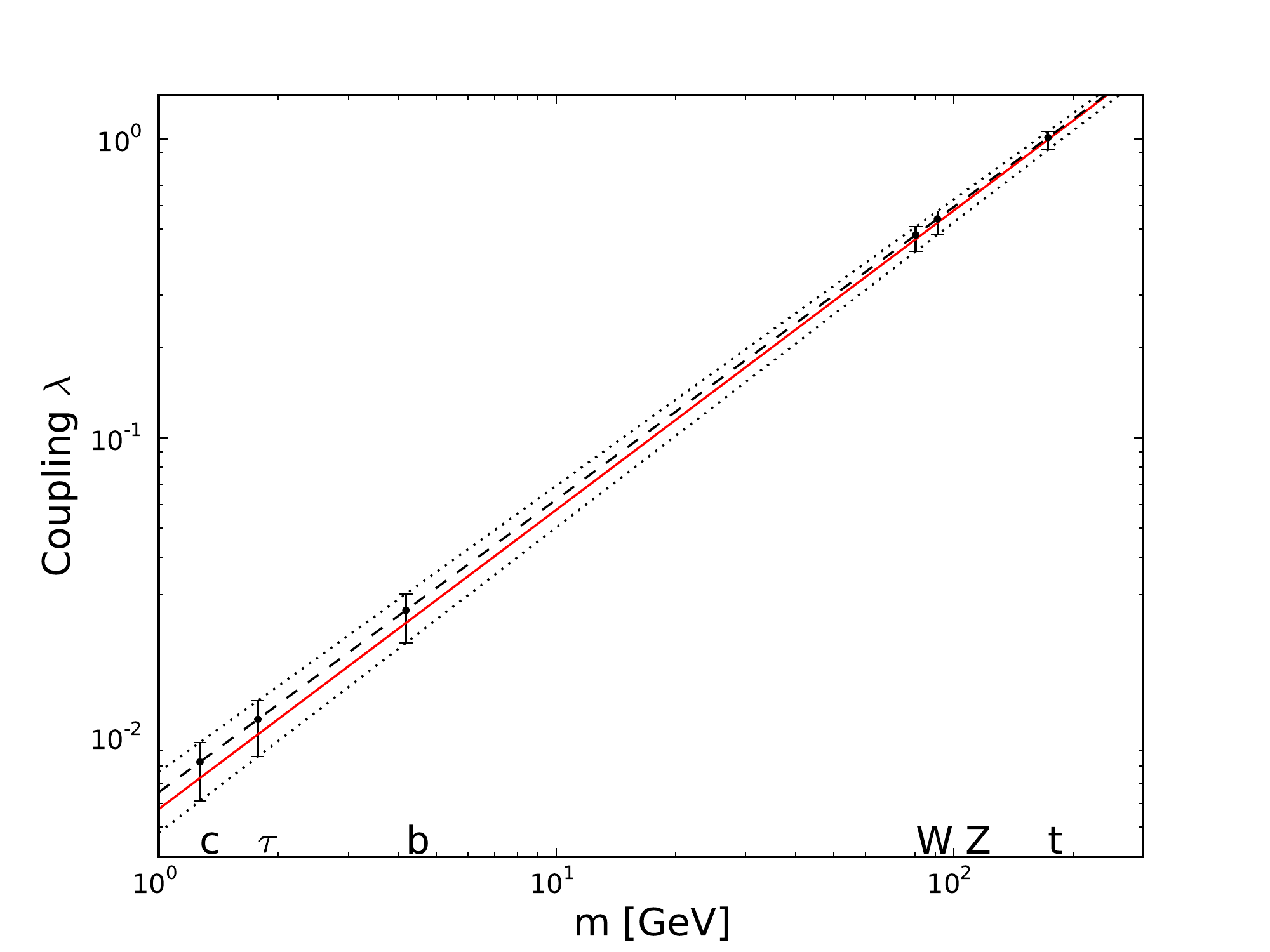}
\end{center}
\caption{\label{fig:ellis} The mass dependence of the couplings of the recently discovered Higgs boson to fermions and gauge bosons, from a two-parameter fit (dashed line) to a combination of the CMS and ATLAS data collected at 7 and 8 TeV in 2011 and 2012, taken from Ref.~\cite{cite:1303.3879}. The dotted lines bound the 68\% C.L. interval. The value of the coupling of the Higgs boson to the c quark shown in the figure is a prediction of the fit. The solid line corresponds to the Standard Model prediction.}
\end{figure}

The Higgs boson with mass around 125 GeV recently discovered by the ATLAS and CMS experiments~\cite{v_Aben_Abi_Abolins_et_al__2012,gicevic_Ero_Fabjan_et_al__2012} at the LHC is found to have properties compatible with the Standard Model predictions~\cite{ATLASHiggsCouplings,Ero_Fabjan_Friedl_et_al__2013}, as shown for example in Fig.~\ref{fig:ellis}~\cite{cite:1303.3879}. Coupled with the absence of any other indication so far for new physics at the LHC, be it either through precision measurements or via direct searches, this fundamental observation seems to push the energy scale of any physics beyond the Standard Model above several hundred GeV. The higher-energy LHC run, which is expected to start in 2015 at $\sqrt{s} \sim 13$-$14$ TeV, will extend the sensitivity by a factor two, in many cases well above 1~TeV. Fundamental discoveries may therefore be made in this energy range by 2017-2018. Independently of the outcome of this higher-energy run, however, there must be new phenomena, albeit at unknown energy scales, as shown by the evidence for non-baryonic dark matter, the cosmological baryon-antibaryon asymmetry and non-zero neutrino masses, which are all evidence for physics beyond the Standard Model. In addition to the high-luminosity upgrade of the LHC, new particle accelerators will be instrumental to understand the physics underlying these observations.

The path towards the choice of the most appropriate machine(s) to analyse these new phenomena may be guided by historical precedents, which reveal the important r\^oles played by lower-energy precision measurements when establishing roadmaps for future discoveries with higher-energy machines. In the late 1970's, precision measurements of neutral currents led to the prediction of the existence of the W and Z bosons, as well as the values of their masses. The W and Z were then discovered in the early 1980's at the CERN S${\rm p\bar p}$S collider with masses in the range predicted. Subsequently, the CERN LEP $\epem$ collider measured the properties of the Z and W bosons with high precision in the 1990's. These precise measurements led to the prediction of the top-quark mass, which was discovered at the FNAL Tevatron with the predicted mass. The measurement of $m_{\rm top}$, together with the precise measurement of the W mass at the Tevatron in the past decade, led in turn to a prediction for the mass of the Higgs boson, which was recently discovered at the LHC within the predicted mass range. 

The details of the optimal strategy for the next large facility after the LHC can only be finalized once the results of the LHC run at 13-14 TeV are known.Depending on these results, a first step in the strategy to look beyond the LHC findings could require a facility that would measure the Z, W, top-quark and Higgs-boson properties with sufficient accuracy to provide sensitivity to new physics at a much higher energy scale. The strategy could then be followed by a second step that would aim at discovering this new physics directly, via access to a much larger centre-of-mass energy. 

For example, new physics at an energy scale of 1 TeV would translate typically into deviations $\delta g_{\rm HXX}$ of the Higgs boson couplings to gauge bosons and fermions, $g_{\rm HXX}$, of up to 5\% with respect to the Standard Model predictions~\cite{ILC:Physics,Gupta_Rzehak_Wells_2012}, with a dependence that is inversely proportional to the square of the new energy scale $\Lambda$:
\begin{equation}
\frac{\delta g_{\rm HXX}}{g_{\rm HXX}^{\rm SM}} \le 5\% \times \left(\frac{1 {\rm TeV}}{\Lambda}\right)^2.
\end{equation}

Therefore the Higgs boson couplings need to be measured with a per-cent accuracy or better to be sensitive to 1~TeV new physics, and with a per-mil accuracy to be sensitive to multi-TeV new physics. Similarly, Electroweak precision measurements made at LEP with $10^7$ Z decays, together with accurate W and top-quark mass measurements from the Tevatron, are sensitive to weakly-coupled new physics at a scale up to 3~TeV. To increase this sensitivity up to 30~TeV, an improvement in precision by two orders of magnitude, i.e., an increase in statistics by four orders of magnitude to at least $10^{11}$ Z decays, would be needed.  At the same time, the current precision of the W and top-quark mass measurements needs to be improved by at least one order of magnitude, i.e., to better than 1~MeV and 50~MeV respectively, in order to match the increased Z-pole measurement sensitivity. These experimental endeavours will also require significant theoretical effort in a new generation of theoretical calculations in order to reap the full benefits from their interpretation.

\begin{figure}[tb]
\begin{center}
\includegraphics[width=0.7\columnwidth]{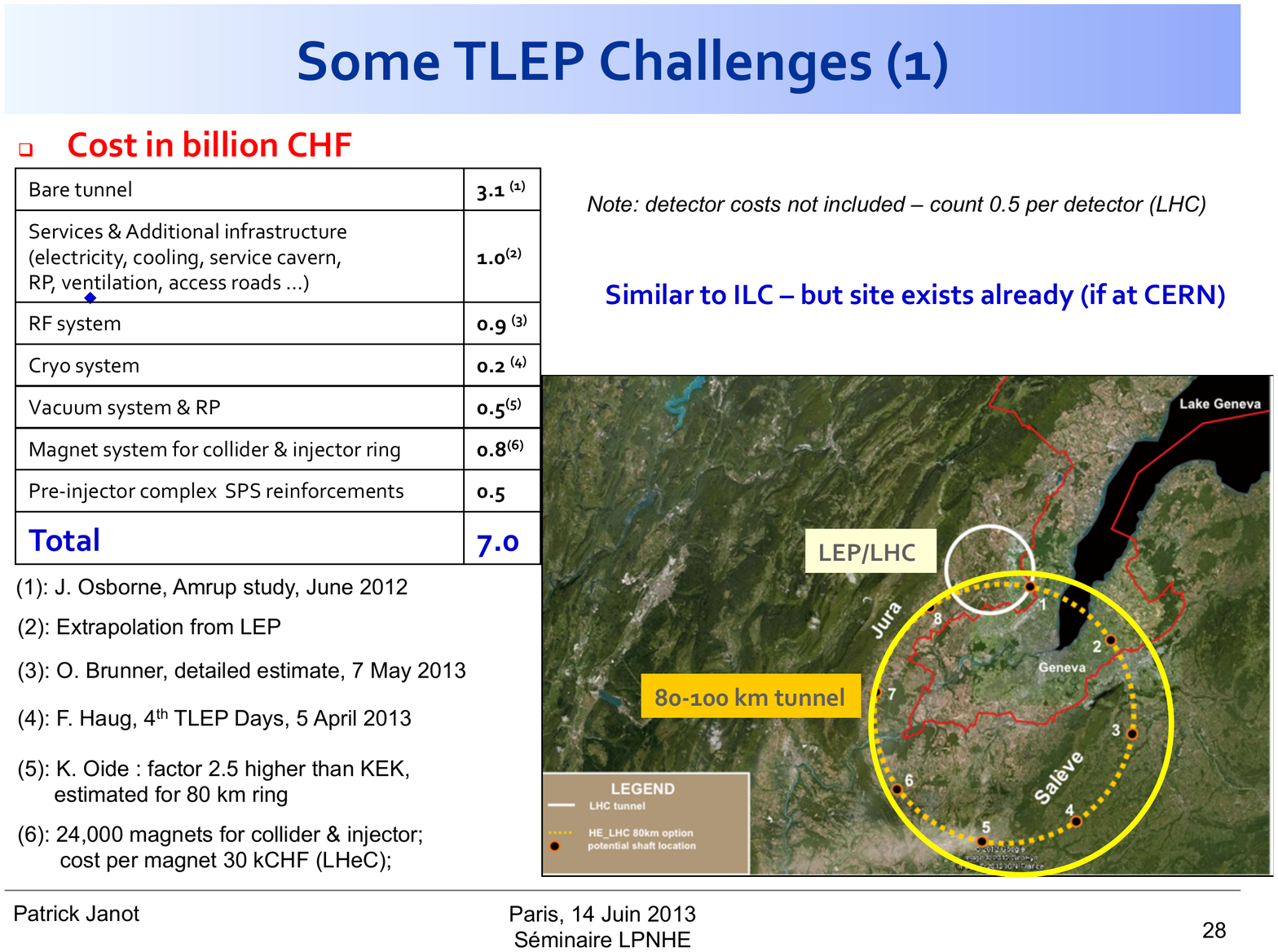}
\end{center}
\caption{\label{fig:TLEP80} A possible implementation of the 80~km tunnel (dashed circle) that would host TLEP and the VHE-LHC in the Geneva area, taken from Ref.~\cite{cite:Osborne}. The 100~km version (full line) is currently under study. }
\end{figure}

Among the various possibilities on the table today (pp colliders, $\epem$ colliders, $\mu^+\mu^-$ colliders and $\gamma\gamma$ colliders), it seems that circular $\epem$ colliders offer the best potential to deliver the integrated luminosities that would be adequate to reach such levels of precision. The proposed TLEP $\epem$ collider~\cite{cite:1305.6498}, which could be hosted in a new 80 to 100 km tunnel~\cite{cite:Osborne} in the Geneva area, as seen in Fig.~\ref{fig:TLEP80}, would be able to produce collisions at centre-of-mass energies from 90 to 350~GeV and beyond, at several interaction points, and make precision measurements at the Z pole, at the WW threshold, at the HZ cross section maximum, and at the $\ttbar$ threshold, with an unequalled accuracy. The same tunnel will be designed to host a hadron collider (called the VHE-LHC), at a centre-of-mass energy of up to 100~TeV, which would give direct access to new physics up to scales of 30~TeV. This visionwas already put forward by the ICFA beam-dynamics workshop~\cite{HF2012} where the design study of a circular Higgs factory was recommended. It is fully in-line with the recent update of the European Strategy, approved at the end of May 2013 by the CERN Council~\cite{cite:Strategy}. In particular, the Council calls upon the Organization to develop a proposal for an ambitious post-LHC accelerator project at the high-energy frontier, and recalls the strong scientific case for an $\epem$ collider that can study the properties of the Higgs boson and other particles with unprecedented precision. This global vision is now being implemented at CERN under the ``Future Circular Colliders'' (FCC) international design study.

This paper is organized as follows. The main characteristics of the TLEP collider relevant for the physics case are summarized in Section~\ref{sec:exp}. In Sections~\ref{sec:Higgs} and~\ref{sec:EWSB}, an overview of the TLEP potential for precise measurements of the Higgs boson properties and of the EWSB parameters is presented. Possible follow-on projects, which include an increase of the TLEP centre-of-mass energy to 500~GeV, and complementing TLEP with a 100~TeV pp collider, the VHE-LHC, are described briefly in Section~\ref{sec:VHE-LHC}. Comparisons with the potential of the high-luminosity LHC upgrade (HL-LHC) and of linear collider projects are made throughout. This paper represents the current, preliminary understanding of the physics potential of TLEP, complemented with mentions of the VHE-LHC reach whenever appropriate. A five-year-long design study -- responding to the recent European Strategy update and part of the CERN medium-term plan~\cite{cite:MTP} for 2014--2018 -- has been launched to refine this understanding, as well as to ascertain the feasibility of TLEP and the VHE-LHC, as input to the next European Strategy update. 

\section{The experimental environment}
\label{sec:exp}

\subsection{Luminosity and energy}
\label{sec:luminosity}

The TLEP collider complex consists of an accelerator ring and a storage ring~\cite{1112.2518}, the former delivering continuous top-up injection to the latter, so that a constant level of luminosity is provided in collisions. The current TLEP working points can be found in Ref.~\cite{cite:1305.6498}, for the four centre-of-mass energies of interest: the Z pole ($\sqrt{s} \sim 91$~GeV); the WW threshold ($\sqrt{s} \sim 161$~GeV); the HZ cross-section maximum ($\sqrt{s} \sim 240$~GeV); and the top-pair threshold ($\sqrt{s} \sim 350$~GeV). The possible upgrade to $\sqrt{s}=500$ GeV is discussed in Section~\ref{sec:VHE-LHC}. The 12 GV RF system is designed to compensate for the energy loss by synchrotron radiation at $\sqrt{s} = 350$ GeV, at which a luminosity of $1.3\times 10^{34}~\cms$ can be delivered at each interaction point (IP), in a configuration with four IPs. At lower centre-of-mass energies, the energy losses decrease steeply like $E^4_{\rm beam}$, and the RF power can be used to accelerate a much larger number of ${\rm e}^\pm$ bunches, from 12 bunches at 350 GeV all the way to 4400 bunches at the Z pole. As a result, the luminosity increases approximately like $1/E^3_{\rm beam}$ when the centre-of-mass energy decreases. (The smaller exponent is a consequence of operating at the beam-beam limit.) The preliminary values of the luminosities expected at each energy are displayed in Table~\ref{tab:lumi}, together with other important parameters of the machine (beam size, RF cavity gradient, number of bunches, and total power consumption), taken from Ref.~\cite{cite:1305.6498}. The last row gives the integrated luminosity expected at each interaction for one year of data taking (1 year = $10^7$ seconds).

\begin{table}
\begin{center}
\caption{\label{tab:lumi} Preliminary values of the luminosity for TLEP in each of the four planned configurations~\cite{cite:1305.6498}. Other parameters relevant for the physics potential of TLEP (beam size, RF cavity gradient, number of bunches, total power consumption and integrated luminosity per year at each IP) are also listed.}
\begin{tabular}{|r|r|r|r|r|}
\hline                         & TLEP-Z & TLEP-W & TLEP-H & TLEP-t \\
\hline\hline $\sqrt{s}$ (GeV) &     90     &    160     &   240      &   350    \\ 
\hline L ($10^{34}~\cms$/IP)  & 56  & 16 & 5 & 1.3 \\
\hline \# bunches & 4400 & 600 & 80 & 12 \\
\hline RF Gradient (MV/m) & 3 & 3 & 10 & 20 \\
\hline Vertical beam size (nm) & 270 & 140 & 140 & 100 \\
\hline Total AC Power (MW)  & 250 & 250 & 260 & 284 \\ 
\hline ${\rm L_{\rm int}}$ ($\inab$/year/IP) & 5.6 & 1.6 & 0.5 & 0.13 \\ \hline
\end{tabular}
\end{center}
\end{table}

\begin{figure}[tb]
\begin{center}
\includegraphics[width=0.8\columnwidth]{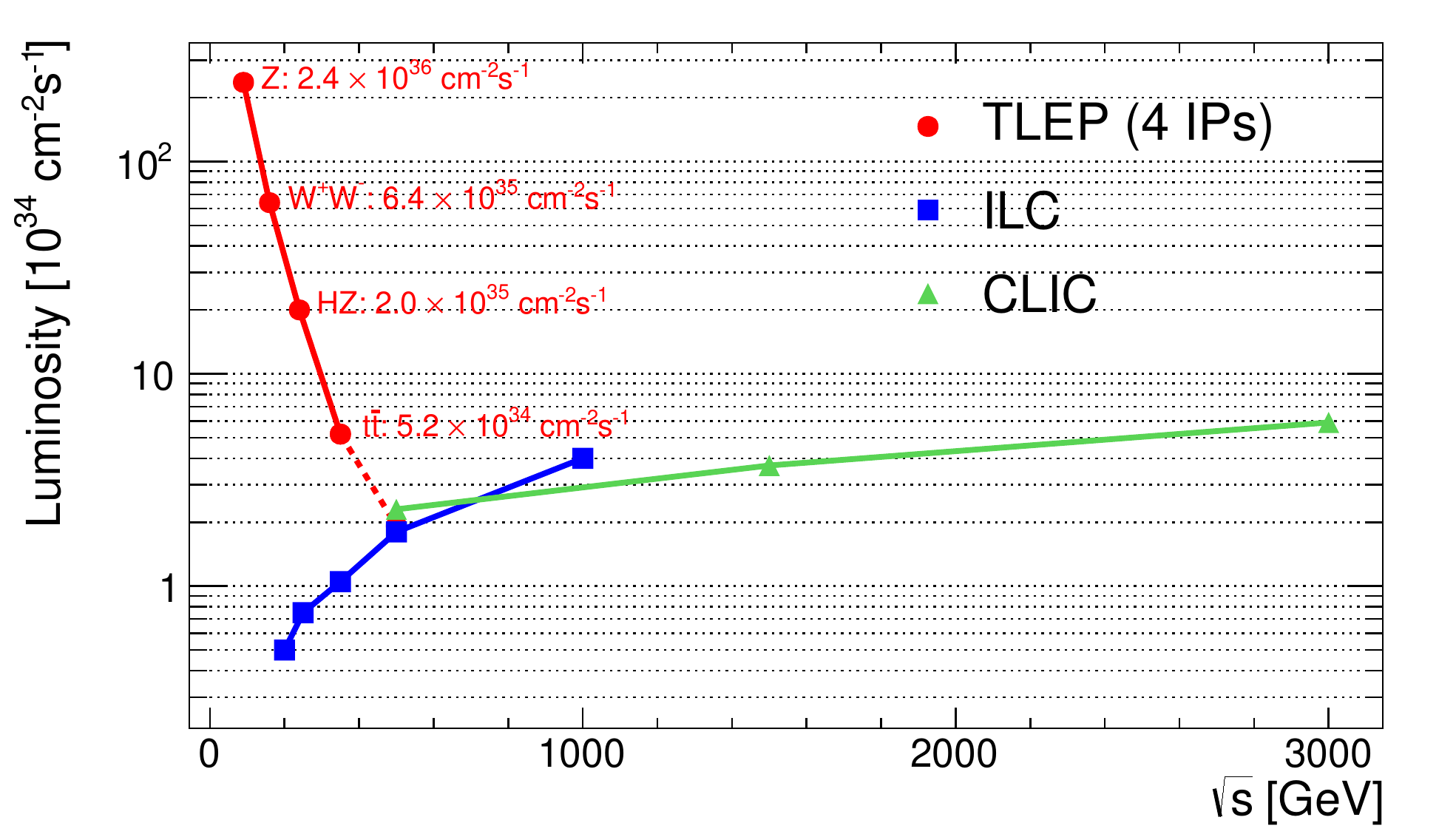}
\end{center}
\caption{\label{fig:lumi} Instantaneous luminosity, in units of $10^{34}~\cms$, expected at TLEP (full red line), in a configuration with  four interaction points operating simultaneously, as a function of the centre-of-mass energy. For illustration, the luminosities expected at linear colliders, ILC (blue line) and CLIC (green line), are indicated in the same graph. As explained in the text, the TLEP luminosity at each interaction point would increase significantly if fewer interaction points were considered. The possible TLEP energy upgrade up to 500 GeV, represented by a dashed line, is briefly discussed in Section~\ref{sec:VHE-LHC}.}
\end{figure}

These luminosity values are obtained in a configuration of the collider with four interaction points, for which the beam-beam parameters can be obtained directly from measurements performed at LEP1 and LEP2 in the 1990's. For this reason, the luminosity summed over the four interaction points, the only relevant quantity when it comes to evaluating the physics potential,  is shown in Fig.~\ref{fig:lumi}. Should TLEP operate with fewer detectors, the larger damping time between collisions would tend to push the beam-beam limit, with the effect of increasing the luminosity at each interaction point by a factor $(4/n_{\rm IP})^{0.4}$~\cite{Assmann:453821}. For example, the use of two detectors instead of four would only reduce the total luminosity by 35\% (as opposed to a naive factor 2 reduction), hence would increase the statistical uncertainties reported in this article by about 20\%. The physics potential of either configuration is summarized in Table~\ref{tab:FitResults} (Section~\ref{sec:FitResults}) and Table~\ref{tab:EW-TLEP} (Section~\ref{sec:EWSB}). Although there is some debate as to the functional dependence of the beam-beam parameter on the damping decrement, any modifications to the formula of Ref.~\cite{Assmann:453821} will have minor effects on the conclusions of this analysis.  

Also displayed in Fig.~\ref{fig:lumi} are the luminosities expected for the two linear collider projects, ILC~\cite{ILC:AcceleratorA,ILC:AcceleratorB} and CLIC~\cite{cite:CLICDR}, as a function of the centre-of-mass energy. It is remarkable that the luminosity expected at TLEP is between a factor 5 and three orders of magnitude larger than that expected for a linear collider, at all centre-of-mass energies from the Z pole to the $\ttbar$ threshold, where precision measurements are to be made, hence where the accumulated statistics will be a key feature. Upgrades aimed at delivering luminosities well beyond the values given above are also being investigated -- although they cannot be guaranteed today.Similar upgrades are also contemplated for the ILC~\cite{1308.3726}. Possibilities for TLEP include beam charge compensation and the use of the {\textquotedblleft}crab-waist{\textquotedblright} collision scheme~\cite{TelnovC, TelnovD}, allowing beamstrahlung effects to be mitigated. Upgrades to higher centre-of-mass energies are discussed in Section~\ref{sec:VHE-LHC}.

\subsection{Beamstrahlung}
\label{sec:beamstrahlung}

\begin{figure}[tb]
\begin{center}
\includegraphics[width=0.7\columnwidth]{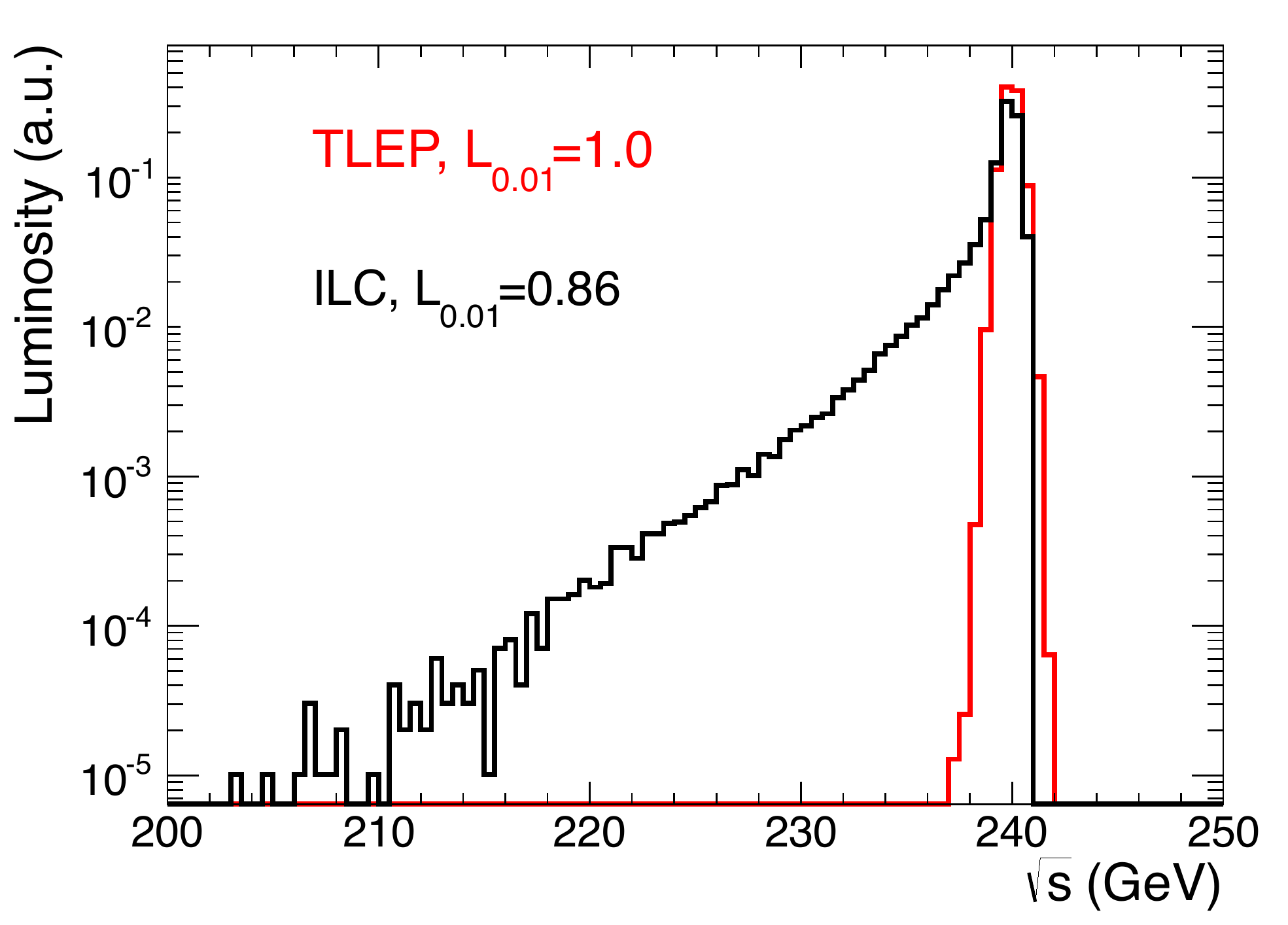}
\end{center}
\caption{\label{fig:beamstrahlung} The beam-energy spectrum for TLEP (red) for $\sqrt{s} = 240$ GeV. For illustration, the beam-energy spectrum expected in presence of beamstrahlung is shown for the ILC (black) at the same centre-of-mass energy. The $L_{0.01}$ value is the fraction of the integrated luminosity produced within 1\% of the nominal centre-of-mass energy. The effect of initial state radiation (common to TLEP and ILC, but physics-process dependent) is not included in this plot.}
\end{figure}

Beamstrahlung is an issue for $\epem$ rings~\cite{Yokoya,TelnovB}, as its effects may cause either the beam lifetime to become prohibitively small, or the beam-energy spread and bunch length to become unacceptably large. Indeed, the continuous loss of even a tiny fraction of the beam at each collision reduces the beam lifetime at the higher TLEP beam energies, and cumulative increases in the energy spread result in significant bunch lengthening, especially at the lower energies. Solutions to mitigate these effects are well known, and are described in Ref.~\cite{TelnovB,cite:1305.6498}. Steadily improved simulations and analytical calculations show that, with the current TLEP parameters at $\sqrt{s} = 350$~GeV~\cite{cite:1305.6498}, a momentum acceptance of 2.0\%, and a ratio of vertical to horizontal emittances of 0.2\%, the luminosity drops by 10\% every minute. With a top-up rate of once per minute, the average luminosity amounts to 95\% of the peak luminosity. Beamstrahlung effects are, on the other hand, benign for the physics performance. For example, the beamstrahlung-induced beam energy spread is expected to be smaller than 0.1\%, as shown in Fig.~\ref{fig:beamstrahlung} for $\sqrt{s}=240$ GeV.

This low level of beamstrahlung provides several advantages, some examples of which are given below.
\begin{itemize}
\item Beamstrahlung is a macroscopic effect that cannot be predicted from first principles, and the resulting beam-energy spectrum needs to be measured in situ, with significant statistical and systematic uncertainties. The measurement of observables relying on a precise beam-energy knowledge (e.g., Z or W masses, Z width, top quark mass, etc.)  therefore profit from the relative absence of beamstrahlung. Similarly, cross sections with a rapid variation as a function of the centre-of-mass energy (e.g., at the Z pole, or at the WW and $\ttbar$ thresholds, as shown for example in Fig.~\ref{fig:ttbar} of Section~\ref{sec:EWSB}) are {\it (i)} maximal; and {\it (ii)} calculable with very good accuracy, leading to small statistical and systematic uncertainties if beamstrahlung effects can be neglected.
\item The forward region of a TLEP detector is free of beamstrahlung photons, which in turn eases both the design of a luminometer and the integrated luminosity measurement. Likewise, the beam-related backgrounds (disrupted beams, photons, $\epem$ pairs) originating from beamstrahlung are small, and so are the parasitic $\gamma\gamma$ collisions. Pile-up of interactions is therefore negligible.
\item Final states with photons (e.g., ${\rm H} \to \gamma\gamma$, ${\rm H \to Z}\gamma$, or $\epem \to {\rm Z}\gamma \to \nnbar\gamma$) can be selected with optimal purity.  
\item The quasi-absence of beamstrahlung photons along the beam axis (in both directions) enables an optimal use of energy and momentum constraints in kinematic fits.
\end{itemize} 
In summary, the known assets of $\epem$ collisions -- cleanliness, calculability, numerous kinematic constraints, and absence of pile-up collisions -- are well preserved at TLEP, mostly because of the absence of beamstrahlung. When it comes to precision measurements, these advantages come in order of importance right after the large integrated luminosity. 

\subsection{Beam polarization}   
\label{sec:beampol}     

\subsubsection{Motivation}     
Polarized beams are useful for several purposes in $\epem$ storage rings. Transverse polarization was used in single beams at LEP for beam energy calibration with 0.1 MeV intrinsic precision~\cite{LEP1Cal, ement_of_the_W_boson_mass_2005}. This precision will be essential for the TLEP measurements of the Z mass and width, and of the W mass, with the required accuracy. Longitudinal polarization was used in collisions at SLC for the measurement of the left-right asymmetry at the Z pole, $\ALR$, with a $10^{-3}$ accuracy~\cite{altay_Band_Barklow_et_al__2000}, which in turn allowed a determination of the weak mixing angle with an accuracy similar to that of the best LEP unpolarized measurements. It is therefore of great interest to establish both transverse and longitudinal polarization with TLEP, and be able to maintain longitudinal polarization in collisions at the Z pole.    

\subsubsection{Transverse polarization}

Transverse beam polarization builds up naturally in a storage ring by the Sokolov-Ternov effect. A transverse polarization in excess of 5-10\%, which was obtained for beam energies up to 61 GeV per beam at LEP, is sufficient for beam energy calibration purposes. It is generally accepted that this upper limit is determined by the energy spread, which becomes commensurate with the fractional part of the spin-tune  $\nu_s = E_{\rm beam} {\rm [GeV]} /0.440665$. Given that the energy spread scales as $E^2_{\rm beam} / \sqrt{\rho}$, where $\rho$ is the ring bending radius, it is expected that beam polarization sufficient for energy calibration should be readily available up to and above  the WW threshold (i.e., 81~GeV per beam) at TLEP. A new machine with a better control of the orbit should, however, be able to increase this limit. For example, a full 3D spin tracking simulation of the electron machine of the Large Hadron-electron Collider (LHeC) project in the 27 km LHC tunnel predicts 20\% polarization at a beam energy of 65 GeV for typical machine misalignments~\cite{1206.2913}. 

At LEP, the natural polarization building time amounted to five hours at the Z peak. This time is predicted to increase like the third power of the ring bending radius, hence will reach the unpractical value of 150 hours at TLEP. Asymmetric "polarization" wigglers were in use in LEP, and their effect on the polarization time and the beam energy spread, as well as other depolarizing sources, is analyzed in Ref.~\cite{cite:Blondel-Jowett-LEP606}. Such polarization wigglers could be used to reduce the polarization time at TLEP, while keeping the energy spread to a reasonable value. As an example, the use of the LEP polarization wigglers in TLEP with a central pole field of 0.6~T would reduce the polarization time to 18 hours at the Z peak, while keeping the beam energy spread below 48~MeV -- a value at which polarization could routinely be obtained in LEP at with a beam energy of 55~GeV. In these conditions, a level of polarization sufficient to perform resonant depolarization could be reached in a couple hours.  Energy calibrations would then be performed every ten minutes if at least twelve bunches of electrons and of positrons were kept "single" (i.e., not colliding) in the machine. For a beam energy of 80~GeV, the polarization time would be 9 hours in TLEP, and the use of wigglers should not be necessary.       

\subsubsection{Longitudinal polarization}

\begin{figure}[tb]
\begin{center}
\includegraphics[width=0.7\columnwidth]{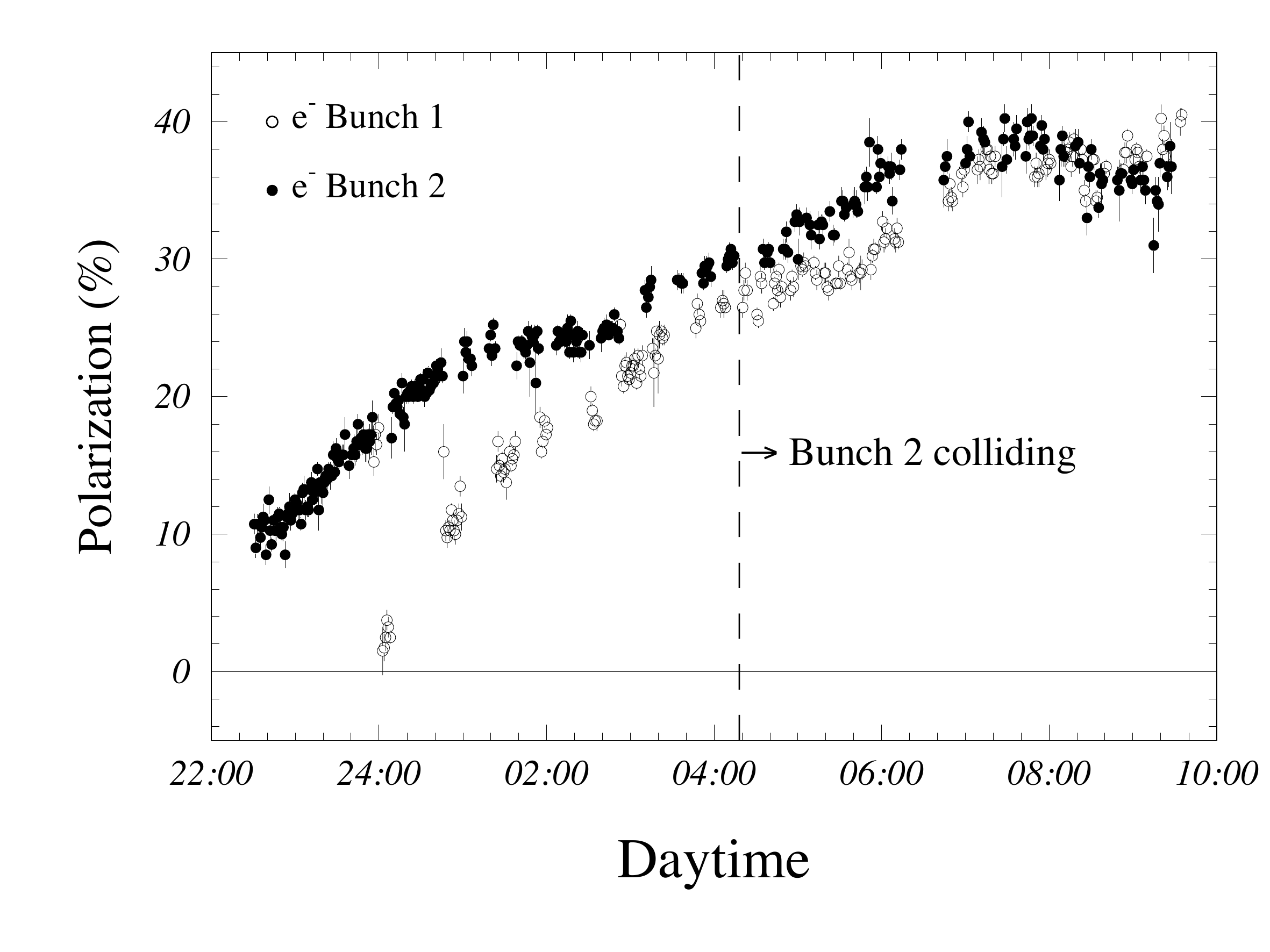}
\end{center}
\caption{\label{fig:polarization} Proof of principle for polarization in collisions around the Z pole energy at LEP ($E_{\rm beam}=$ 44.71 GeV). The measured transverse polarization of the two electron bunches is plotted as function of time. One of the two electron bunches was brought into collisions with a positron bunch at 4:10 am, and remained polarized at the same level as the non-colliding bunch for more than five hours afterwards.}
\end{figure}

Measurements with longitudinal polarization require maintaining polarization of both ${\rm e}^+$ and ${\rm e}^-$ beams in collisions. At LEP, transverse beam polarization of 40\% was observed and maintained in collisions for more than five hours at Z pole energies ($\sim 45$ GeV per beam) with one collision point, a beam-beam tune shift of 0.04, and a single bunch luminosity of $10^{30}~\cms$~\cite{cite:LEP-beam-beam-pol}. The polarization levels measured during this experiment are displayed in Fig.~\ref{fig:polarization} as a function of time. With the smaller value of $\beta_y^{*}$ and the larger number of bunches, similar polarization levels could be envisioned in collisions at the Z pole with TLEP with a luminosity reduced to around $10^{35}~\cms$, for the same total beam-beam tune shift. A suitable working point will have to be found to optimize the benefits from the much reduced top-up rate, and the adverse effects of the beamstrahlung and the required polarization wigglers on the energy spread.

Movable spin rotators as designed for HERA~\cite{cite:HERA-beams} would therefore allow a program of longitudinally polarized beams at the Z peak. (The spin rotator design foreseen for LEP requires tilting the experiments and is unpractical for TLEP.) For the same level of polarization in collisions as that observed at LEP, and assuming that a fraction of the bunches can be selectively depolarized, a simultaneous measurement~\cite{Blondel:1987wr} of the beam polarization and of the left-right  asymmetry $\ALR$ can be envisioned at TLEP. 

\subsubsection{Polarization at higher energies}

\begin{figure}[tb]
\begin{center}
\includegraphics[width=0.7\columnwidth]{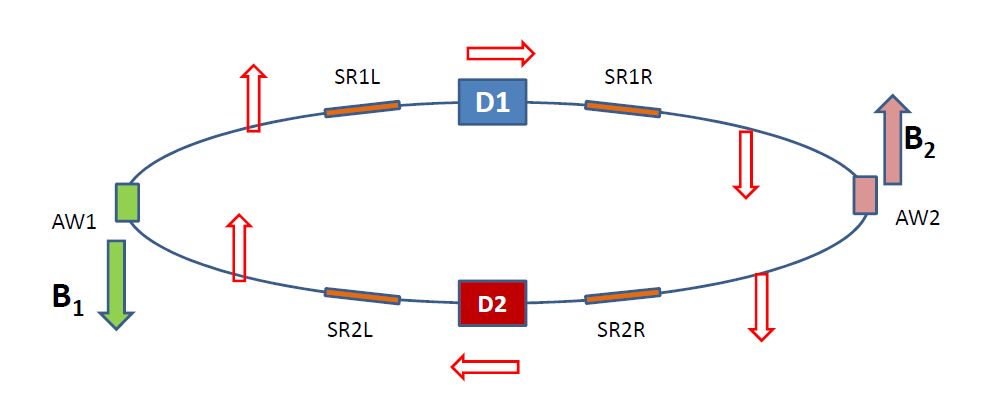}
\end{center}
\caption{\label{fig:longitudinal} A possible scheme to obtain longitudinal beam polarization at high energies  ($E_{\rm beam} \gg m_{\rm Z}/2$ ) with TLEP. Taking advantage of the low magnetic field in the arcs, the polarization is generated dominantly by strong asymmetric wigglers of opposite polarities (AW1 and AW2) in two halves of the ring. The transverse polarization obtained this way is rotated to longitudinal in the experimental straight sections  in detector D1,  by 90 degrees spin rotators (SR1L, etc.), and brought back to vertical (but reversed) in the following arc, and similarly for the next experimental straight section, D2. The scheme easily generalizes to the situation with four IPs. This scheme generates a spin transport with an integer part of the spin tune equal to zero. The spin polarization of the electrons is shown.  Given separated beam pipes for the  ${\rm e}^+$ and ${\rm e}^-$ beams, they can be exposed to wigglers of opposite polarity, enabling  positron polarization parallel to that of the electrons. In this way highly polarized $\epem$ systems at the collision point can be obtained. Polarization can be reversed by reversing the wiggler polarity. The possibility of depolarizing a fraction of the bunches in this scheme, to provide a normalization of polarimetry from the measured cross-sections, is being investigated. }
\end{figure}

As mentioned above, the maximum level of polarization is limited by the increase of the beam energy spread when the beam energy increases. The establishment of longitudinal polarization at higher energies therefore requires a cancellation of depolarizing effects, by reducing the spin-tune spread associated with the energy spread. Siberian snake solutions \cite{cite:Wienans-TLEP4} invoking combinations of spin rotators situated around the experiments and polarization wigglers are being discussed. They take advantage of the fact that the TLEP arcs have very low fields, which can be overruled by polarization wigglers suitably disposed around the ring. An example is displayed in Fig.~\ref{fig:longitudinal}.  These schemes need to be worked out and simulated before the feasibility of longitudinal polarization in high-energy collisions can be asserted.

\subsection{Beam energy measurement}

As mentioned in Section~\ref{sec:beampol}, transverse polarization can be naturally established at TLEP at the Z pole and at the WW threshold. A technique unique to $\epem$ rings, called resonant spin depolarization~\cite{on_Blondel_Assmann_Dehning_1992}, can therefore be used to measure the beam energy with high precision. This technique was developed and successfully used during the LEP1 programme, and allowed the average beam energy to be known with a precision of 1 MeV. The intrinsic precision of the method, 0.1 MeV or better~\cite{LEP1Cal}, was not fully exploited  at LEP1 because no attempt was made to perform this measurement during collisions. Instead, regular measurements were performed by separating the beams at the end of physics fills, and it was soon realized that the energy actually drifted with time because of, e.g., tides and stray currents from nearby train tracks, so that an extrapolation had to be made to {\textquotedblleft}predict{\textquotedblright} the beam energy during collisions. This extrapolation is the dominant contributor to the current systematic uncertainty of 2 MeV on the Z mass and width~\cite{LEP1Cal}.

At TLEP, instead, it will be possible to keep a few non-colliding bunches out of the 4400 (Z pole) or 600 (WW threshold) bunches without significant loss of luminosity, and apply regular resonant spin depolarization on those. This technique will allow continuous beam energy measurements, in the exact same conditions as for the colliding bunches, with an accuracy of 100 keV or so for each measurement, hence with an accuracy of 100 keV/$\sqrt{N}$ for $N$ measurements. With the statistics foreseen to be available at the Z pole, a precision better than 0.1 MeV will therefore be at hand for the Z mass and width measurements. Similarly, at the WW threshold, the beam energy uncertainty should translate to a systematic uncertainty smaller than 0.1 MeV on the W mass.

If polarization cannot be established at higher centre-of-mass energies, the beam energy can be determined from the precise knowledge of the Z and W masses and the use of the energy-momentum conservation in kinematic fits of the $\epemto {\rm Z}\gamma$~\cite{cite:0506115}, $\epemto {\rm ZZ}$, and $\epemto {\rm WW}$ processes. These three processes should allow the average beam energy (and its spread) to be determined at $\sqrt{s} = 240$ and $350$ GeV with a precision sufficient for all practical purposes.

\subsection{Integrated luminosity measurement}

The experimental conditions at TLEP will be similar to those of LEP, with the additional bonus of very stable beam conditions brought by the continuous top-up injection. Nevertheless, there will be a number of notable differences, as exemplified below.
\begin{itemize}
\item The smaller value of $\beta^*_y$~\cite{cite:1305.6498} increases the beam divergence at the interaction point to the extent that it may have a sizeable effect on the acceptance of low angle detectors used for the luminosity measurement. The better stability of the TLEP beams will help to keep the uncertainty on the beam divergence to a level similar to that evaluated at LEP. 
\item The strong final-focus quadrupoles will generate large amounts of synchrotron radiation, which need to be simulated and against which appropriate shielding must be provided. 
\item An increased amount of beamstrahlung may lead to a somewhat larger background of electromagnetic radiation produced in the interaction region. As mentioned in Section~\ref{sec:beamstrahlung}, it is nevertheless several orders of magnitude smaller than the level expected in a linear collider environment. 
\item The repetition rate in multi-bunch operations will reach 20 MHz at the Z pole. This specificity has to be taken into account in the design of the detectors. 
\end{itemize}

To the extent that the aforementioned issues are properly addressed and solved, there should be no significant difficulty to achieve luminosity measurements with an experimental precision similar to that obtained at LEP, typically a few times $10^{-4}$. At the Z peak it would be of interest to achieve even better precision, e.g., for the measurement of the invisible width hence the number of light neutrinos, which will require a more precise construction of the luminometers. The main limitation on the luminosity measurement, however,  would presently come from the theoretical calculation of the low angle Bhabha cross section. Clearly, progress in this aspect would pay great dividends.

\subsection{Detectors}

The detector designs developed for the ILC~\cite{ILC:Detectors} or for CLIC~\cite{cite:CLICDR} include a highly granular calorimetry, called imaging calorimetry, for particle-flow purposes. The 3D granularity allows hadron showers to be tracked individually, towards an optimally efficient neutral hadron identification, hence a better energy resolution for jets. This technical choice, however, poses power dissipation and cooling challenges. The solution of pulsed electronics, chosen for linear colliders, cannot be exploited at circular colliders because of the large repetition rate.

While the use of imaging calorimetry will be included in the forthcoming design study, more conservative choices have therefore been made so far in the evaluation of the TLEP physics case potential. For example, a study -- carried out in Ref.~\cite{cite:1208.1662} with full simulation of the CMS detector at $\sqrt{s} = 240$ GeV -- demonstrated that the Higgs coupling accuracy is close to being optimal  even with a more conventional detector. The underlying reason is that the precise measurement of jet energies is most often not a key factor in $\epem$ collisions: for events with no or little missing mass, jet energies can be determined with high precision from their directions, making use of energy-momentum conservation. 

The TLEP design study will aim, in particular, at defining the minimal detector performance needed to measure the Higgs boson couplings and the EWSB parameters with the desired precision. In the meantime, the  choice made in Ref.~\cite{cite:1208.1662} was adopted in this note too to make a conservative estimate of the TLEP potential: the performance of the CMS detector is assumed throughout. The only exceptions are {\it (i)} the vertex detector, for which performance similar to that of a linear collider detector is needed, with lifetime-based c-tagging capabilities; and {\it (ii)} a precision device for luminosity measurement with Bhabha scattering, obviously absent in the CMS design. The estimates presented in this note are based on the simultaneous operation of four of these detectors. As mentioned in Section~\ref{sec:luminosity}, a configuration with only two such detectors would lead to a moderate 20\% increase of all statistical uncertainties presented here, as summarized in Tables~\ref{tab:FitResults} and~\ref{tab:EW-TLEP}. 

A specificity of TLEP is the possibility to run at the Z pole with a luminosity of $5\times 10^{35}\,\cms$ at each interaction point, corresponding to a trigger rate of 15 kHz for Z decays in the central detector, and 60 kHz for Bhabha scattering in the luminometer. This rate is of the same order of magnitude as that proposed for the LHCb upgrade~\cite{Bediaga}, with events of a size similar or larger than the size of the TLEP events. In addition, the events will be as ``clean'' as at LEP, with no pile-up interactions and negligible beam backgrounds. No insurmountable difficulty is therefore expected in this respect, but the design study will need to ascertain the data analysis feasibility, and to assess the needs for online and offline computing resources with such trigger rates.

\subsection{Possible timescale and physics programme}
The design study is expected to deliver its conclusion in 2018, in time for the next update of the European Strategy. The TLEP and the VHE-LHC design studies  will be conducted in close coordination, with the aim of providing maximum flexibility for the installation of the two machines and possible concurrent (but not simultaneous) operation. Should the case be still as strong as today, a go-ahead decision could be taken immediately and the tunnel excavation could start at the beginning of the next decade, for a duration of four to eight years, with the simultaneous operation of up to three drilling machines~\cite{OsbornePrivate}. The construction and installation of the collider and the detectors would then proceed in parallel with the HL-LHC running for another four to five years. It could thus be technically envisioned, setting aside political, financial, etc., considerations, to start commissioning for the first TLEP physics run as early as in 2030. It will take between a couple months (as at LEP2) and a couple years (as at LEP1) to achieve the design luminosity.

Typically, the baseline physics programme of TLEP would consist of 
\begin{itemize}
\item two years at the Z pole (of which one year with the design luminosity of $5.6~\inab$ at each IP, and one year with longitudinal polarization at reduced luminosity), with resonant depolarization of single bunches at intervals of around 20 minutes, for beam energy calibration;
\item one or two years at the WW threshold -- with $1.6~\inab$ per year at each IP -- with periodic returns at the Z peak (in the TLEP-W conditions) for detector calibration, and with resonant depolarization of single bunches at intervals of around 20 minutes, for beam energy calibration; 
\item five years at 240 GeV as a Higgs factory -- with $500~\infb$ per year at each IP -- with periodic returns at the Z peak (in the TLEP-H conditions); 
\item and five years at the $\ttbar$ threshold -- with $130~\infb$ per year at each IP -- with periodic returns at the Z peak (in the TLEP-t conditions).
\end{itemize} 
The effective duration of the running at each energy as well as the appropriate order will be defined according to the physics needs and the collider capacities as more knowledge is acquired. Possible luminosity and energy upgrades are not included in this baseline programme. In this aggressive schedule, the VHE-LHC would be installed in the 2040's, and its physics programme could start in 2050 or thereabout. 

\subsection{Elements of costing}
One of the aims of the design study is to produce a detailed costing of the TLEP project. Not surprisingly, the main cost drivers for the whole complex are expected to be the tunnel, the shafts and the related services and infrastructure (including access roads). The corresponding cost, however, is considered as general CERN infrastructure to serve both TLEP and VHE-LHC, and possibly other projects as well. The length of the tunnel will be optimized on the basis of geological and accessibility criteria. For example, a tunnel of 100 km (also shown in Fig.~\ref{fig:TLEP80}, and for which a feasibility assessment is ongoing) might be more cost-effective than the 80~km version~\cite{OsbornePrivate}. 

Besides, the cost of the accelerator and collider rings, dominated by the 600-m-long RF system and the 80~km of low-field magnets -- possibly recyclable for the VHE-LHC injector -- was found in a very prelimimary estimate to be smaller than the LHC cost (Table~\ref{tab:cost}). In view of the large number of Higgs bosons, Z and W bosons, and top quarks to be analysed in very clean experimental conditions, TLEP is therefore expected to be exceedingly competitive. 

\begin{table}
\begin{center}
\caption{Indicative costs for the main cost drivers of the TLEP collider.\label{tab:cost}}
\begin{tabular}{|r|r|}
\hline Item                          & Cost (Million CHF) \\ \hline 
\hline RF system & 900 \\
\hline Cryogenics system & 200 \\
\hline Vacuum system & 500 \\
\hline Magnets systems for the two rings & 800 \\
\hline Pre-injector complex & 500 \\ \hline 
{\bf Total} & {\bf 2,900} \\ \hline
\end{tabular}
\end{center}
\end{table}


\section{Precise measurements of the Higgs boson properties}
\label{sec:Higgs}
The primary goal of a Higgs factory is to measure the Higgs boson properties with the best possible precision as to be sensitive to physics beyond the Standard Model at the highest possible scale. Tree-level couplings of the Higgs boson to fermions and gauge bosons are expected to be modified with respect to the standard-model prediction, with a magnitude rapidly decreasing with the new physics scale $\Lambda$, typically like $1/\Lambda^2$. For $\Lambda = 1$ TeV, departures up to 5\% are expected~\cite{ILC:Physics,Gupta_Rzehak_Wells_2012}. To discover new physics through its effects on the Higgs boson couplings with a significance of 5$\sigma$, it is therefore necessary to measure these couplings to fermions and gauge bosons with a precision of at least 1\%, and at the per-mil level to reach sensitivity to $\Lambda$ larger than 1 TeV, as suggested at by the negative results of the searches at the LHC. 

\begin{figure}[tb]
\begin{center}
\includegraphics[width=0.9\columnwidth]{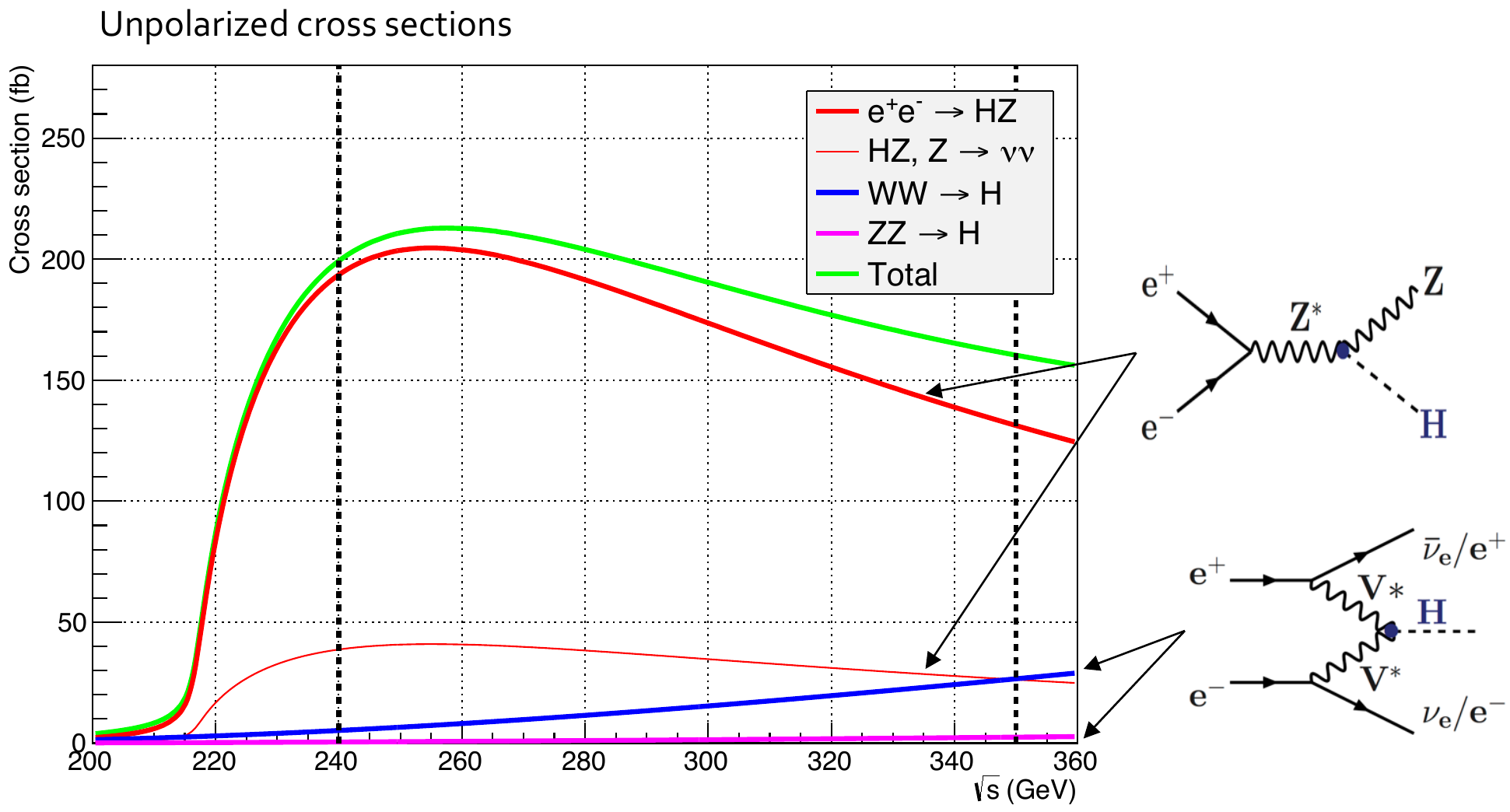}
\end{center}
\caption{\label{fig:HiggsCross} The Higgs boson production cross section as a function of the centre-of-mass energy in unpolarized $\epem$ collisions, as predicted by the {\tt HZHA} program~\cite{cite:HZHA}. The thick red curve shows the cross section expected from the Higgs-strahlung process $\epemto {\rm HZ}$, and the thin red curve shows the fraction corresponding to the ${\rm Z}\to \nu\bar\nu$ decays. The blue and pink curves stand for the WW and ZZ fusion processes (hence leading to the ${\rm H}\nu_{\rm e}\bar\nu_{\rm e}$ and ${\rm H}\epem$ final states), including their interference with the Higgs-strahlung process. The green curve displays the total production cross section. The dashed vertical lines indicate the centre-of-mass energies at which TLEP is expected to run for five years each, $\sqrt{s} = 240$ GeV and $\sqrt{s} \sim 2m_{\rm top}$.}
\end{figure}

The number of Higgs bosons expected to be produced, hence the integrated luminosity delivered by the collider, are therefore key elements in the choice of the right Higgs factory for the future of high-energy physics:  a per-mil accuracy cannot be reached with less than a million Higgs bosons. The Higgs production cross section (obtained with the {\tt HZHA} generator~\cite{cite:HZHA}), through the Higgs-strahlung process $\epemto {\rm HZ}$ and the ${\rm WW}$ or ${\rm ZZ}$ fusion processes, is displayed in Fig.~\ref{fig:HiggsCross}. A possible operational centre-of-mass energy is around 255 GeV, where the total production cross section is maximal and amounts to 210~fb. 

\begin{figure}[tb]
\begin{center}
\includegraphics[width=0.7\columnwidth]{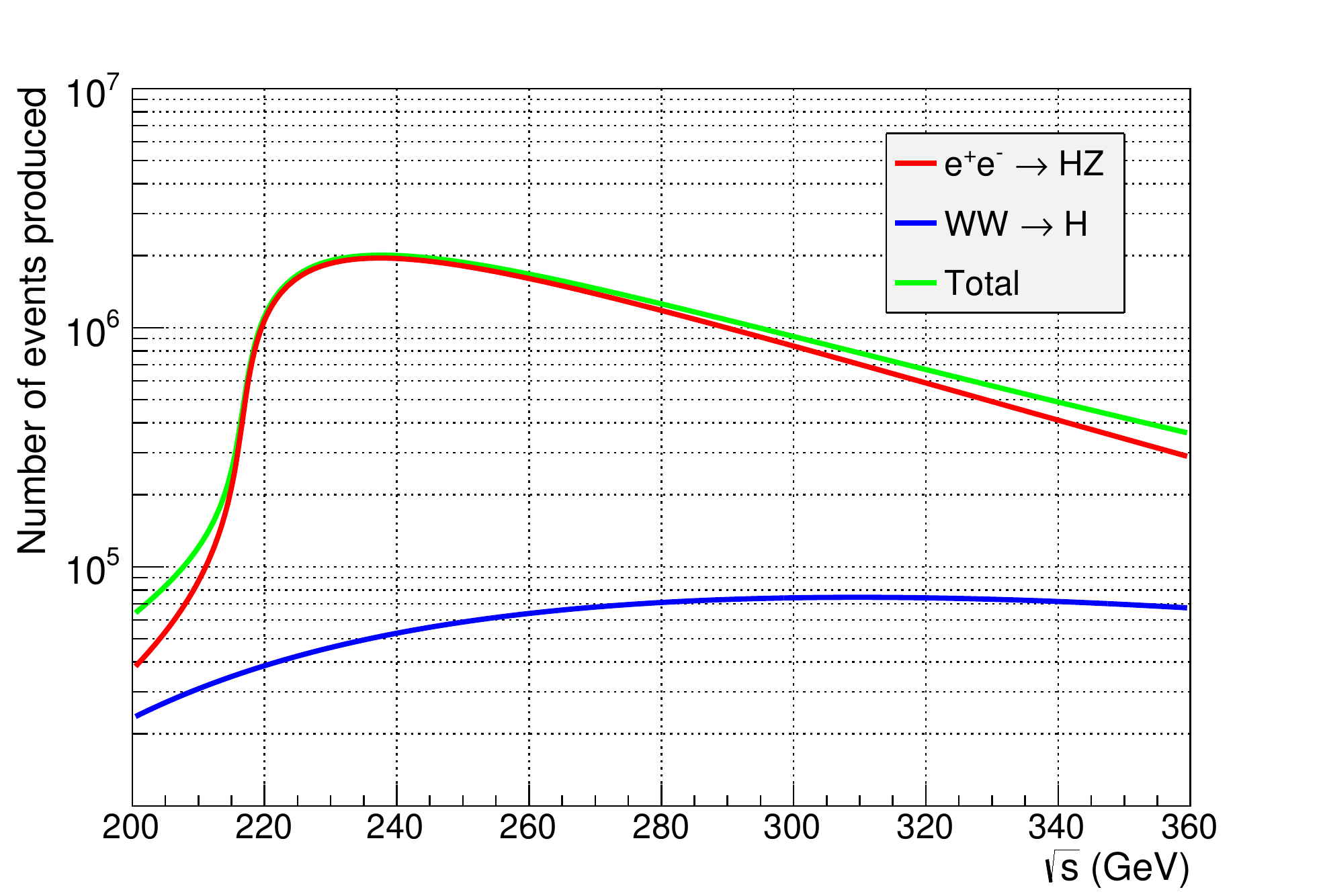}
\end{center}
\caption{\label{fig:NumberOfHiggsEvents} Number of Higgs bosons produced at TLEP as a function of the centre-of-mass energy (green curve), as obtained from a five-year running period with the TLEP luminosity profile of Fig.~\ref{fig:lumi} delivered to four interaction points, and the Higgs production cross section of Fig.~\ref{fig:HiggsCross}. The number of events from the Higgs-strahlung process $\epemto {\rm HZ}$ is displayed in red, and the number of events from WW fusion is displayed in blue.}
\end{figure}

The luminosity profile of TLEP as a function of the centre-of-mass energy (Fig.~\ref{fig:lumi}) leads to choose a slightly smaller value, around $240$ GeV, where the total number of Higgs bosons produced is maximal, as displayed in Fig.~\ref{fig:NumberOfHiggsEvents}. The number of WW fusion events has a broad maximum for centre-of-mass energies between 280 and 360 GeV. It is therefore convenient to couple the analysis of the WW fusion with the scan of the $\ttbar$ threshold, at $\sqrt{s}$ around 350 GeV, where the background from the Higgs-strahlung process is smallest and most separated from the WW fusion signal.

\subsection{Measurements at $\sqrt{s}=240$ GeV}

At $\sqrt{s} = 240$ GeV, the TLEP luminosity is expected to be $5\times 10^{34}~\cms$ at each interaction point, in a configuration with four IPs. The total integrated luminosity accumulated in five years, assuming running for $10^7$ seconds per year, is shown in Table~\ref{tab:HiggsCross}, together with the corresponding numbers of Higgs bosons produced. 

\begin{table}
\begin{center}
\caption{Integrated luminosity and number of Higgs bosons produced with TLEP at $\sqrt{s} = 240$ GeV (summed over four IPs), for the Higgs-strahlung process and the WW fusion. For illustration, the corresponding numbers are also shown  for the baseline ILC programme~\cite{ILC:Summary} at $\sqrt{s} = 250$~GeV, with beams polarized at a level of 80\% for electrons and 30\% for positrons.\label{tab:HiggsCross}}
\begin{tabular}{|r|r|r|}
\hline & TLEP 240 & ILC 250 \\ \hline 
\hline Total Integrated Luminosity ($\inab$) & {\bf 10} & 0.25 \\ 
\hline Number of Higgs bosons from $\epemto {\rm HZ}$ & {\bf 2,000,000} & 70,000 \\
\hline Number of Higgs bosons from boson fusion & {\bf 50,000} & 3,000 \\ \hline 
\end{tabular}
\end{center}
\end{table}

\begin{figure}[tb]
\begin{center}
\includegraphics[width=0.7\columnwidth]{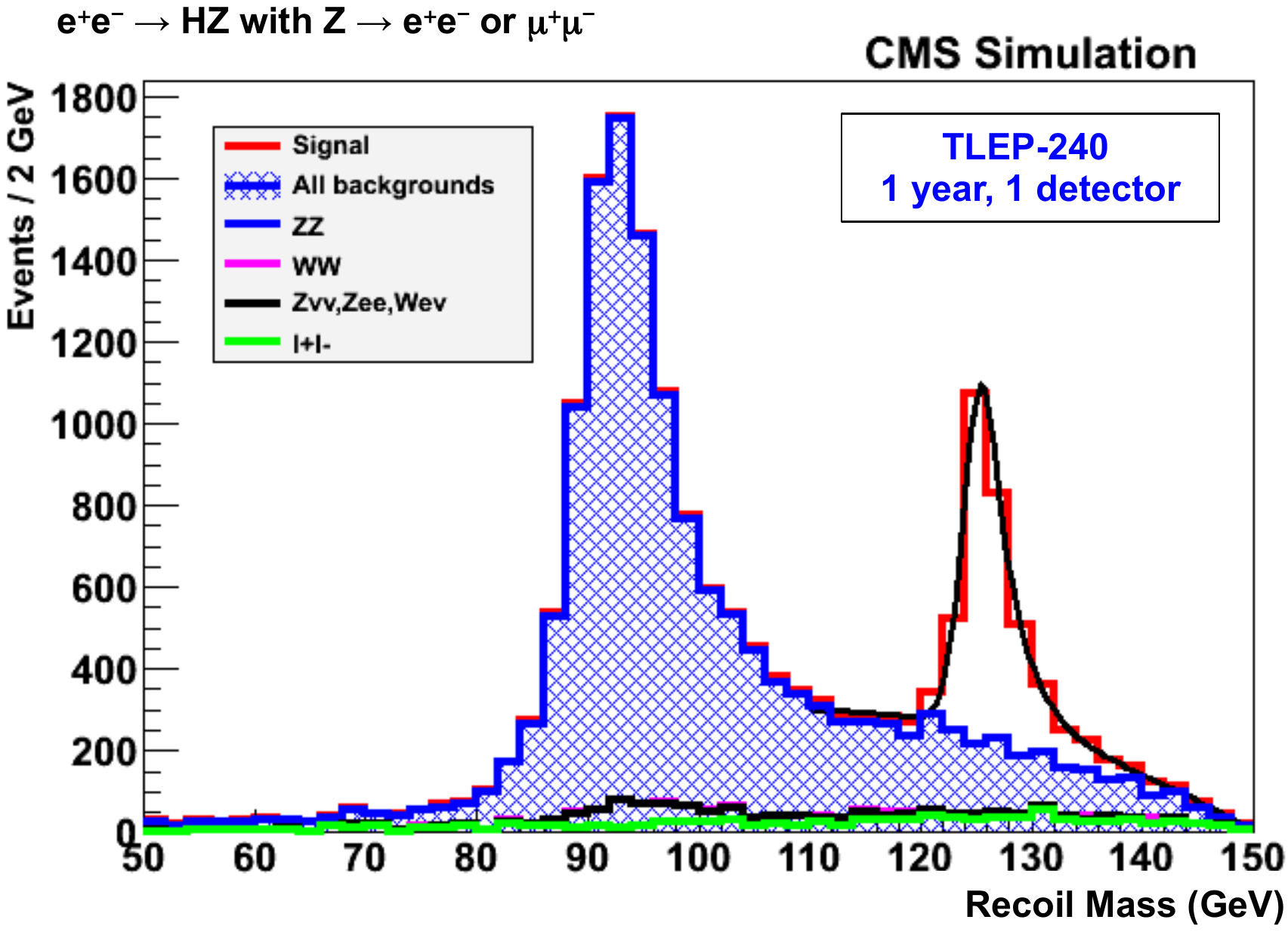}
\end{center}
\caption{\label{fig:Recoil}
Distribution of the mass recoiling against the lepton pair in the $\epemto {\rm HZ}$ channel, in the ${\rm Z} \to \ell^+\ell^-$ final state ($\ell = {\rm e}, \mu$), taken from Ref.~\cite{cite:1208.1662}, for an integrated luminosity equivalent to one year of data taking with one TLEP detector (assumed to be the CMS detector). The number of Higgs boson events (the red histogram) obtained from a fit of this distribution is proportional to the inclusive HZ cross section, $\sigma_{\rm HZ}$. }
\end{figure}

From the sole reading of this table, it becomes clear that TLEP is in a position to produce enough Higgs bosons in a reasonable amount of time to aim at the desired sub-per-cent precision for Higgs boson coupling measurements. Detailed simulations and simple analyses have been carried out in Ref.~\cite{cite:1208.1662}  to ascertain the claim, with an integrated luminosity of $500~\infb$ (representing only one year of data taking at $\sqrt{s} = 240$ GeV in one of the TLEP detectors), fully simulated in the CMS detector. For example, the distribution of the mass recoiling against the lepton pair in the $\epem{\rm H}$ and $\mpmm{\rm H}$ final states, independently of the Higgs boson decay, is shown in Fig.~\ref{fig:Recoil}, taken from Ref.~\cite{cite:1208.1662}, for one year of data taking in the CMS detector. The number of Higgs boson events obtained from a fit to this distribution of the signal and background contributions allows the total $\epemto {\rm HZ}$ cross section to be measured with a precision of 0.4\% at TLEP. As pointed out in Ref.~\cite{1305.5251}, the measurement of the total $\epemto {\rm HZ}$ cross section is a sensitive probe of possible new physics that can reduce the fine-tuning of the Higgs boson mass. Such new physics would also renormalize the Higgs couplings by a universal factor, and the TLEP measurement of the $\epemto {\rm HZ}$ cross section with a precision of 0.4\% would be sensitive to new particles that could not be meaningfully probed in any other way.

A summary of the statistical precision of the measurements presented in Ref.~\cite{cite:1208.1662} for $\sqrt{s} = 240$ GeV -- extrapolated to the TLEP luminosity and to four detectors -- is given in Table~\ref{tab:HiggsBranching}. In this table, a few numbers are added with respect to Ref.~\cite{cite:1208.1662}. First, the precision for $\sigma_{\rm HZ} \times {\rm BR(H\to c\bar c)}$ and $\sigma_{\rm HZ} \times {\rm BR(H\to gg)}$ is extrapolated from the ILC prediction, as would be obtained if the CMS detector were upgraded with a vertex detection device with adequate c-tagging performance. Secondly, the precision for  $\sigma_{\rm HZ} \times {\rm BR(H\to ZZ)}$ is obtained from an almost background-free dedicated search for ZZZ final states including four leptons, recently developed for that purpose.

\begin{table}
\begin{center}
\caption{Statistical precision for Higgs measurements obtained from  the proposed TLEP programme at $\sqrt{s} = 240$ GeV only (shown in Table~\ref{tab:HiggsCross}). For illustration, the baseline ILC figures at $\sqrt{s}=250$ GeV, taken from Ref.~\cite{ILC:Physics}, are also given. The order-of-magnitude smaller accuracy expected at TLEP in the ${\rm H}\to\gamma\gamma$ channel is the threefold consequence of the larger luminosity, the superior resolution of the CMS electromagnetic calorimeter, and the absence of background from Beamstrahlung photons.\label{tab:HiggsBranching}}
\begin{tabular}{|r|r|r|}
\hline & TLEP 240 & ILC 250 \\ \hline
\hline $\sigma_{\rm HZ}$ & {\bf 0.4\%} & 2.5\% \\ 
\hline $\sigma_{\rm HZ} \times {\rm BR(H\to b\bar b)}$ & {\bf 0.2\%} & 1.1\% \\
\hline $\sigma_{\rm HZ} \times {\rm BR(H\to c\bar c)}$ & {\bf 1.2\%} & 7.4\% \\
\hline $\sigma_{\rm HZ} \times {\rm BR(H\to gg)}$ & {\bf 1.4\%} & 9.1\% \\
\hline $\sigma_{\rm HZ} \times {\rm BR(H \to WW)}$ & {\bf 0.9\%} & 6.4\% \\
\hline $\sigma_{\rm HZ} \times {\rm BR(H} \to \tau\tau{\rm )}$ & {\bf 0.7\%} & 4.2\% \\
\hline $\sigma_{\rm HZ} \times {\rm BR(H\to ZZ)}$ & {\bf 3.1\%} & 19\% \\
\hline $\sigma_{\rm HZ} \times {\rm BR(H}\to \gamma\gamma{\rm )}$ & {\bf 3.0\%} & 35\% \\
\hline $\sigma_{\rm HZ} \times {\rm BR(H}\to \mu\mu{\rm )}$ & {\bf 13\%} & 100\% \\ \hline 
\end{tabular} 
\end{center}
\end{table}

The latter measurement has an important consequence for the determination of the total Higgs decay width. In $\epem$ collisions, it is not possible to directly observe the width of the Higgs boson if it is as small as the Standard Model prediction of 4 MeV. However, the total width of the Higgs boson is given by $\Gamma_{\rm tot} = \Gamma {\rm(H \to ZZ)}/{\rm BR(H \to ZZ)}$. As the partial decay width $\Gamma{\rm(H \to ZZ)}$ is directly proportional to the inclusive cross section $\sigma_{\rm HZ}$, $\Gamma_{\rm tot}$ can be  measured with the same precision as the ratio $\sigma_{\rm HZ}^2/\sigma_{\rm HZ} \times {\rm BR(H\to ZZ)}$. Therefore, with the sole 240 GeV data, TLEP is able to determine the Higgs boson decay width with a precision of the order of 3.1\% from this channel. The ${\rm H \to b\bar b}\nu\bar\nu$ final state produced via WW fusion can also be used for that purpose, as described in more detail in the next section. 

Finally, the $\ell^+\ell^-{\rm H}$ final state and the distribution of the mass recoiling against the lepton pair can also be used to directly measure the invisible decay width of the Higgs boson, in events where the Higgs boson decay products escape undetected. With the TLEP data at 240 GeV, the Higgs boson invisible branching fraction can be measured with an absolute precision of 0.25\%. If not observed, a 95\% C.L. upper limit of 0.5\% can be set on this branching fraction. 

\subsection{Measurements at $\sqrt{s}=350$ GeV}

At $\sqrt{s} = 350$ GeV, the TLEP luminosity is expected to amount to $1.3\times 10^{34}~\cms$ at each IP.  The total integrated luminosity accumulated in five years is shown in Table~\ref{tab:HiggsCross350}, together with the corresponding numbers of Higgs bosons produced. 

\begin{table}
\begin{center}
\caption{Integrated luminosity and numbers of Higgs bosons produced with TLEP (summed over four IPs) at $\sqrt{s} = 350$ GeV, in the Higgs-strahlung process and in WW fusion. For illustration, the corresponding numbers are also shown  for the baseline ILC programme at the same centre-of-mass energy,  with beams polarized at a level of 80\% for electrons and 30\% for positrons.\label{tab:HiggsCross350}}
\begin{tabular}{|r|r|r|}
\hline & TLEP 350 & ILC 350 \\ \hline
\hline Total Integrated Luminosity ($\inab$) & {\bf  2.6} & 0.35\\ 
\hline Number of Higgs bosons from $\epemto {\rm HZ}$ & {\bf 340,000} & 65,000 \\
\hline Number of Higgs bosons from boson fusion & {\bf 70,000} & 22,000 \\ \hline 
\end{tabular}
\end{center}
\end{table}

\begin{figure}[tb]
\begin{center}
\includegraphics[width=0.5\columnwidth]{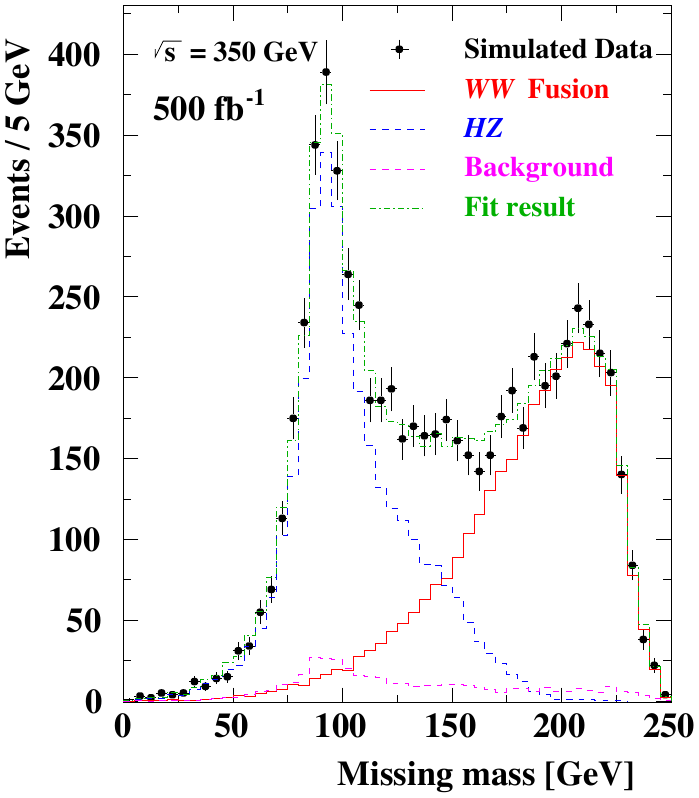}
\end{center}
\caption{\label{fig:MissingMass} Distribution of the mass recoiling against the ${\rm b\bar b}$ system in the ${\rm b\bar b}\nu\bar\nu$ final state, from Higgs-strahlung (blue) and WW-fusion (red) production for $500~\infb$ at $\sqrt{s} = 350$~GeV, taken from Ref.~\cite{cite:durig}.}
\end{figure}

The additional events from the Higgs-strahlung process at 350 GeV allow the statistical precision for all the aforementioned measurements to be improved by typically 5\% for TLEP with respect to the sole 240 GeV data. The large number of Higgs bosons produced by boson fusion allows a measurement of the total width, most straightforwardly done in the copious ${\rm b\bar b}\nu\bar\nu$ final state. At $\sqrt{s}=350$~GeV, both the Higgs-strahlung process (when the Z decays to a neutrino pair) and the WW fusion contribute to this final state with a similar cross section (Fig.~\ref{fig:HiggsCross}), and with a small interference term. The mass recoiling against the ${\rm b \bar b}$ system (also called missing mass), however, peaks at $m_{\rm Z}$ for the Higgs-strahlung and the interference term, but clusters around $\sqrt{s}-m_{\rm H}$ for the WW fusion. A fit of the HZ and WW fusion contributions to the distribution of this missing mass, shown in Fig.~\ref{fig:MissingMass} from Ref.~\cite{cite:durig}, allows  $\sigma_{{\rm WW \to H}} \times {\rm BR(H \to b\bar b)}$ to be obtained with a relative precision of 0.6\% at TLEP.

This measurement can be performed in a very similar manner with the data at $\sqrt{s} = 240$ GeV, albeit with a reduced discrimination between the HZ and the WW fusion contributions. The statistical precision with which $\sigma_{{\rm WW \to H}} \times {\rm BR(H \to b\bar b)}$ can be measured at both centre-of-mass energies is displayed in Table~\ref{tab:WWFusion}. 

\begin{table}
\begin{center}
\caption{Statistical precision of the TLEP measurement of  $\sigma_{{\rm WW \to H}} \times {\rm BR(H \to b\bar b)}$. For illustration, the ILC potential at the same centre-of-mass energies is also indicated.\label{tab:WWFusion}}
\begin{tabular}{|r|r|r|}
\hline $\sqrt{s}$ (GeV) & TLEP & ILC \\ \hline
\hline 240 - 250  & {\bf 2.2\%} & 10.5\% \\
\hline 350   & {\bf 0.6\%} & 1.0\% \\ \hline 
\end{tabular} 
\end{center}
\end{table}
These measurements can also be used to determine the total Higgs decay width in a way similar to that described in the previous section. Indeed, the total Higgs boson width is given by $\Gamma_{\rm tot} = \Gamma {\rm(H \to WW)}/{\rm BR(H \to WW)}$. The partial decay width $\Gamma{\rm(H \to WW)}$ is directly proportional to the inclusive cross section $\sigma_{\rm WW \to H}$. The Higgs boson branching ratios to WW and to ${\rm b\bar b}$ are in turn obtained from the measurements performed at $\sqrt{s} = 240$ GeV, the precision of which can be inferred from Table~\ref{tab:HiggsBranching}. With the 350 (240) GeV data, TLEP is therefore able to determine the Higgs boson decay width with a precision of the order of 1.2\% (2.4\%) with WW fusion. When combined with the ZZZ final state, the precision on the total Higgs boson width from TLEP is estimated to be 1.0\%. These numbers are summarized in Table~\ref{tab:Width}.
\begin{table}
\begin{center}
\caption{Statistical precision of the total Higgs boson width measurements with TLEP at $\sqrt{s}=$ 240 and 350 GeV. For illustration, the ILC potential at the same centre-of-mass energies is also indicated.\label{tab:Width}}
\begin{tabular}{|r|r|r|}
\hline Process and final state & TLEP & ILC \\ \hline
\hline $\epemto {\rm HZ}$ with ${\rm H \to ZZ}$ & {\bf 3.1\%} & 20\% \\
\hline ${\rm WW \to H}$ with ${\rm H \to b\bar b}$ at 240 GeV  & {\bf 2.4\%} & 12\% \\
\hline ${\rm WW \to H}$ with ${\rm H \to b\bar b}$ at 350 GeV  & {\bf 1.2\%} & 7\% \\ \hline
\hline {\bf Combined} & {\bf 1.0\%} & 6.0\% \\ \hline 
\end{tabular} 
\end{center}
\end{table}

\subsection{Global fit for Higgs boson couplings}
\label{sec:FitResults}

\begin{figure}[tb]
\begin{center}
\includegraphics[width=0.7\columnwidth]{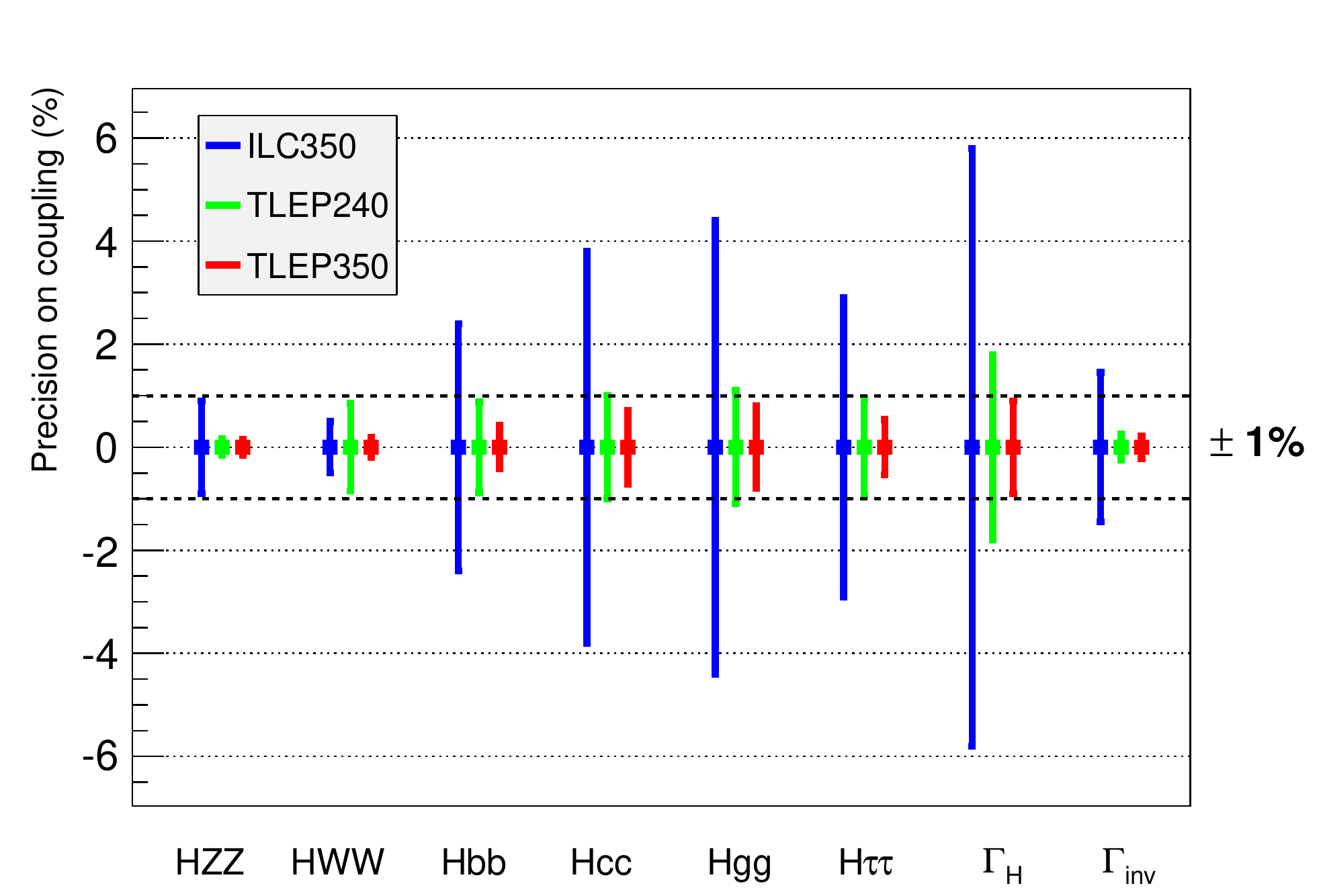}
\end{center}
\caption{\label{fig:FitResults}  Relative statistical uncertainty on the Higgs boson couplings from a truly model-independent fit, as expected from two five-year-long running periods at $\sqrt{s} = $ 240-250 and 350 GeV for TLEP and ILC. The red and blue bars correspond to the combination of the data at 240-250 GeV and 350 GeV, while the green bars hold for the sole period of TLEP at 240 GeV. The dashed lines show the $\pm 1\%$ band, relevant for sensitivity to multi-TeV new physics. Also indicated are the expected uncertainties on the total decay width and on the invisible decay width. The H$\mu\mu$ and H$\gamma\gamma$ coupling uncertainties, which do not fit in the $\pm 6\%$ scale of the figure for ILC, can be read off Table~\ref{tab:FitResults}.}
\end{figure}

The accuracies on the Higgs boson couplings are obtained here from a fit to all observables reported in Tables~\ref{tab:HiggsBranching} and~\ref{tab:WWFusion} for TLEP at $\sqrt{s}=$ 240 and 350 GeV. The fit closely follows the logic presented in Ref.~\cite{cite:1207.2516}, and indeed reproduces the results presented therein for the combination of the ILC and LHC projections. Here, the results of standalone fits, i.e., without combination with LHC sensitivities, are given so as to compare the LHC, ILC and TLEP relative performance in terms of Higgs boson coupling and width measurements. The other two assumptions made in Ref.~\cite{cite:1207.2516} consist in  {\it (i)} bounding from above the couplings to the Z and the W to the Standard Model couplings; and {\it (ii)} saturating the exotic decay width by the sole invisible Higgs boson decays. These assumptions introduce some model dependency which are not called for when it comes to measure the Higgs boson properties in a truly model-independent manner. These two assumptions were therefore removed from the fit, the results of which are presented in the first three columns of Table~\ref{tab:FitResults} and in Fig.~\ref{fig:FitResults}. For completeness, and for direct comparison with Ref.~\cite{cite:1207.2516}, the results of the fit with these two assumptions are also given in the last two columns of the same table. 

\begin{table}
\begin{center}
\caption{Relative statistical uncertainty on the Higgs boson couplings, as expected from the physics programme at $\sqrt{s} = 240$ and $350$ GeV at TLEP. (The first column indicates the expected precision at TLEP when the sole 240 GeV data are considered. The substantial improvement with the inclusion of the 350 GeV data -- in the second column -- mostly stems from the precise total Higgs boson width measurement, which constrains all couplings simultaneously.) The numbers between brackets indicates the uncertainties expected with two detectors instead of four. For illustration, the uncertainties expected from the ILC baseline programme at 250 and 350 GeV are also given. The first three columns give the results of a truly model-independent fit, while the last two include the two assumptions made in Ref.~\cite{cite:1207.2516} on the W/Z couplings and on the exotic decays, for completeness and easier comparison. The column labelled "TLEP-240" holds for the sole period at 240 GeV for TLEP. The last line gives the {\it absolute} uncertainty on the Higgs boson branching fraction to exotic particles (invisible or not).\label{tab:FitResults}}
\begin{tabular}{|r|r|rl|r|rl|r|}
\hline  & \multicolumn{4}{c|}{Model-independent fit} & \multicolumn{3}{c|}{Constrained fit} \\ 
\hline Coupling & TLEP-240 & \multicolumn{2}{c|}{TLEP} & ILC  & \multicolumn{2}{c|}{TLEP} & ILC\\ \hline
\hline $g_{\rm HZZ}$  & 0.16\% & {\bf 0.15\%} & (0.18\%) & 0.9\% &  0.05\% & (0.06\%) & 0.31\% \\ 
\hline $g_{\rm HWW}$  & 0.85\% & {\bf 0.19\%} & (0.23\%) & 0.5\% &  0.09\% & (0.11\%) & 0.25\% \\ 
\hline $g_{\rm Hbb}$  & 0.88\% & {\bf 0.42\%} & (0.52\%) & 2.4\% &  0.19\% & (0.23\%) & 0.85\% \\ 
\hline $g_{\rm Hcc}$  & 1.0\% & {\bf 0.71\%}  & (0.87\%) & 3.8\% &  0.68\% & (0.84\%) & 3.5\% \\ 
\hline $g_{\rm Hgg}$  & 1.1\% & {\bf 0.80\%}  & (0.98\%) & 4.4\% &  0.79\% & (0.97\%) & 4.4\% \\ 
\hline $g_{{\rm H}\tau\tau}$  & 0.94\% & {\bf 0.54\%} & (0.66\%) & 2.9\% &  0.49\% & (0.60\%) & 2.6\% \\ 
\hline $g_{{\rm H}\mu\mu}$  & 6.4\% & {\bf 6.2\%} & (7.6\%) & 45\% &  6.2\% & (7.6\%) & 45\% \\ 
\hline $g_{{\rm H}\gamma\gamma}$ &1.7\% & {\bf 1.5\%} & (1.8\%) & 14.5\% &  1.4\% & (1.7\%) & 14.5\% \\ \hline
\hline ${\rm BR}_{\rm exo}$ & 0.48\% & {\bf 0.45\%} & (0.55\%) & 2.9\% &  0.16\% & (0.20\%) & 0.9\% \\ \hline
\end{tabular} 
\end{center}
\end{table}

As is clearly visible from Table~\ref{tab:FitResults} and Fig.~\ref{fig:FitResults}, a model-independent precision better than 1\% for all couplings (and at times approaching the per-mil level), required for these measurements to become sensitive to (multi-)TeV new physics, can be obtained with the TLEP high-statistics data samples. 

\begin{figure}[tb]
\begin{center}
\includegraphics[width=0.7\columnwidth]{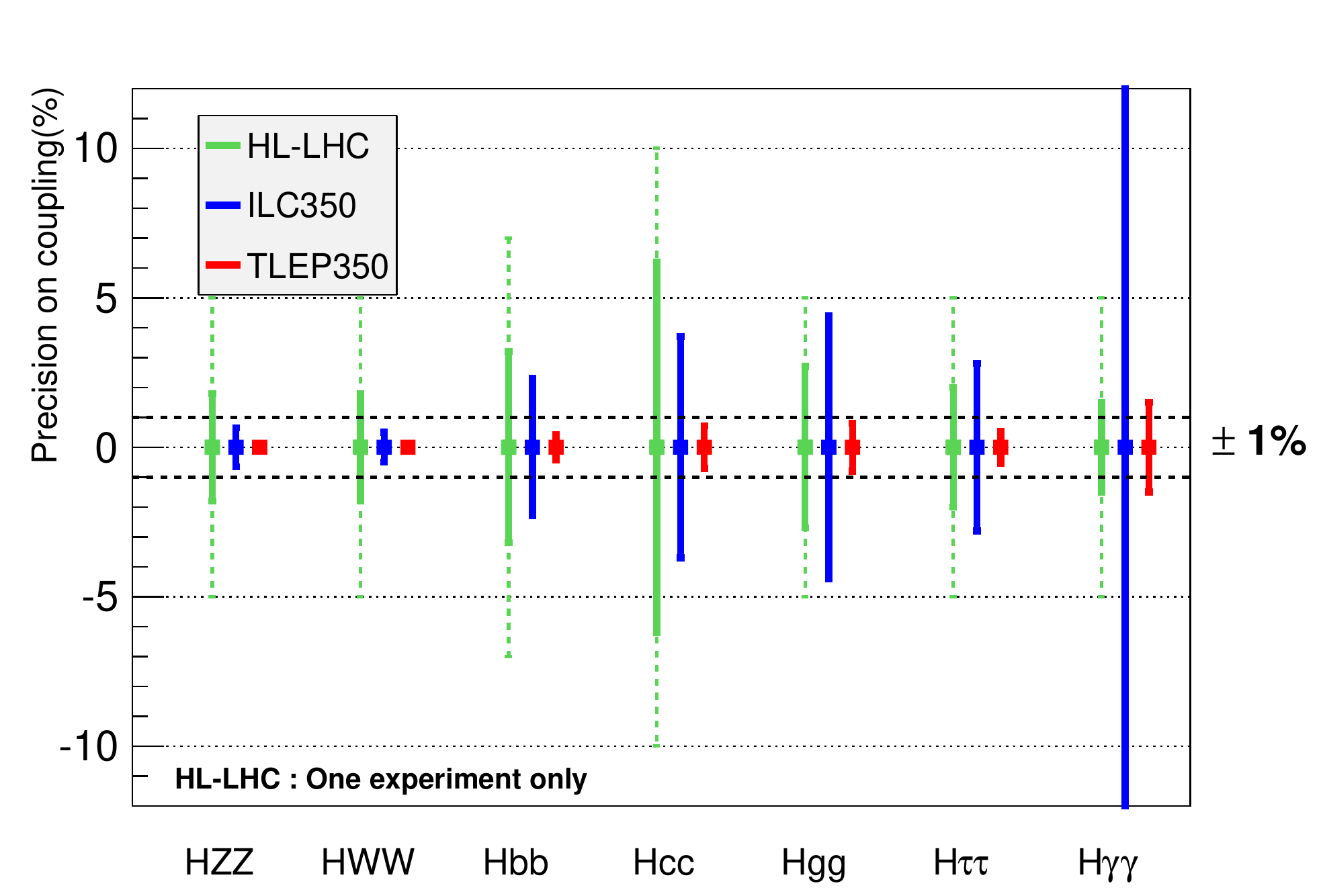}
\end{center}
\caption{\label{fig:HLLHC} Comparison between the projections of the HL-LHC (green) and of $\epem$ Higgs factories (blue: ILC, red: TLEP) for the Higgs boson coupling relative uncertainties. For the HL-LHC projections, the dashed bars represent CMS Scenario 1 and the solid bars represent CMS Scenario 2, for one experiment only~\cite{1307.7135}. For the Higgs factories, the data up to $\sqrt{s} = 350$ GeV are combined. The dashed horizontal lines show the $\pm 1\%$ band, relevant for sensitivity to multi-TeV new physics. }
\end{figure}

It is also important to compare the projections of TLEP to those from the HL-LHC, as to evaluate the added value of a circular $\epem$ Higgs factory after $3~\inab$ of proton-proton collision data. A truly model-independent fit cannot be performed from proton-proton collision data: the total decay width cannot be easily determined with the sole LHC measurements and the ${\rm H \to c\bar c}$ decay likely cannot be isolated by the LHC detectors -- although new ideas are emerging on these two fronts~\cite{1208.1533,1305.3854,1306.5770B}. Additional assumptions thus need to be made for a meaningful comparison with $\epem$ Higgs factories. Here, constraints similar to those used in Ref.~\cite{Barger_Ishida_Keung_2012} are applied: it is assumed that no Higgs boson exotic decays take place, and that deviations of the charm and top couplings are correlated.  The CMS report~\cite{1307.7135} submitted to the recent Snowmass process contains estimates of the CMS projected performance with $3~\inab$, with similar hypotheses, in two scenarios: Scenario 1 with all systematic uncertainties unchanged, and Scenario 2, with experimental systematic uncertainties scaling like $1/\sqrt{L}$ and theoretical errors halved. These estimates are displayed in Fig.~\ref{fig:HLLHC} and compared to a fit of the TLEP projectionsextracted with the same assumptions about the theoretical uncertainties in Higgs boson decays.

Within the mildly model-dependent assumptions used in the fit -- no exotic decays, and correlated up-type-quark couplings -- the projections for HL-LHC in Scenario 2 are truly impressive, and will further improve by including the other detector (ATLAS projections are available in Ref.~\cite{1307.7292}) and additional dedicated analyses in the combination. In this challenging context, TLEP data collected at 250 and 350 GeV would enable very significant improvements on these coupling measurements well beyond the HL-LHC projected precision, and with an accuracy adequate to become sensitive to multi-TeV new physics. The interest of $\epem$ collision data at centre-of-mass energies above 350 GeV for Higgs boson physics is briefly discussed in Section~\ref{sec:VHE-LHC}. 

\subsection{Sensitivity to new physics and theory uncertainties}

As examples of new physics models that would be probed with precision Higgs measurements at TLEP, supersymmetric models that are compatible with current measurements, including the non-observation of supersymmetric particles at the LHC, are considered. These models are simplified, in that they assume universal supersymmetry-breaking masses for squarks and sleptons, and for gauginos, at a high scale. In the case of the CMSSM, this assumption is extended to include the supersymmetric Higgs bosons, but this assumption is relaxed in the NUHM1 model~\cite{Isidori_Marrouche_et_al__2012}. A global frequentist analysis of the present data found two CMSSM fits that yield very similar values of the global $\chi^2$ function, with lower and higher sparticle masses respectively, whilst the best NUHM1 fit is qualitatively similar to the low-mass CMSSM fit. These fits have not been excluded by the 2012 LHC run at 8 TeV, but lie within the potential reach of the forthcoming LHC 13/14 TeV run. On the other hand, the high-mass CMSSM point is likely to lie beyond the reach of the LHC. Thus, these models represent different potential challenges for the TLEP precision physics programme: verify predictions of new physics models at the quantum level, or find indirect evidence for new physics beyond the reach of the LHC.

\begin{figure}[tb]
\begin{center}
\includegraphics[width=0.6\columnwidth]{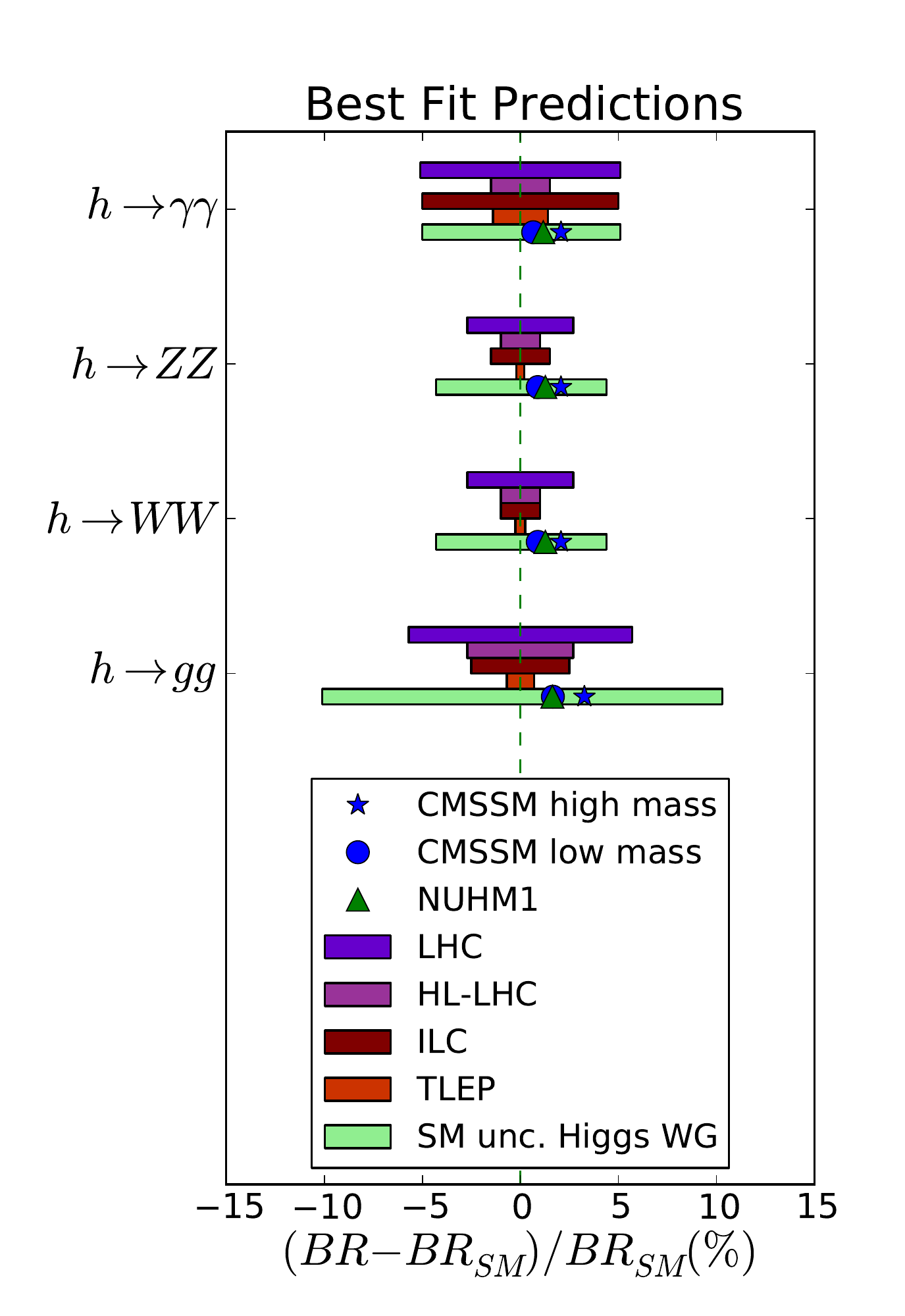}
\end{center}
\caption{\label{fig:SUSYbenchmarks} A compilation of prospective experimental errors in measurements of the principal Higgs decay branching ratios at the LHC, HL-LHC, ILC and TLEP (all with baseline luminosities and energies), compared with current estimates of the uncertainties in the Standard Model predictions~\cite{1307.1347} and the deviations from the Standard Model calculated in various supersymmetric models described in the text, and in more detail in~Ref.~\cite{Isidori_Marrouche_et_al__2012}.}
\end{figure}

Figure~\ref{fig:SUSYbenchmarks} displays the deviations from the Standard Model predictions for some principal Higgs decay branching ratios, calculated in these CMSSM and NUHM1 models. Also shown are the potential measurement uncertainties attainable with the LHC programme that is currently approved, with HL-LHC, with the ILC and with TLEP. Only TLEP has measurement errors that are expected to be significantly smaller than the deviations of the supersymmetric model predictions from the central values of the Standard Model predictions, thereby offering the possibilities of a check of the predictions of the low-mass models at the quantum level, and of indirect evidence for the high-mass CMSSM.

It can also be noted from Fig.~\ref{fig:SUSYbenchmarks}, however, that the uncertainties in the Standard Model predictions for the Higgs decay branching ratios stated by the LHC Higgs cross section Working Group~\cite{1307.1347} are considerably larger than the deviations of the supersymmetric models from the Standard Model predictions, and also larger than the projected experimental errors. This means that the TLEP programme of high-precision Higgs measurements must be accompanied by a substantial theoretical effort to reduce the uncertainties in the theoretical calculations of Higgs properties.

\section{Precise measurements of the EWSB parameters}
\label{sec:EWSB}

Electroweak loops have the remarkable property of being sensitive to the existence of weakly-coupled particles, even if they cannot be directly produced or observed in current experiments. For example, the measurements of the Z resonance line-shape parameters, undertaken  at LEP during a dedicated scan in 1993, led to a prediction of the top quark mass $m_{\rm top}$ of $172\pm 20$~GeV by the time of the Moriond conference in March 1994~\cite{cite:Pietrzyk}. The uncertainty on $m_{\rm top}$ was dominated by the range of assumptions for the Higgs boson mass, varied from 60 to 1000~GeV. When the top quark was discovered at the Tevatron in 1995, and its mass measured with precision of a few~GeV within one standard deviation of the prediction, the Electroweak fits of the LEP data became sensitive to the only remaining unknown quantity in the Standard Model, the Higgs boson mass $m_{\rm H}$, predicted to be $m_{\rm H} = 99^{+28}_{-23}$~GeV~\cite{_Lys_Murayama_Wohl_et_al__2012}. It is remarkable that the observation of the H(126) particle at the LHC falls, once again, within one standard deviation of this prediction. 

These two historical examples are specific of the Standard Model, with its particle content -- and nothing else. Now that the Higgs boson mass is measured with a precision of a fraction of a~GeV, and barring accidental or structural cancellations, these fits rule out the existence of any additional particle that would have contributed to the Electroweak loop corrections in a measurable way. As emphasized in Ref.~\cite{_Lys_Murayama_Wohl_et_al__2012}, the corrections to the W and Z masses do not necessarily decouple when the mass of new additional particles increase (contrary to the corrections to, e.g., $(g-2)_{\mu}$). For example, the top-quark loop correction scales like $(m^2_{\rm top} - m^2_{\rm b}) / m^2_{\rm W} $. The Electroweak loop corrections are also delicately sensitive to the details of the Electroweak Symmetry Breaking Mechanism. 

As summarized in Section~\ref{sec:exp}, the TLEP physics programme offers the potential of  considerable improvements in the precision of a large number of Electroweak observables. The outstandingly large luminosity, the precise energy definition, the absence of energy bias due to beamstrahlung, and an accurate energy calibration with resonant depolarization, are among the unique characteristics of TLEP towards an unparalleled precision for most of the measurements. 

In the following, the potential of TLEP for precise measurements at or around the Z pole, at the W pair threshold, and the top quark pair threshold, is briefly described. A set of the most important measurements is given in Table~\ref{tab:EW-TLEP}. When combined with the precision measurements of the Higgs boson properties (reviewed in Section~\ref{sec:Higgs}), TLEP could offer definitive investigations on the Electroweak Symmetry breaking, and on the possible existence of weakly interacting particles beyond those already known, with a precision sufficient for discovery. It will be the task of the upcoming design study to examine the requirements and the possible difficulties in turning this potential into reality.

\begin{sidewaystable}
\caption{Selected set of precision measurements at TLEP.
The statistical errors have been determined with {\it (i)} a one-year scan of the Z resonance with 50\%~data at the peak, leading to $7\times 10^{11}$ Z visible decays, with resonant depolarization of single bunches for energy calibration at O(20min) intervals; {\it (ii)} one year at the Z peak with 40\% longitudinally-polarized beams and a luminosity reduced to 20\% of the nominal luminosity; {\it (iii) } a one-year scan of the WW threshold (around 161 GeV), with resonant depolarization of single bunches for energy calibration at O(20min)  intervals; and {\it (iv)} a five-years scan of the $\ttbar$ threshold (around 346 GeV). The statistical errors expected with two detectors instead of four are indicated between brackets.
The systematic uncertainties indicated below are only a ``first look'' estimate and will be revisited in the course of the design study.\hfill\break\label{tab:EW-TLEP}}
\begin{center}
{\small
\begin{tabular}{|l|c|c|c|c|c|c|c|}
\hline  Quantity &      Physics & Present  &  Measured  &Statistical  & Systematic  &   Key &   Challenge \\
  &     & precision &  from & uncertainty &       uncertainty  &   & \\
\hline \hline
$m_{\rm Z} $ (keV)  & Input & $ 91187500 \pm 2100 $ &   Z Line shape scan &     $5$ ($6$) keV &
        $<100$ keV      & $E_{\rm beam}$ calibration &  QED corrections\\
\hline
 $\Gamma_{\rm Z} $ (keV) & $ \Delta\rho$ (not $\Delta\alpha_{\rm had}$)& $      2495200  \pm 2300$ &    Z Line shape scan & $8$ ($10$) keV &$<100$ keV     & $E_{\rm beam}$ calibration &  QED corrections \\
\hline
$R_{\ell}$      & $\alpha_{\rm s} , \delta_{\rm b}$ & $ 20.767 \pm 0.025$&      Z Peak& $       0.00010~(12)$&$       <0.001$&        Statistics      &QED corrections\\ 
\hline
$N_{\nu}$ &     PMNS Unitarity, ... & $ 2.984 \pm 0.008$ &      Z Peak &        $0.00008~(10)$ &     $<0.004 $ &     &       Bhabha scat.\\ 
\hline
$N_{\nu}$ &     ... and sterile $\nu$'s & $  2.92 \pm 0.05$ &        ${\rm Z}\gamma$, 161 GeV &      $0.0010~(12)$ &       $<0.001 $ & Statistics          &       \\ 
\hline
$R_{\rm b}$     &$ $ $\delta_{\rm b}$ & $       0.21629  \pm 0.00066$&  Z Peak&$0.000003~(4)$&$<0.000060$&  Statistics, small IP&   Hemisphere correlations\\
\hline
$\ALR$& $\Delta\rho$, $\epsilon_3$, $\Delta\alpha_{\rm had}$ & $       0.1514
\pm 0.0022$&    Z peak, polarized&$     0.000015~(18)$&$     <0.000015$&     4 bunch scheme, 
 2exp&  Design experiment \\
\hline
$m_{\rm W}$
(MeV) & $\Delta\rho$ , $\epsilon_3$, $\epsilon_2$, $\Delta\alpha_{\rm had}$ &   $80385
 \pm 15$&       WW threshold scan&$     0.3$ ($0.4$)MeV&$      <0.5$ MeV&    $E_{\rm beam}$,  
Statistics&     QED corrections\\
\hline
$m_{\rm top}$ (MeV)      &Input&$173200 \pm 900$& $\ttbar$ threshold scan  & $10$ ($12$) MeV& $<10$ MeV &        Statistics & Theory interpretation\\
\hline
\end{tabular}
}
\end{center}
\end{sidewaystable}

\subsection{Measurements with TeraZ}

With a continuous luminosity of $5.6\times 10^{35}~\cms$ per IP at a centre-of-mass energy of 91 GeV, TLEP is a Z factory able to deliver over $20~\inab$ of data, i.e., $7\times 10^{11}$ visible Z decays for one year of running (hence the ``Tera Z'' appellation), with very clean experimental conditions, centre-of-mass energy known to a fraction of MeV, and the possibility of longitudinally polarized beams, with which the following experiments can be carried out: 
\begin{itemize}
\item a high-statistics line-shape scan of the Z resonance, allowing an extremely precise determination of the Z mass and width;
\item high-statistics data collection at the Z peak, for the measurement of the Z partial widths, the determination of the number of light neutrinos, and the detection of rare decays;
\item high-statistics data taking with longitudinally-polarized beams, for a very precise determination of the weak mixing angle.
\end{itemize}
An extensive description of Electroweak measurements performed at LEP and SLC in 1988-1998 can be found in Ref.~\cite{ements_on_the_Z_resonance_2006}. It is beyond the scope of this article to revisit all the measurements in view of establishing the improvements potentially brought about by TLEP. Only a brief account of a few key measurements is given here. Typically, TLEP will bring a factor $10^5$ to the statistics accumulated at LEP, which corresponds to statistical uncertainties reduced by a factor 300. With such a huge improvement, it is clear that a detailed consideration of experimental systematic uncertainties will be essential before a precise conclusion be drawn on the ultimately achievable precisions. Above all, uncertainties in the theoretical interpretation will need to be revisited, which implies a significant new programme of calculations of higher-order Electroweak corrections.

\subsubsection{The Z mass and width}

The Z mass was determined at LEP from the line shape scan to be  $91187.5 \pm 2.1$~MeV. The statistical error of 1.2~MeV would be reduced below 5 keV at TLEP. The systematic uncertainty was dominated by the error pertaining to the beam energy calibration (1.7~MeV). As seen in Section~\ref{sec:exp}, a continuous measurement with resonant depolarization of single bunches should allow a reduction of this uncertainty to well below 100 keV. Other errors include the theoretical uncertainties in the calculation of initial state radiation ($\le 100$~keV), in the production of additional lepton pairs ($\le 300$~keV), and in the theoretical line-shape parametrization ($\le 100$~keV). It is clear that revisiting the QED corrections will be a high priority item when embarking in a new program of precision measurements at TLEP. \\
\textbf{\textit{An overall uncertainty of 100~keV or better is therefore a reasonable target for the Z mass precision at TLEP}}.

The Z width was also determined from the line shape scan at LEP to be  $2495.2 \pm 2.3$ MeV. The statistical error of 2 MeV would be reduced to less than 10 keV at TLEP. The systematic uncertainty from the LEP energy calibration was 1.2 MeV, clearly dominated by the reproducibility issues of the beam energy calibration. Again, this uncertainty is expected be reduced to below 100 keV at TLEP. 
The theory systematic uncertainties on $\Gamma_{\rm Z}$ were estimated at the level of 200 keV and should be revisited.\\
\textbf{\textit{An overall uncertainty of 100 keV or better is a reasonable target for the Z width precision at TLEP}}.

\subsubsection{The Z hadronic and leptonic partial widths}
\label{sec:Rell}

Determination of the Z partial widths requires measurements of branching ratios at the Z peak -- in particular the ratio of  branching fractions of the Z boson into lepton and into hadrons -- and the peak hadronic cross section. The hadronic-to-leptonic ratio was measured at LEP to be 
\begin{equation}
\label{eq:Rl} R_\ell = \frac{\Gamma_{\rm had}}{\Gamma_{\ell}}= 20.767\pm 0.025,
\end{equation}
with a systematic uncertainty of 0.007. The experimental uncertainty was dominated by the statistics of leptonic decays, and other uncertainties related to the event selection will tend to decrease with statistics. The remaining systematic uncertainties were related to the $t$-channel contribution in the electron channel (which would vanish by the sole use of the muon channel) and to the detailed modelling of final-state radiation or emission of additional lepton pairs. Here, theory should be considerably helped by the large sample of leptonic Z decays available for the study of these rare processes. The measurements of the partial widths into electron, muon and tau pairs will also allow tests of the lepton universality in Z decays with considerably improved precision with respect to what was achieved LEP.\\ 
\textbf{\textit{A relative precision of $\boldsymbol{\it 5 \times 10^{-5}}$ is considered to be a reasonable target for the ratio of the Z hadronic-to-leptonic partial widths at TLEP, as well as for the ratios of the Z leptonic widths (as a test of lepton universality).}}\\

\subsubsection{The leptonic weak mixing angle}

Determinations of the weak mixing angle $\sintw$ are made from a variety of measurements, such as the leptonic and hadronic forward-backward asymmetries or the $\tau$ polarization in ${\rm Z} \to \tau\tau$ decays. These measurements will be performed with high statistics at the occasion of the line-shape scan without polarized beams.  
\\ 

The single most precise measurement, however, comes from the inclusive left-right beam-pola\-ri\-zation asymmetry $\ALR$. This quantity can be measured from the total cross-section asymmetry upon reversal of the polarization of the $\epem$ system. For the same level of polarization in collisions as that observed at LEP, and assuming that a fraction of the bunches can be selectively depolarized, a simultaneous measurement~\cite{Blondel:1987wr} of the beam polarization and of the left-right  asymmetry $\ALR$ can be envisioned at TLEP. For one year of data taking with a luminosity of $10^{35}~\cms$, a precision on $\ALR$ of the order of $10^{-5}$ -- or a precision on $\sintw$ of the order of $10^{-6}$ -- is achievable. Other beam polarization asymmetries for selected final states, like for example $A_{\rm FB}^{\rm pol,f}$, would allow precise measurements of the Electroweak couplings, and become an interesting tool for flavour selection.\\
\textbf{\textit{A precision of $\boldsymbol{\it 10^{-6}}$ on $\boldsymbol{\it \sin^2\theta^{eff}_W}$  is a reasonable goal for the measurement of the leptonic weak mixing angle at TLEP.}}

\subsubsection{The ${\rm Z \to \bbbar}$ partial width}

An Electroweak correction of great interest is the vertex correction to the ${\rm Z} \to \bbbar$ partial width. This correction affects the total Z width $\Gamma_{\rm Z}$, the leptonic branching fraction $R_\ell$, the peak hadronic cross section $\sigma^{\rm peak}_{\rm had}$, and most sensitively, $R_{\rm b} \equiv \Gamma_{\rm Z \to \bbbar}/\Gamma_{\rm had}$. At LEP and SLC, $R_{\rm b}$ was measured by tagging the presence of one b-quark jet, and the efficiency was controlled by the ``double tag'' method. The present experimental value, $R_{\rm b} = 0.21629 \pm  0.00066$, has a roughly equal sharing between systematic and statistical uncertainties. 

Because the double b-tagging method is self-calibrating, its accuracy is expected to improve with accumulated statistics. The SLD detector at SLC had the best efficiency for this selection, by the twofold effect of a more granular vertex detector and a smaller beam spot, which allowed a more precise determination of the impact parameter of secondary hadrons. While the experimental conditions at TLEP are expected to be similar to those at LEP, the beam spot size will be very significantly smaller in all dimensions than at SLC, and a next-generation vertex detector will be used.  The b-tagging capabilities should therefore be similar to or better than those of SLD.\\
\textbf{\textit{A precision of $\boldsymbol{\it 2}$ to $\boldsymbol{\it 5 \times 10^{-5}}$ seems therefore to be a reasonable goal for the measurement of $\boldsymbol{\it R_b}$  at TLEP.}}

\subsubsection{Rare decays}

The very large statistics accumulated at TLEP, including $3 \times 10^{10}$ tau pairs or muon pairs, and more than $2 \times 10^{11}$ b quarks or c quarks, should allow a new range of searches for rare phenomena and tests of conservation laws that remain to be investigated. As an illustration, more than 20,000  ${\rm B_s \to \tau^+\tau^-}$ decays would be produced, according to the Standard Model prediction: the few thousand events observed will bring stringent constraints on new physics, which may change this branching fraction by large factors. It will also be possible to probe small flavour-changing-neutral-current couplings of the Z to quarks and leptons with very high accuracy. (Flavour-changing-neutral-current couplings of the top quark can also be probed both in production and in decays by running at the $\ttbar$ threshold and above.) It will be the purpose of the upcoming design study to examine and develop further the immense physics potential of TLEP in the search for rare decays and their theoretical interpretation.  

\subsection{Measurements with OkuW}

With more than $2\times 10^8$ W pairs produced at centre-of-mass energies at the WW threshold and above (hence the ``OkuW'' appelation), of which $2.5\times 10^7$ W pairs at $\sqrt{s} \sim 161$~GeV, TLEP will be a W factory as well. Because the quantity of data expected at the WW production threshold is $10^5$ times larger than that produced at LEP, the measurements to be performed by TLEP at this centre-of-mass energy need to be thoroughly reviewed by the starting design study. Here, only brief accounts of the W mass measurement and the determination of the number of active neutrinos are given. A precise measurement of the strong coupling constant can also be done when the large WW event samples expected at $\sqrt{s}=240$ and $350$~GeV are exploited too. 

\subsubsection{The W mass}

The safest and most sensitive measurement of the W mass can be performed at threshold. At LEP~\cite{1302.3415}, this measurement was done at a unique centre-of-mass energy of 161.3~GeV. A more thorough scan, including a point below threshold for calibration of possible backgrounds, should probably be envisioned to provide the redundancy necessary for a precise measurement at TLEP. The measurement is essentially statistics dominated and the only relevant uncertainties are those associated with the definition of the centre-of-mass energy, as described in Section~\ref{sec:exp}. The precision achieved at LEP on $m_{\rm W}$ was about 300~MeV per experiment. A statistical error of 1~MeV on the W mass should therefore be achievable at TLEP per experiment (i.e., 0.5~MeV from a combination of four experiments).  

As energy calibration with resonant depolarization will be available at TLEP at least up to 81~GeV per beam, the threshold scan should involve beam energies close to the point of maximum $m_{\rm W}$ sensitivity and situated at the half-integer spin tune, $\nu_s = 182.5$ and $183.5$, i.e.,  $E_{\rm beam}= 80.4$ and $80.85$~GeV. Because the beam-energy spread and the beamstrahlung are negligibly small at TLEP, this measurement is not sensitive to the delicate understanding of these two effects. A more careful analysis may reveal systematic uncertainties that are relevant at this  level of precision. They should, however, be somewhat similar to those involved in the Z mass measurement from the resonance line shape, i.e., dominated by the uncertainties on the initial state QED corrections and the theoretical parameterization of the WW threshold cross section.  With the same logic as above, these uncertainties should be reducible to a level below 100~keV on $m_{\rm W}$.\\
\textbf{\textit{An overall, statistics-dominated, uncertainty of 500~keV is therefore considered as a reasonable target for the W mass precision at TLEP.}}

\begin{figure}[tb]
\begin{center}
\includegraphics[width=0.6\columnwidth]{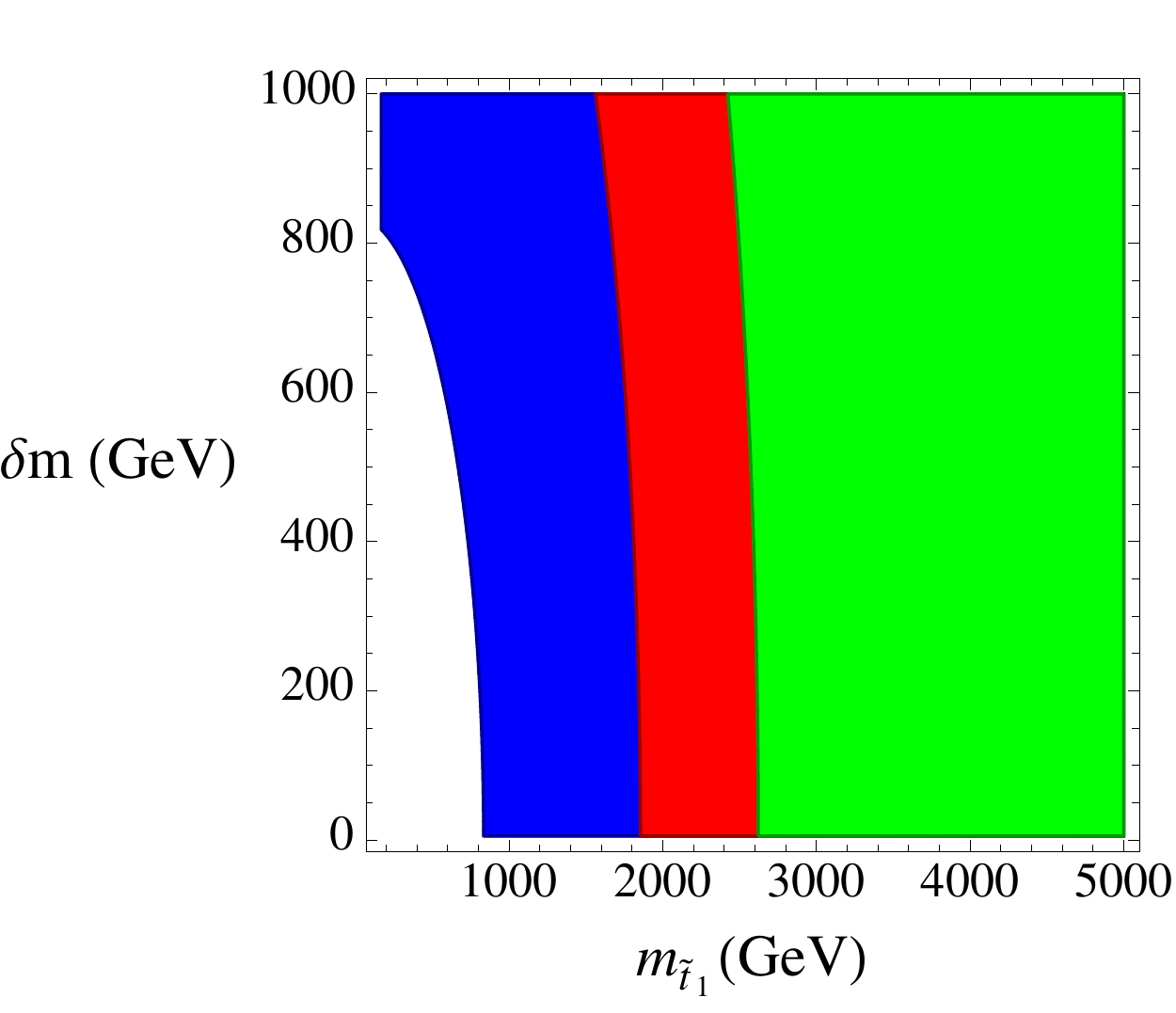}
\end{center}
\caption{\label{fig:veronica} Sensitivity of the W mass measurement to the mass $m_{\tilde t_1}$ of the lighter supersymmetric partner of the top quark (horizontal axis) as a function of the difference $\delta m$ between the masses of the two stop squarks (vertical axis), from the analysis of Ref.~\cite{1207.7355}. The colours indicate that measurements of the W mass with a precision smaller than 5~MeV (blue), 1~MeV (red) and 500~keV (green) would be sensitive to a stop mass of 850~GeV, 1.9~TeV and 2.6~TeV, respectively, independently of the stop decay modes.}
\end{figure}

This sole measurement would already be a very sensitive probe of new physics, able to provide indirect evidence for the existence of particles that could not be observed directly at the LHC. One example is provided by the supersymmetric partners of the top quark, from the analysis of Ref.~\cite{1207.7355}, as illustrated in Fig.~\ref{fig:veronica}. The TLEP precision of 500~keV on the W mass would give sensitivity to a stop squark of about 3~TeV, far heavier than could be detected at the HL-LHC, and independently of the stop decay mode. This is another example of how the unparallelled TLEP precision could give access to physics beyond the Standard Model.

\subsubsection{The Z invisible width and the number of neutrinos}

The measurement of the Z decay width into invisible states is of great interest as it constitutes a direct test of the unitarity of the PMNS matrix -- or of the existence of sterile neutrinos, as pointed out in Ref.~\cite{Jarlskog_1990}. It can be performed at the Z pole from the peak hadronic cross section or at larger centre-of-mass energies with radiative return to the Z~\cite{Barbiellini}. As explained below, at TLEP the latter is likely to be more accurate than the former.

The measurement of the peak hadronic cross-section at the Z pole is indeed already dominated by theoretical systematics today, related to the understanding of the low-angle Bhabha-scattering cross section (used for the integrated luminosity determination).  The present measurement, expressed in terms of a number of active neutrinos,
\begin{equation}
N_\nu = 2.984 \pm 0.008,
\end{equation} 
is two standard deviations below the SM value of 3.00. The experimental conditions at TLEP will be adequate to improve the experimental uncertainty considerably, but, to make this measurement worthwhile, a commensurate effort would have to be invested in the theoretical calculations of the small-angle Bhabha-scattering cross section used for normalization. A desirable goal would be to reduce the uncertainty on $N_\nu$ down to 0.001, but it is not clear that it can be achieved from Z peak measurements.

Above the Z peak, the $\epemto {\rm Z}\gamma$ process provides a very clean photon-tagged sample of on-shell Z bosons, with which the Z properties can be measured. From the WW threshold scan alone, the cross section of about 5~pb~\cite{cite:Abbiendi2000hh,cite:Heister2002ut,cite:Achard2003tx,cite:Abdallah2003np} ensures that 10 million ${\rm Z}\gamma$ events will be produced in each TLEP experiment with a ${\rm Z} \to \nnbar$ decay and a high-energy photon in the detector acceptance. The three million ${\rm Z}\gamma$ events with leptonic Z decays will in turn provide a direct measurement of the ratio $\Gamma_{\rm Z}^{\rm inv}/\Gamma_{\rm Z}^{\rm lept}$, in which uncertainties associated with absolute luminosity and photon detection efficiency cancel. The 40 million ${\rm Z}\gamma$ events with either hadronic or leptonic Z decays will also provide a cross check of the systematic uncertainties and backgrounds related to the QED predictions for the energy and angular distributions of the high-energy photon. The invisible Z width will thus be measured with a statistical error corresponding to 0.001 neutrino family. Systematic uncertainties are expected to be at the same level or smaller. 

The data taken at $\sqrt{s} = 240$ and $350$ GeV will contribute to further reduce this uncertainty with the $\epemto {\rm Z}\gamma$ process, and to perform independent cross checks and redundant $\Gamma_{\rm Z}^{\rm inv}$ measurements with {\rm ZZ} and maybe {\rm HZ} production. It is to be determined by the design study whether a dedicated run at a somewhat lower centre-of-mass energy -- with both a larger luminosity and a larger ${\rm Z}\gamma$ cross section -- is more appropriate for this important measurement.\\
\textbf{\textit{An overall, statistics-dominated, uncertainty smaller than 0.001 of a SM neutrino partial width is therefore considered as a reasonable target for the Z invisible width at TLEP.}}

\subsubsection{The strong coupling constant}

The prospective TLEP precisions on the EWSB parameters call for a similar improvement of the strong coupling constant accuracy, which would otherwise become a leading systematic uncertainty in the theoretical interpretation of the TLEP measurements, and in particular in the determination of the top quark mass from the measurement of the $\ttbar$ production threshold cross section. Complementary determinations of the strong coupling constant, $\alpha_{\rm s}$, may be obtained both at the Z pole and at energies at the WW threshold and above, with similar accuracies.

The precise experimental measurement of the inclusive hadronic Z decay rate at the Z pole is sensitive to $\alpha_{\rm s}$. The theoretical prediction for such an inclusive observable is known with ${\rm N_3}{\rm LO}$ QCD corrections~\cite{Baikov_Chetyrkin_Kuhn_2008, Chetyrkin_Kuhn_Rittinger_2012}, with strongly suppressed non-perturbative effects. Some caveat is in order since Electroweak corrections can in principle be sensitive to the particle content of the Electroweak theory. The extraction of $\alpha_{\rm s}$ may therefore not be completely free of model dependence of Electroweak nature. A good way around this caveat is to constrain radiative-correction effects with other Electroweak measurements at the Z pole or elsewhere. In the case at stake here, the hadronic partial width is sensitive to new physics through the ``oblique'' Electroweak corrections known as $\epsilon_1 (\equiv \Delta\rho)$ and $\epsilon_3$, and through the vertex correction $\delta_{\rm b}$ to the $ {\rm Z} \to \bbbar$ partial width. The $\Delta \rho$ sensitivity cancels when taking the ratio $R_\ell$ with the leptonic partial width, and the $\epsilon_3$ corrections can be strongly constrained by the determination of $\sintw$ from leptonic asymmetries or from $\ALR$. The b-vertex contribution can be constrained by the direct extraction of $R_{\rm b}$, hence is not expected to be a limitation.

The ratio $ R_\ell$ has been used for the determination of $\alpha_{\rm s}$ at LEP. Up to a few years ago, when only NNLO QCD predictions were available, and the Higgs boson mass was still unknown, this measurement was translated to~\cite{Bethke_2004}
\begin{equation}
\scriptstyle
\alpha_{\rm s} (m^2_{\rm Z})= 0.1226 \pm 0.0038 \ ({\rm exp}) {\ }^{+0.0028\ (\mu = 2.00 m_{\rm Z})}_{-0.0005\ (\mu = 0.25 m_{\rm Z})}
{\ }^{+0.0033\ (m_{\rm H} = 900\ {\rm GeV})}_{-0.0000\ (m_{\rm H} = 100\ {\rm GeV})}
{\ }^{+0.0002\ (m_{\rm top} = 180\ {\rm GeV})}_{-0.0002\ (m_{\rm top} = 170\ {\rm GeV})} \pm 0.0002\ ({\rm th}).
\end{equation}

Now that {\it (i)} the uncertainty due to the Higgs boson mass dependence is no longer relevant; {\it (ii)} the uncertainty due to the top-quark mass dependence is negligible; and {\it (iii)} the pQCD scale uncertainty from the latest ${\rm N_3}{\rm LO}$ calculations~has dropped to 0.0002, this method potentially allows access to a high-precision measurement of $\alpha_{\rm s}$. As shown in Eq.~\ref{eq:Rl}, $R_\ell$ was measured at LEP with a relative uncertainty of 0.12\%. As mentioned in Section~\ref{sec:Rell}, this precision is expected to improve to $5 \times 10^{-5}$ with TLEP. The LEP experimental error of 0.0038 on $\alpha_s (m^2_{\rm Z})$ will scale accordingly to 0.00015 at TLEP, becoming of the same order as the theory uncertainty.\\
\textbf{\textit{A reasonable target for the measurement of $\boldsymbol{\alpha_s (m^2_Z)}$ with a run at the Z pole with TLEP is therefore a precision of 0.0002.}}

Beyond the measurement of $R_\ell$ at the Z pole, another interesting possibility for the $\alpha_{\rm s}$ determination is to use the W hadronic width as measured from W-pair events at and above 161 GeV. The quantity of interest is the branching ratio $B_{\rm had}= \Gamma_{\rm W \to hadrons}/\Gamma^{\rm tot}_{\rm W}$, which can be extracted by measuring the fractions of WW events to the fully leptonic, semi-leptonic and fully hadronic final states: 
\begin{eqnarray}
{\rm BR}({\rm W^+W^-} \to \ell^+ \nu \ell{'}^- \bar{\nu}) & = & (1-B_{\rm had})^2,\\
{\rm BR}({\rm W^+W^-} \to \ell^+ \nu {\rm q\bar{q}^\prime}) & = & (1-B_{\rm had})\times B_{\rm had},\\
{\rm BR}({\rm W^+W^-} \to {\rm q\bar{q}^\prime q^{\prime\prime}\bar{q}^{\prime\prime\prime}}) & = & B_{\rm had}^2.
\end{eqnarray}

The LEP2 data taken at centre-of-mass energies ranging from 183 to 209 GeV led to $B_{\rm had} = 67.41 \pm 0.27$~\cite{1302.3415}, a measurement with a 0.4\% relative precision. This measurement was limited by WW event statistics of about $4\times 10^4$ events. With over $2 \times 10^8$ W pairs expected at TLEP at $\sqrt{s} =$ 161, 240 and 350~GeV, it may therefore be possible to reduce the relative uncertainty on $B_{\rm had}$ by a factor $\sim 70$, down to $5\times 10^{-5}$, and thus the absolute uncertainty on $\alpha_{\rm s}$ to $\pm 0.00015$. 

This measurement is both competitive with and complementary to that performed with the Z hadronic width, because the sensitivity to Electroweak effects is completely different in $B_{\rm had}$ and in $R_\ell$. In particular, the coupling of the W to pairs of quarks and leptons is straightforwardly given by the CKM matrix elements with little sensitivity to any new particles.\\
\textbf{\textit{A reasonable target for the measurement of $\boldsymbol{\alpha_s (m^2_W)}$ with the runs at and above 161~GeV with TLEP is therefore a precision better than 0.0002. When combined with the measurement at the Z pole, a precision of 0.0001 is within reach for $\boldsymbol{\alpha_s (m^2_Z)}$.}}

As another example of the importance of precision measurements, the LEP determination of $\alpha_{\rm s}(m_{\rm Z})$ was already able, in association with $\sintw$, to distinguish between supersymmetric and non-supersymmetric models of grand unification~\cite{Ellis:1990,Amaldi:1991,Langacker:1991,Giunti:1991}. The prospective TLEP accuracies on these quantities would take this confrontation between theory and experiments to a completely new level.

\subsection{Measurements with MegaTop}

With an integrated luminosity of  the order of $130~\infb$ per year and per experiment, TLEP will be a top factory as well, with over one million $\ttbar$ pairs produced in five years (hence the ``MegaTop'' appellation) at $\sqrt{s} \sim 345$~GeV. The precise measurement of the cross section at the $\ttbar$ production threshold is sensitive to the top-quark pole mass, $m_{\rm top}$, the total top-quark decay width, $\Gamma_{\rm top}$, as well as to the Yukawa coupling of the top quark to the Higgs boson, $\lambda_{\rm top}$, through the virtual exchange of a Higgs boson between the two top quarks. 

\begin{figure}[tb]
\begin{center}
\includegraphics[width=0.7\columnwidth]{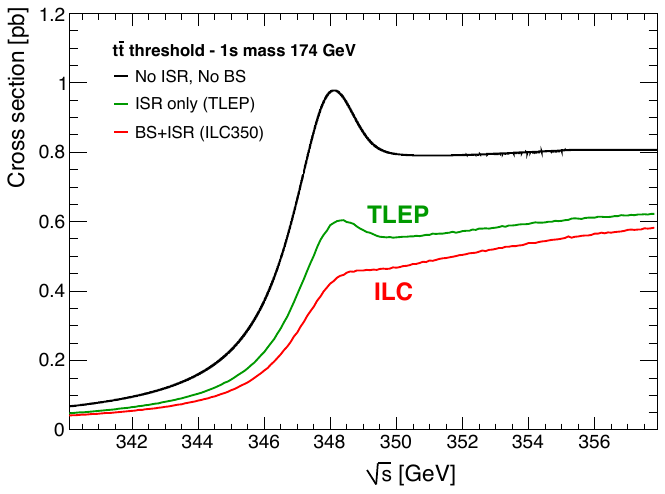}
\end{center}
\caption{\label{fig:ttbar} The $\ttbar$ cross section at the production threshold, for a top quark mass of 174 GeV, as a function of the centre-of-mass energy, taken from Ref.~\cite{cite:1303.3758}. (Note: the measured top quark mass from Tevatron and LHC is approximately 1 GeV smaller. The 1s peak is therefore around 346 GeV instead of 348 GeV as shown here.) The black curve is the next-to-next-to-leading-order QCD-corrected cross section. The green curve shows the effect of photon emission exclusively by initial state radiation (ISR), as is expected in TLEP collisions. For illustration, the red curve includes in addition the effects of the ILC beamstrahlung at ${\sqrt{s}=350}$ GeV.}
\end{figure}

The production cross section at threshold~\cite{g_Manohar_Stewart_Teubner_2001}, corrected for QCD effects up to the next-to-next-to-leading order, is displayed in Fig.~\ref{fig:ttbar} for $m_{\rm top} = 174$~GeV, with and without the effects of initial-state radiation (present at all $\epem$ colliders) and of beamstrahlung (only affecting linear colliders). As mentioned in Section~\ref{sec:beamstrahlung}, the absence of beamstrahlung at TLEP slightly increases the steepness, hence the sensitivity to the top-quark mass, and absolute value of the cross-section profile at the $\ttbar$ threshold. The corresponding numbers of events expected at TLEP are given in Table~\ref{tab:TopCross}.

\begin{table}
\begin{center}
\caption{Integrated luminosity and total number of $\ttbar$ pairs produced with TLEP at $\sqrt{s} \sim 345$~GeV (where the sensitivity to the top quark mass is maximal). For illustration, the corresponding numbers are also indicated for the baseline ILC programme at $\sqrt{s} \sim 350$ GeV.\label{tab:TopCross}}
\begin{tabular}{|l|r|r|}
\hline & TLEP & ILC \\ \hline \hline
Total Integrated Luminosity ($\inab$) & {\bf 2.6} & 0.35 \\ \hline
Number of $\ttbar$ pairs  & {\bf 1,000,000} & 100,000 \\ \hline
\end{tabular}
\end{center}
\end{table}

The most thorough study of the $\ttbar$ threshold measurements was done in the context of the TESLA project in Ref.~\cite{Martinez_Miquel_2003}, the parameters of which are very close to those of the ILC. The study makes use of a multi-parameter fit of $m_{\rm top}$, $\Gamma_{\rm top}$, $\lambda_{\rm top}$ and $\alpha_{\rm s}$ to the top cross section, the top momentum distributions, and the forward-backward asymmetry. When constraining the value of $\alpha_s(m_{\rm Z})$ to its currently measured value, the study obtained statistical uncertainties of $\Delta m_{\rm top} = 31$~MeV, $\Delta\Gamma_{\rm top} = 34$~MeV, and a relative uncertainty on the Yukawa coupling $\lambda_{\rm top}$ of the order of 40\%. The dominant experimental systematic uncertainties on the mass stem from the knowledge of $\alpha_s(m_{\rm Z})$ ($\pm$30~MeV per unit of $\pm$0.0007, the current uncertainty on this quantity), and from the knowledge of the beam-energy spectrum:  a 20\% uncertainty of the RMS width of the main luminosity peak would result in top mass uncertainties of approximately 75~MeV, far in excess of the statistical uncertainty~\cite{cite:1303.3758}.

The expected TLEP statistical uncertainties are summarized in Table~\ref{tab:TopMass}. In addition to the ten-fold increase in the number of $\ttbar$ events at TLEP, which reduces the statistical uncertainties by a factor of three, the much better knowledge of the beam-energy spectrum, and the precise measurement of the strong coupling constant with TeraZ and OkuW are bound to reduce the main experimental systematic uncertainties by one order of magnitude, hence below the statistical uncertainties. The starting design study plans to demonstrate fully the TLEP potential in this respect. A specific effort to reduce the theoretical Electroweak uncertainties on the cross section by one order of magnitude will also be needed.

\begin{table}
\begin{center}
\caption{Expected statistical uncertainties for $m_{\rm top}$, $\Gamma_{\rm top}$ and $\lambda_{\rm top}$  for TLEP, obtained from a five-years scan of $\ttbar$ threshold at $\sqrt{s} \sim 350$ GeV. The dominant experimental systematic uncertainties on the top quark mass are expected to be of the order of or smaller than the statistical uncertainties for TLEP. Also indicated is the baseline ILC potential for these measurements.\label{tab:TopMass}}
\begin{tabular}{|l|r|r|r|}
\hline  &  $m_{\rm top}$ & $\Gamma_{\rm top}$ & $\lambda_{\rm top}$ \\ \hline \hline
TLEP & {\bf 10~MeV} & {\bf 11~MeV} & ${\bf 13\%}$\\ \hline
ILC  & 31~MeV & 34~MeV & $40\%$ \\ \hline
\end{tabular}
\end{center}
\end{table}

\noindent\textbf{\textit{An overall experimental uncertainty of 10 to 20~MeV is therefore considered to be a reasonable target for the top-quark mass measurement at TLEP.}}

\subsection{ Reducing the theory uncertainties}

The unprecedented precision in Higgs, W, Z and top measurements at TLEP will require significant theoretical effort in a new generation of theoretical calculations in order to reap the full benefits from their interpretation, as illustrated in Section~\ref{sec:globalFit}. In their absence, a few considerations are given here, based on calculations made in the context of GigaZ and MegaW studies at the ILC~\cite{Ferroglia_Sirlin_2013}. The current measurements of $m_{\rm H}$, $m_{\rm Z}$, $\alpha_{\rm em}$, $m_{\rm top}$ and $\alpha_{\rm s}$ may be used to estimate $m_{\rm W}$ and $\sintw$,
\begin{eqnarray}
m_{\rm W} & = & 80.361 \pm 0.006 \pm 0.004 \; {\rm GeV},\\
\sintw & = & 0.23152 \pm 0.00005 \pm 0.00005,
\label{eq:uncertainties}
\end{eqnarray}
where in each case the first error is the parametric uncertainty and the second is the estimated uncertainty due to higher-order Electroweak corrections. 

In both cases~\cite{1007.5232}, the dominant parametric uncertainty is due to the experimental error in the top mass, $\delta m_{\rm top} \sim 1$~GeV, responsible for $\delta m_{\rm W} \sim 6$~MeV and $\delta \sintw \sim 3 \times 10^{-5}$. A measurement of $m_{\rm top}$ with a statistical precision of 10 to 20 MeV, as discussed above, could in principle reduce these parametric uncertainties to $\delta m_{\rm W} \sim 0.1$~MeV and $\delta\sintw < 10^{-6}$, respectively. However, there is currently a theoretical uncertainty in $m_{\rm top}$ associated with non-perturbative QCD, of the order of $\sim 100$~MeV or more, which would need to be understood better. Other important parametric uncertainties are those due to $\delta m_{\rm Z}$, responsible for $\delta m_{\rm W} \sim 2.5$~MeV and $\delta \sintw \sim 1.4 \times 10^{-5}$. The projected measurement of $m_{\rm Z}$ with an error $\delta M_Z \sim 0.1$~MeV would reduce these two parametric uncertainties to $\delta m_{\rm W} \sim 0.1$~MeV and $\delta\sintw \sim 10^{-6}$ as well. Other important parametric uncertainties are those associated with $\alpha_{\rm em}(m_{\rm Z})$, which are currently $\delta m_{\rm W} \sim 1$~MeV and $\delta \sintw \sim 1.8 \times 10^{-5}$. The exploitation of the full power of TLEP would require reducing $\delta\alpha_{\rm em}(m_{\rm Z})$ by almost an order of magnitude, which will require significant improvements not only in lower-energy measurements of $\epemto$ hadrons, but also in the theoretical understanding of radiative corrections~\cite{aQEDD,aQEDC,aQEDB,aQEDA}.

These prospective reductions in the parametric errors of Eq.~\ref{eq:uncertainties} will need to be accompanied by order-of-magnitude reductions in the uncertainties associated with Electroweak corrections. This will require a new generation of Electroweak calculations to higher order in Electroweak perturbation theory, that are perhaps beyond the current state of the art, but within reach on the time scale required by TLEP.

\subsection{Global fit of the EWSB parameters}
\label{sec:globalFit}

Once the Higgs boson mass is measured and the top quark mass determined with a precision of a few tens of MeV, the Standard Model prediction of a number of observables sensitive to Electroweak radiative corrections will become absolute with no remaining additional parameters. Any deviation will be a demonstration of the existence of new, weakly interacting particle(s). As was seen in the previous chapters, TLEP will offer the opportunity of measurements of such quantities with  precisions between one and two orders of magnitude better than the present status of these measurements. The theoretical prediction of these quantities with a matching precision will be a real challenge -- as discussed in the next section -- but the ability of these tests of the completeness of the Standard Model to discover new weakly-interacting particles beyond those already known is real. 

\begin{figure}[tb]
\begin{center}
\includegraphics[width=0.7\columnwidth]{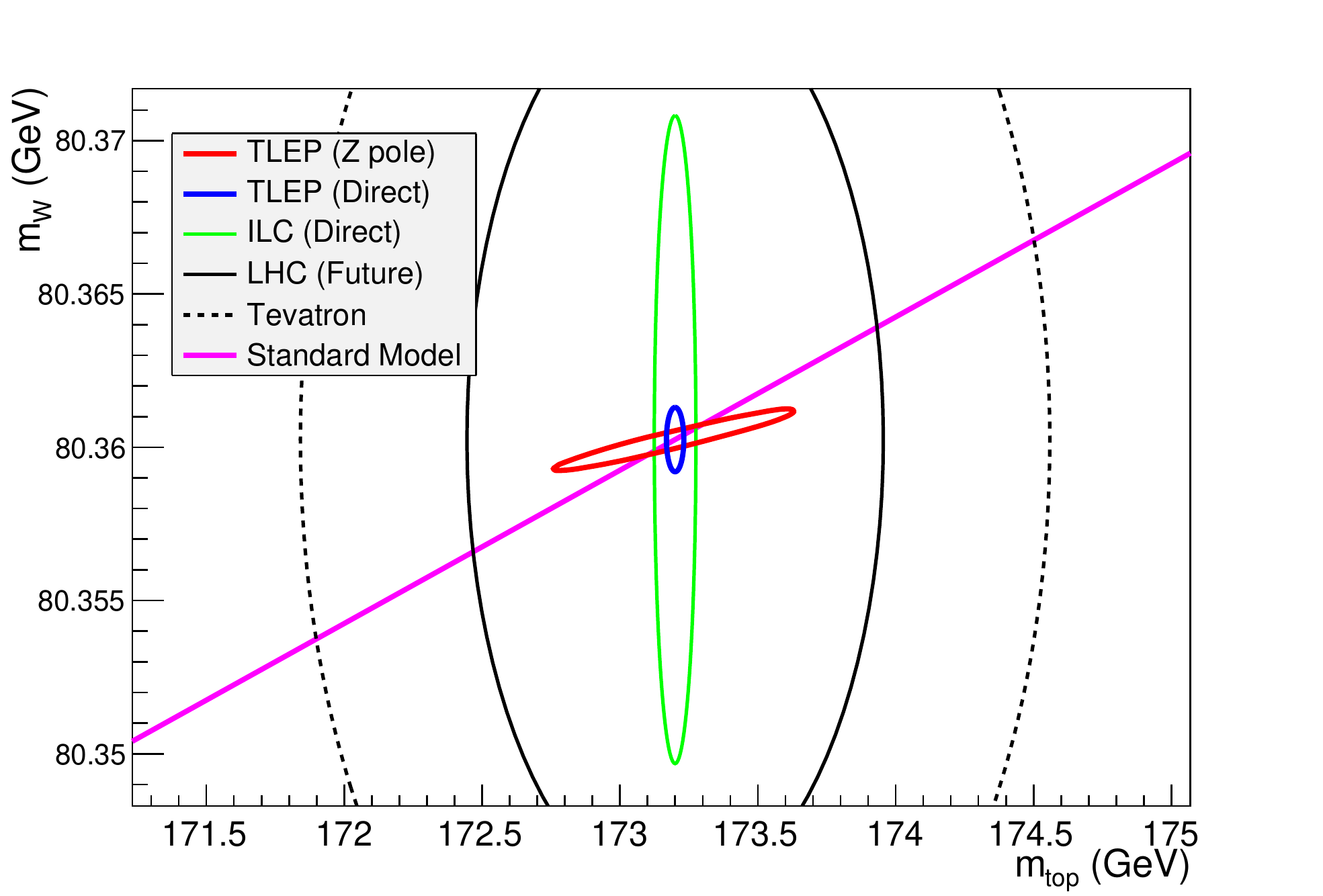}
\end{center}
\caption{\label{fig:GFitter1} The 68\% C.L. contour from the fit of all Electroweak precision measurements from TLEP-Z  (red curve) in the $(m_{\rm top},m_{\rm W})$ plane, should the relevant theory uncertainties be reduced to match the TLEP experimental uncertainties, compared to the direct W and top mass precisions (blue curve) expected at TLEP-W and TLEP-t. For illustration, the LHC (black curve) and ILC (green curve) projections for the direct $m_{\rm W}$ and $m_{\rm top}$ precisions are also indicated, as well as the current precision of the Tevatron measurements (dashed curve). The value of the Tevatron W mass was modified in this figure to match the SM prediction for $m_{\rm top}=$ 173.2~GeV. The purple line shows the prediction from the Standard Model for $m_{\rm H}=$ 125~GeV. (For the LHC or the ILC on their own, the thickness of this line would need to be increased by at least the error stemming from the Z mass measured at LEP, i.e., about $\pm 2$~MeV on the W mass. This error disappears in the case of TLEP.)  No theory error was included in this line. }
\end{figure}

\begin{figure}[tb]
\begin{center}
\includegraphics[width=0.7\columnwidth]{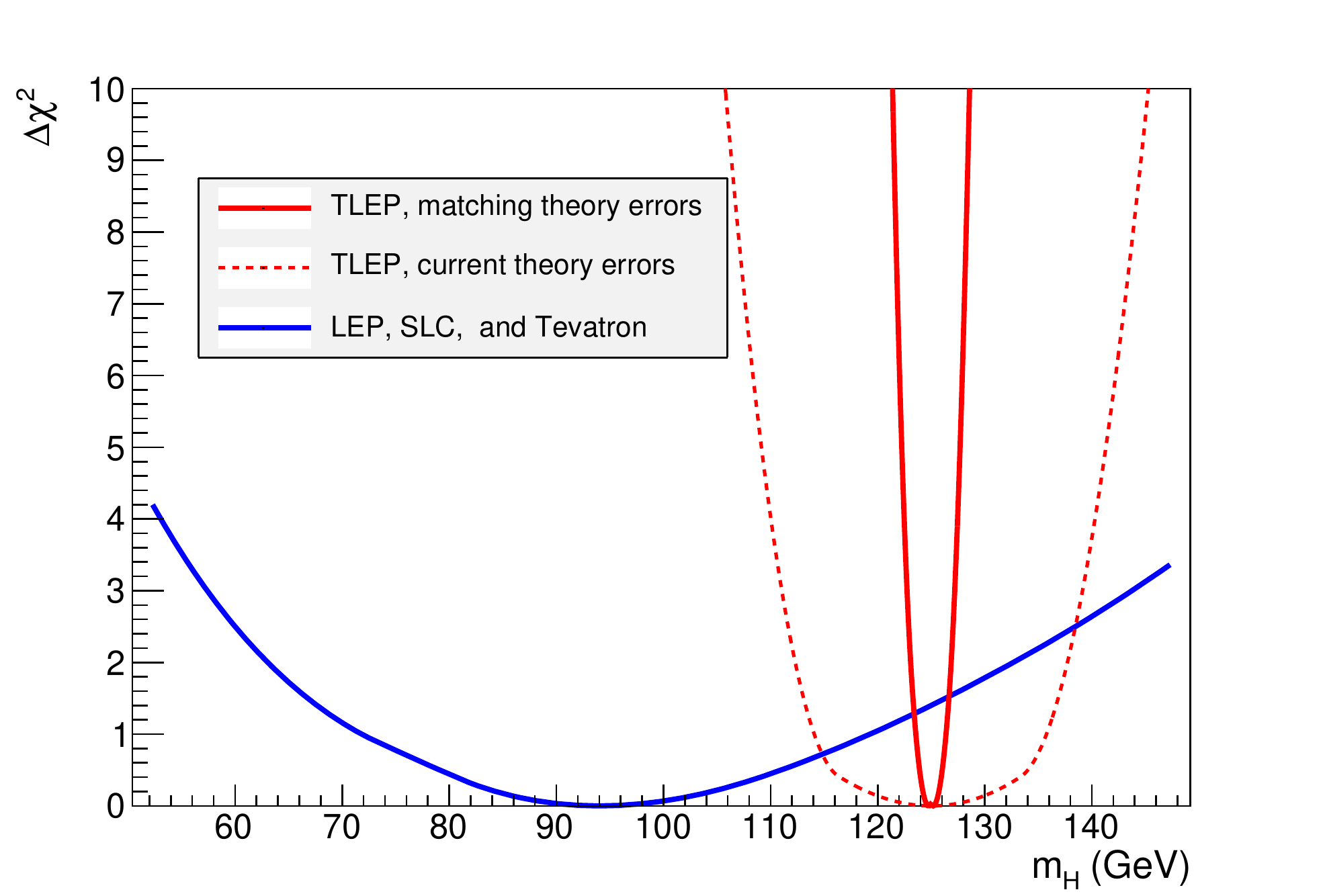}
\end{center}
\caption{\label{fig:GFitter2} The $\Delta\chi^2$ of the Standard-Model Higgs boson mass fit to the projected TLEP precision measurements (red curve) (with the exception of the direct Higgs boson mass measurement), compared to the $\Delta\chi^2$ of the current fit to the LEP, SLC and Tevatron measurements (blue curve). A precision of 1.4~GeV can be obtained on $m_{\rm H}$, should the relevant theory uncertainties be reduced to match the TLEP experimental uncertainties. The dashed curve shows the result of the fit with the current theory uncertainties, as implemented in Ref.~\cite{gler_Monig_Schott_Stelzer_2012}.}
\end{figure}

As an illustration, the result of the fit of the Standard Model to all the Electroweak measurements foreseen with TLEP-Z, as obtained with the GFitter program~\cite{gler_Monig_Schott_Stelzer_2012} under the assumptions that all relevant theory uncertainties can be reduced to match the experimental uncertainties and that the error on $\alpha_{\rm em}(m_{\rm Z})$ can be reduced by a factor 5, is displayed in Fig.~\ref{fig:GFitter1} as 68\% C.L. contours in the $(m_{\rm top},m_{\rm W})$ plane. This fit is compared to the direct $m_{\rm W}$ and $m_{\rm top}$ measurements expected from \mbox{TLEP-W} and \mbox{TLEP-t}. For illustration, a comparison with the precisions obtained with the current Tevatron data, as well as from LHC and ILC projections, is also shown. Among the many powerful tests that will become available with TLEP data, an inclusive, albeit unidimensional, test is commonly proposed by the most popular fitting programmes, namely the comparison of the Higgs boson mass prediction from all Electroweak observables with the mass actually measured. Figure~\ref{fig:GFitter2} shows the $\Delta\chi^2$ of the Higgs boson mass fit, obtained from GFitter under the same assumptions, to the TLEP Electroweak precision measurements. A precision of 1.4~GeV on $m_{\rm H}$ is predicted if all related theory uncertainties can be reduced to match the experimental uncertainties. If the theory uncertainties were kept as they are today~\cite{gler_Monig_Schott_Stelzer_2012}, the precision on $m_{\rm H}$ would be limited to about 10~GeV, as shown also in Fig.~\ref{fig:GFitter2}.


\section{High-energy upgrades}
\label{sec:VHE-LHC}

The European Strategy update recalls the strong physics case of an $\epem$ collider for the measurement of the Higgs boson and other particle properties with unprecedented precision. As demonstrated in Sections~\ref{sec:Higgs} and~\ref{sec:EWSB}, the TLEP project superbly qualifies for this purpose. The projected precisions are sufficient to achieve sensitivities to new physics up to 5~TeV if it couples to the scalar sector, and up to 30~TeV for weakly-coupled new physics. The European Strategy update also states that the project must be upgradeable to higher energies. It is therefore important to evaluate the scientific relevance of a possible energy upgrage of TLEP in the context of the FCC project, especially when compared to (multi-)TeV $\epem$ colliders.

Both $\epem$ Higgs factories discussed in Section~\ref{sec:Higgs} (TLEP and ILC) have high-energy upgrade options. In the case of TLEP, the centre-of-mass energy  can be increased to $\sqrt{s} = 500$~GeV by tripling the RF length from 600 to 1700~m, thereby increasing the total RF voltage from 12 to 35~GV to compensate for the 31~GeV lost per turn by synchrotron radiation in the 100~km ring. In the case of the ILC, its length can be doubled to reach a centre-of-mass energy of 500 GeV. 

With a 2.5\% momentum acceptance at each interaction point, TLEP-500 would have a one-minute beam lifetime, which would allow for an average luminosity of 90\% of the peak luminosity with the baseline TLEP top-off injection scheme.  With these parameters, a luminosity of $0.5\times10^{34}~\cms$ would be delivered at each interaction point with a beam-beam tune shift of 0.1, for a total luminosity of $2 \times 10^{34}~\cms$ when summed over the four IPs, as displayed in Fig.~\ref{fig:lumi}. Although not included in the TLEP baseline programme at this time, the design study will investigate the feasibility of such an option and define the maximum reachable centre-of-mass energy under reasonable assumptions. 

The possibility of further increasing the centre-of-mass energy of the ILC by another factor of two to $\sqrt{s}=$ 1 TeV has also been considered. The other linear collider project, CLIC~\cite{cite:CLICDR}, could provide a higher-energy physics programme all the way to $\sqrt{s}=$ 3 TeV. It would require, however, considerably more electrical power, estimated at $\sim 600$~MW.

The ultimate energy-frontier option for TLEP, however, is of a very different and more ambitious nature. In the context of the FCC,  it would consist of using the 80~to~100~km tunnel to host a very-high-energy large hadron collider, the VHE-LHC. If equipped with magnets of 15~T, pp collisions could be produced at a centre-of-mass energy of 80 to 100 TeV, giving access to the direct production of new coloured particles with masses of up to 30~TeV. (For completeness, we also note that pp collisions with a centre-of-mass energy of 33 TeV could be obtained by re-using the LHC tunnel for a pp collider using 20~T magnets, the high-energy large hadron collider, HE-LHC.)

\subsection{Higgs physics in $\epem$ collisions at $\sqrt{s} = 500$~GeV}
\label{sec:ee500}

The TLEP physics potential at this centre-of-mass energy would be similar to that of the linear colliders ILC and CLIC, which have nominal luminosities that are comparable at $\sqrt{s}=500$~GeV. The ILC TDR~\cite{ILC:Physics} shows that the addition of 500~$\infb$ at 500~GeV to the baseline programme with 250~$\infb$ at 250~GeV and 350~$\infb$ at 350 GeV would improve the precision on all Higgs boson couplings to light fermions and gauge bosons by less than a factor 1.5 (still  far from the sub-per-cent precision provided by TLEP at 240~GeV), and by a negligible amount at TLEP. The measurement of the invisible width of the Higgs boson would not be improved in either case.   

On the other hand, the opening of the $\epemto \ttbar{\rm H}$ process allows the Htt coupling to be measured directly, typically with a precision of 10 to 15\%. However, the improvement with respect to the TLEP measurement at the $\ttbar$ threshold, which has an accuracy of 13\%, is marginal. More importantly, these precisions are not competitive with the HL-LHC projections~\cite{1307.7135,1307.7292}. For example, the CMS collaboration would be able to measure the Htt coupling with an accuracy of 4\%~\cite{petruc} with an integrated luminosity of 3~$\inab$. 

Similarly, the opening of the $\epemto {\rm ZHH}$ and $\nnbar{\rm HH}$ processes at $\sqrt{s} = 500$~GeV enables a ``measurement'' of the triple Higgs-boson self-coupling, $\lambda_{\rm H}$, with 50 to 80\% precision. Again, these accuracies are not competitive with the HL-LHC projections, for which a 30\% accuracy on $\lambda_{\rm H}$ is envisioned. 

At this stage of the study, it appears that once sufficient $\epem$ data are collected at 250~and 350~GeV, the potential gain in Higgs physics alone is not enough to justify an upgrade to a centre-of-mass energy of 500~GeV. On the other hand, as discussed below, the appearance of some threshold for new physics above 350~GeV could change the picture entirely.

\subsection{Higgs physics at higher energy}

\subsubsection{The Htt coupling}

As mentioned in Section~\ref{sec:ee500}, a centre-of-mass energy of 500 GeV cannot compete with the HL-LHC for the Htt coupling measurement. To reach an accuracy in $\epem$ collisions similar to the HL-LHC (less than 4\%), the upgrade of either ILC up to $\sqrt{s}=1$~TeV or CLIC up to $\sqrt{s}=3$~TeV is needed. A precision of 4\% on the Htt coupling would be achieved with an integrated luminosity of 1~$\inab$ (ILC-1000) or 2~$\inab$ (CLIC). On the other hand, an integrated luminosity of 3~$\inab$ with pp collisions at either the HE-LHC or the VHE-LHC would allow the precision on the Htt coupling to be significantly improved to a couple of per-cent or a fraction of a per-cent, respectively, making the FCC project quite appealing in this respect.

\subsubsection{The HHH coupling}

\begin{figure}[tb]
\begin{center}
\includegraphics[width=0.7\columnwidth]{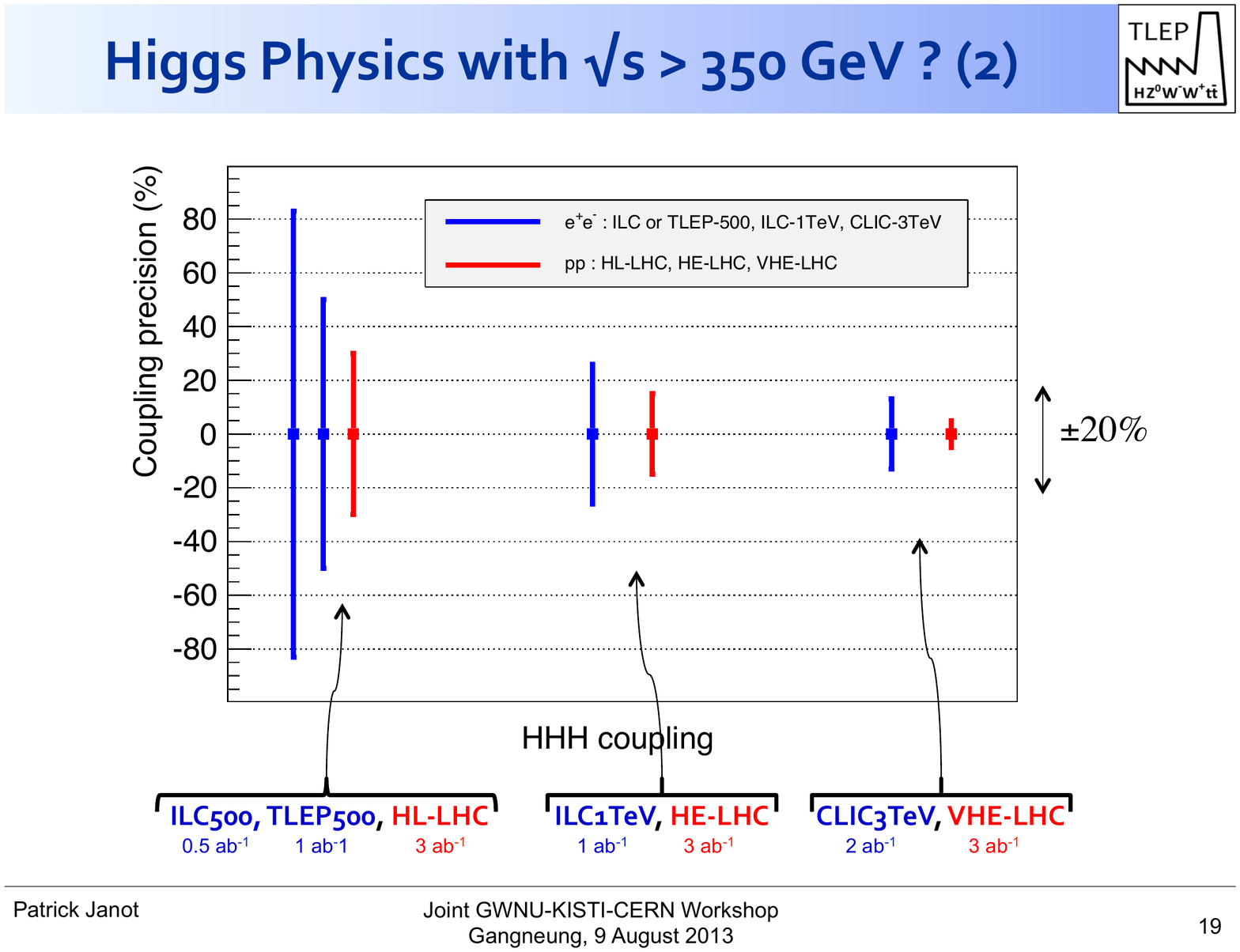}
\end{center}
\caption{\label{fig:VHELHC} Expected relative statistical accuracy in \% on the trilinear Higgs self-coupling for $\epem$ (blue) and pp (red) colliders at the high-energy frontier. The accuracy estimates are given, from left to right, for ILC500, TLEP500, HL-LHC, ILC1000, HE-LHC, CLIC and VHE-LHC, for integrated luminosities of 0.5, 1, 3, 1, 3, 2, and 3~$\inab$, respectively. }
\end{figure}

The measurement of the trilinear Higgs self-coupling $\lambda_{\rm H}$ would benefit substantially from higher energy, because of the fast increase of the double-Higgs-boson production cross section, in both $\epem$ and proton-proton collisions. Studies exist, albeit with different levels of maturity, for the sensitivity of the ILC~\cite{ILC:Physics}, CLIC~\cite{cite:CLICDR}, and HL-LHC~\cite{cite:ATLAS,cite:CMS} to this coupling. From the HL-LHC estimates and from the known HH production cross-section increase at higher energies~\cite{Mangano_Rojo_2012}, extrapolations for $3~\inab$ of pp collision data at the HE-LHC and the VHE-LHC can be inferred~\cite{Snowmass}. An executive summary of the achievable precisions is displayed in Fig.~\ref{fig:VHELHC}. 

A measurement of the trilinear Higgs self-coupling with a significance of at least 5$\sigma$ can only be done at the HE-LHC, CLIC or the VHE-LHC, with projected precisions in the Standard Model of 15\%, 10 to 16\%, and 5\%, respectively. Since deviations in the HHH coupling arising from new physics effects are expected to be smaller than $\pm 20\%$ with respect to the Standard Model prediction~\cite{cite:1305.6397}, such new physics effects could only be probed at the VHE-LHC. The VHE-LHC is also the only machine that could have a say on the quartic self-coupling~\cite{QuarticHiggs}, needed to fully understand Electroweak Symmetry Breaking. 

\noindent In summary, the potential of the FCC project for Higgs physics cannot be challenged by any other projects on the market.

\subsection{Direct search for new physics}

As seen above, the case for $\epem$ collisions with centre-of-mass energy of 500 GeV and above is not compelling for the study of the H(126) particle alone. A stronger motivation would exist if a new particle were found (or inferred) at LHC during the next run at 13-14 TeV, if and only if $\epem$ collisions could bring substantial new information about it. 

Typically, $\epem$ colliders can pair-produce new particles with masses up to half the centre-of-mass energy, if they are either electrically charged or have a non-vanishing coupling to the Z. The reach of ILC500, ILC1000 and CLIC is therefore limited to particles lighter than 250, 500 and 1500~GeV, respectively. The lowest threshold for new particles could be that for pair-production of dark matter particles, such as the lightest neutralinos of supersymmetric models, through their Z or Higgs couplings, in association with an initial-state-radiation photon. This search was performed at LEP, but was limited by the kinematic reach and the large background from conventional neutrinos. Similar searches are performed at the LHC (mono-photon, mono-jet, accompanied with missing energy), but are competitive with astrophysical searches only for very small dark-matter particle masses. The high luminosity of TLEP up to centre-of-mass energies of 350 to 500~GeV, associated with the absence of photon background from beamstrahlung,  may provide a promising opportunity to extend the sensitivity of such single-photon searches for dark matter. 

The absence of new phenomena at the LHC so far has reduced the prospects for direct new physics discovery in $\epem$ collisions below 1~TeV in the centre of mass (with few exceptions like the aforementioned possible observation of light dark matter). The next LHC run at 13-14 TeV, to start in 2015, will bring clarity in this respect. Discovery of a new particle lighter than 1.5~TeV in the 13-14 TeV LHC data would rejuvenate the proposal of CLIC at $\sqrt{s} =$ 3 TeV.  A 100 TeV proton-proton collider,in the context of the FCC project, would instead be able to produce new coloured particles up to several tens of~TeV, thus opening a unique window at high energy. A detailed study of the VHE-LHC physics case has started in this context, in order to have relevant answers ready for the next European Strategy update, to take place around 2018.

\section{Conclusion}
\label{sec:conclusion}

The discovery at the LHC of a particle that resembles strongly the long-sought Higgs boson of the Standard Model has placed studies for the next large machine for high-energy physics in a new perspective. The prospects for the next decade already look quite promising: the HL-LHC is an impressive Higgs factory, with great potential for measuring many Higgs couplings with accuracies of a few per-cent. The LHC run at 13-14 TeV may well discover something else, and it would be premature to mortgage the future of high-energy physics before knowing what it reveals. In the meantime new ideas are emerging for possible future Higgs factories.

In view of the financial, technical and personnel resources needed for the next large high-energy physics instrument, it is essential to choose a strategy that provides complementarity to the LHC, with optimal capabilities beyond what can be achieved with HL-LHC, in both precision measurements and/or discovery potential. 

In our view, TLEP,  a large $\epem$ circular collider in a tunnel with 80 to 100~km circumference, would best complement the LHC, as it would provide {\it (i)} per-mil precision in measurements of Higgs couplings, {\it (ii)} unique precision in measurements of Electroweak Symmetry-Breaking parameters and the strong coupling constant, {\it (iii)} a measurement of the Z invisible width equivalent to better than 0.001 of a conventional neutrino species, and {\it (iv)} a unique search programme for rare Z, W, Higgs, and top decays. We emphasize that circular $\epem$ colliders use a mature technology that has been developed during the construction and operation of successive $\epem$ machines over 50 years, and in particular in a very similar regime at LEP2. Many of the key technical advances that make TLEP possible will be demonstrated by SuperKEKB, which has many parameters similar to TLEP. Experience with SuperKEKB will make more reliable the cost estimates, power evaluations, and luminosity predictions for TLEP. Moreover, TLEP would be a stepping-stone towards a 100 TeV pp collider in the same tunnel, and therefore provides a unique long-term vision for high-energy physics. The FCC project -- namely the combination of TLEP and the VHE-LHC -- offers, for a great cost effectiveness, the best precision and the best search reach of all options presently on the market.

The design study of TLEP has now started, in close collaboration with the VHE-LHC design study, with worldwide collaboration from Asia, USA and Europe, and with full support from the CERN Council. The study is now included in the approved CERN Medium-Term Plan for the years 2014-2018. The first proposed step is a design study report in 2015, to be followed by a conceptual design report and a detailed cost estimate in 2018-2019. In this paper, we have taken a first look at a potentially very rich TLEP physics programme, which can serve as a baseline for a comprehensive exploration of its possibilities during this period. An informed decision on the FCC project could then be taken in full knowledge of the LHC results at 13-14 TeV and operational experience with SuperKEKB. Technically, and if given the necessary financial and political support, TLEP could be ready for physics in 2030.

\section*{Acknowledgements}

We are indebted to Andreas Hoecker for his help with the GFitter fitting program (http://cern.ch/gfitter), and for his patient explanations of the underlying physics. We would like to acknowledge the contributions of all participants in the design study and in the first five TLEP workshops. Writing this article was greatly eased by the use of the online collaborative editor, Authorea (https://www.authorea.com/), ceaselessly improved by the founders, Nathan Jenkins and Alberto Pepe. The support of the CERN Director for Accelerators and Technology and of the PH department, of the European Commission under the FP7 Research Infrastructures project EuCARD, grant agreement no. 227579 ( http://cern.ch/eucard) and under the FP7 Capacities project EuCARD-2, grant agreement no. 312453 (http://cern.ch/eucard2), and of the Swiss National Foundation under the grant 200021-144133, are gratefully acknowledged. The work of J.E. was supported in part by the London Centre for Terauniverse Studies (LCTS), using funding from the European Research Council via the Advanced Investigator Grant 267352.

\bibliography{converted_to_latex.bib}

\end{document}

%% file: authors.tex
\noindent{\bf The TLEP Design Study Working Group}\\ \\
M. Bicer\\
{\it\small (Faculty of Science, Ankara University, Ankara, Turkey)}\\
H. Duran Yildiz\\
{\it\small (IAT, Ankara University, Ankara, Turkey)}\\
I. Yildiz\\
{\it\small (Middle East Technical University, Ankara, Turkey)}\\
G.~Coignet, M.~Delmastro\\
{\it\small (Laboratoire d'Annecy-Le-Vieux de Physique des Particules, IN2P3/CNRS, Annecy-Le-Vieux, France)}\\
T.~Alexopoulos\\
{\it\small (National Technical University of Athens, Athens, Greece)}\\
C.~Grojean\\
{\it\small (Institucio Catalana de Recerca i Estudis, Barcelona, Spain)}\\
S. Antusch\\
{\it\small (Universität Basel, Basel, Switzerland)}\\
T. Sen\\
{\it\small (Fermilab, Batavia IL, United States)}\\
H.-J. He\\
{\it\small (Tsinghua University, Beijing, China)}
% ...


%% file: JHEP.bbl
\begin{thebibliography}{10}
\expandafter\ifx\csname url\endcsname\relax
  \def\url#1{\texttt{#1}}\fi
\expandafter\ifx\csname urlprefix\endcsname\relax\def\urlprefix{URL }\fi
\providecommand{\bibinfo}[2]{#2}
\providecommand{\eprint}[2][]{\url{#2}}

\bibitem{v_Aben_Abi_Abolins_et_al__2012}
\bibinfo{author}{Aad, G.} \emph{et~al.}
\newblock \bibinfo{title}{{Observation of a new particle in the search for the
  Standard Model Higgs boson with the ATLAS detector at the LHC}}.
\newblock \emph{\bibinfo{journal}{Physics Letters B}}
  \textbf{\bibinfo{volume}{716}}, \bibinfo{pages}{1--29}
  (\bibinfo{year}{2012}).
\newblock \emph{\bibinfo{journal}{ArXiv e-prints} \eprint{1207.7214}}.

\bibitem{gicevic_Ero_Fabjan_et_al__2012}
\bibinfo{author}{Chatrchyan, S.} \emph{et~al.}
\newblock \bibinfo{title}{{Observation of a new boson at a mass of 125 GeV with
  the CMS experiment at the LHC}}.
\newblock \emph{\bibinfo{journal}{Physics Letters B}}
  \textbf{\bibinfo{volume}{716}}, \bibinfo{pages}{30--61}
  (\bibinfo{year}{2012}).
\newblock \emph{\bibinfo{journal}{ArXiv e-prints} \eprint{1207.7235}}.

\bibitem{ATLASHiggsCouplings}
\bibinfo{author}{ATLAS Collaboration}.
\newblock \bibinfo{title}{{Combined coupling measurements of the Higgs-like
  boson with the ATLAS detector using up to 25 ${\rm fb}^{-1}$ of proton-proton
  collision data}}.
\newblock \bibinfo{type}{Tech. Rep.} \bibinfo{number}{ATLAS-CONF-2013-034}
  (\bibinfo{year}{2013}).
\newblock \urlprefix\url{http://cds.cern.ch/record/1528170}.
\newblock\bibinfo{author}{ibid}.
\newblock\bibinfo{title}{{Measurements of Higgs boson production and
    couplings in diboson final states with the ATLAS detector at the
    LHC}}.
\newblock \emph{\bibinfo{journal}{Physics Letters B}}
  \textbf{\bibinfo{volume}{726}}, \bibinfo{pages}{88--119}
  (\bibinfo{year}{2013}).
\newblock \emph{\bibinfo{journal}{ArXiv e-prints} \eprint{1307.1427}}.

\bibitem{Ero_Fabjan_Friedl_et_al__2013}
\bibinfo{author}{Chatrchyan, S.} \emph{et~al.}
\newblock \bibinfo{title}{{Observation of a new boson with mass near 125 GeV in
  pp collisions at $ \sqrt{s}=7 $ and 8 TeV}}.
\newblock \emph{\bibinfo{journal}{Journal of High Energy Physics}}
  \textbf{\bibinfo{volume}{6}}, \bibinfo{pages}{81} (\bibinfo{year}{2013}).
\newblock \urlprefix\url{http://dx.doi.org/10.1007/JHEP06(2013)081}.

\bibitem{cite:1303.3879}
\bibinfo{author}{{Ellis}, J.} \& \bibinfo{author}{{You}, T.}
\newblock \bibinfo{title}{{Updated global analysis of Higgs couplings}}.
\newblock \emph{\bibinfo{journal}{Journal of High Energy Physics}}
  \textbf{\bibinfo{volume}{6}}, \bibinfo{pages}{103} (\bibinfo{year}{2013}).
\newblock \emph{\bibinfo{journal}{ArXiv e-prints} \eprint{1303.3879}}.

\bibitem{ILC:Physics}
\bibinfo{author}{{Baer}, H.} \emph{et~al.}
\newblock \bibinfo{title}{{The International Linear Collider Technical Design
  Report - Volume 2: Physics}}.
\newblock \emph{\bibinfo{journal}{ArXiv e-prints} \eprint{1306.6352}} (\bibinfo{year}{2013}).
\newblock \urlprefix\url{http://arxiv.org/abs/arXiv:1306.6352}.

\bibitem{Gupta_Rzehak_Wells_2012}
\bibinfo{author}{Gupta, R.~S.}, \bibinfo{author}{Rzehak, H.} \&
  \bibinfo{author}{Wells, J.~D.}
\newblock \bibinfo{title}{{How well do we need to measure Higgs boson
  couplings?}}
\newblock \emph{\bibinfo{journal}{Physical Review D}}
  \textbf{\bibinfo{volume}{86}} (\bibinfo{year}{2012}).
\newblock \urlprefix\url{http://dx.doi.org/10.1103/PhysRevD.86.095001}.

\bibitem{cite:1305.6498}
\bibinfo{author}{{Koratzinos}, M.} \emph{et~al.}
\newblock \bibinfo{title}{{TLEP: A High-Performance Circular $e^+e^-$ Collider
  to Study the Higgs Boson}} 
\newblock \emph{\bibinfo{journal}{ArXiv e-prints} \eprint{1305.6498}}
(\bibinfo{year}{2013}).
\newblock \urlprefix\url{http://arxiv.org/abs/1305.6498}.

\bibitem{cite:Osborne}
\bibinfo{author}{Osborne, J.~A.} \& \bibinfo{author}{Waaijer, C.~S.}
\newblock \bibinfo{title}{{Contribution to the Open Symposium of the European
  Strategy Preparatory Group: Pre-Feasability Assessment for an 80 km Tunnel
  Project at CERN}} (\bibinfo{year}{2012}).
\newblock
  \urlprefix\url{https://indico.cern.ch/contributionDisplay.py?contribId=165&confId=175067}.

\bibitem{HF2012}
\bibinfo{author}{Blondel, A.}, \bibinfo{author}{Chao, A.},
\bibinfo{author}{Chou, W.}, \bibinfo{author}{Gao, J.},
\bibinfo{author}{Schulte, D.},\bibinfo{author}{ {\it et al.}}
\newblock \bibinfo{title}{{Report of the ICFA Beam Dynamics Workshop 
``Accelerators for a Higgs Factory: Linear vs. Circular'' (HF2012)}}
(\bibinfo{year}{2012}).
\newblock \emph{\bibinfo{journal}{ArXiv e-prints} \eprint{1302.3318}}
\newblock \urlprefix\url{http://arxiv.org/abs/1302.3318}.

\bibitem{cite:Strategy}
\bibinfo{author}{CERN-Council}.
\newblock \bibinfo{title}{{The European Strategy for Particle Physics}}
  (\bibinfo{year}{2013}).
\newblock
  \urlprefix\url{http://council.web.cern.ch/council/en/EuropeanStrategy/ESParticlePhysics.html}.

\bibitem{cite:MTP}
\bibinfo{author}{CERN}.
\newblock \bibinfo{title}{{Medium-Term Plan for the period 2014-2018 and Draft
  Budget of the Organization for the sixtieth financial year 2014}}.
\newblock \bibinfo{type}{Tech. Rep.} (\bibinfo{year}{2013}).
\newblock \urlprefix\url{http://cds.cern.ch/record/1557135}.

\bibitem{1112.2518}
\bibinfo{author}{{Blondel}, A.} \& \bibinfo{author}{{Zimmermann}, F.}
\newblock \bibinfo{title}{{A High Luminosity $e^+e^-$ Collider in the LHC
  tunnel to study the Higgs Boson}}.
\newblock \emph{\bibinfo{journal}{ArXiv e-prints} \eprint{1112.2518}}  
(\bibinfo{year}{2011}).
\newblock \urlprefix\url{http://arxiv.org/abs/1112.2518}.

\bibitem{Assmann:453821}
\bibinfo{author}{A{\ss}mann, R.} \& \bibinfo{author}{Cornelis, K.}
\newblock \bibinfo{title}{{The Beam-Beam Interaction in the Presence of Strong
  Radiation Damping}} \bibinfo{pages}{4 p} (\bibinfo{year}{2000}).
\newblock \urlprefix\url{http://cds.cern.ch/record/453821/}.

\bibitem{ILC:AcceleratorA}
\bibinfo{author}{{Adolphsen}, C.} \emph{et~al.}
\newblock \bibinfo{title}{{The International Linear Collider Technical Design
  Report - Volume 3.I: Accelerator R{\&}D in the Technical Design Phase}}.
\newblock \emph{\bibinfo{journal}{ArXiv e-prints} \eprint{1306.6353}}  (\bibinfo{year}{2013}).
\newblock \urlprefix\url{http://arxiv.org/abs/arXiv:1306.6353}.

\bibitem{ILC:AcceleratorB}
\bibinfo{author}{{Adolphsen}, C.} \emph{et~al.}
\newblock \bibinfo{title}{{The International Linear Collider Technical Design
  Report - Volume 3.II: Accelerator Baseline Design}}.
\newblock \emph{\bibinfo{journal}{ArXiv e-prints} \eprint{1306.6328}}  (\bibinfo{year}{2013}).
\newblock \urlprefix\url{http://arxiv.org/abs/arXiv:1306.6328}.

\bibitem{cite:CLICDR}
\bibinfo{author}{Aicheler, M.} \emph{et~al.}
\newblock \bibinfo{title}{{A Multi-TeV linear collider based on CLIC
  technology: CLIC Conceptual Design Report}}.
\newblock \bibinfo{type}{Tech. Rep.} \bibinfo{number}{CERN-2012-007}
  (\bibinfo{year}{2012}).
\newblock
  \urlprefix\url{http://project-clic-cdr.web.cern.ch/project-CLIC-CDR/CDR_Volume1.pdf}.

\bibitem{1308.3726}
\bibinfo{author}{{Harrison}, M.}, \bibinfo{author}{{Ross}, M.} \&
  \bibinfo{author}{{Walker}, N.}
\newblock \bibinfo{title}{{Luminosity Upgrades for ILC}}
\newblock \emph{\bibinfo{journal}{ArXiv e-prints} \eprint{1308.3726}}  (\bibinfo{year}{2013}).
\newblock \urlprefix\url{http://arxiv.org/abs/1308.3726}.

\bibitem{TelnovC}
\bibinfo{author}{{Telnov}, V.~I.}
\newblock \bibinfo{title}{{Limitation on the luminosity of e+e- storage rings
  due to beamstrahlung}}.
\newblock \emph{\bibinfo{journal}{ArXiv e-prints} \eprint{1307.3915}}  (\bibinfo{year}{2013}).
\newblock \urlprefix\url{http://arxiv.org/abs/1307.3915}.

\bibitem{TelnovD}
\bibinfo{author}{Telnov, V.~I.}
\newblock \bibinfo{title}{{Problems of charge compensation in a ring $e^+e^-$
  Higgs factory, Talk given at the 5th TLEP Workshop}} (\bibinfo{year}{2013}).
\newblock
  \urlprefix\url{https://indico.fnal.gov/getFile.py/access?contribId=6&resId=0&materialId=slides&confId=6983}.

\bibitem{Yokoya}
\bibinfo{author}{Yokoya, K.}
\newblock \bibinfo{title}{{Scaling of High-Energy $e^+e^-$ Ring Colliders, KEK Accelerator Seminar}} 
(\bibinfo{year}{2012}).\break
\newblock
  \urlprefix\url{http://accl.kek.jp/seminar/file/AccPhys_Yokoya_RingColliderScaling-AcceleratorSeminar-2012-0315-Yokoya.pptx}.

\bibitem{TelnovB}
\bibinfo{author}{{Telnov}, V.~I.}
\newblock \bibinfo{title}{{Restriction on the Energy and Luminosity of
  e$^{+}$e$^{-}$ Storage Rings due to Beamstrahlung}}.
\newblock \emph{\bibinfo{journal}{Physical Review Letters}}
  \textbf{\bibinfo{volume}{110}}, \bibinfo{pages}{114801}
  (\bibinfo{year}{2012}).
\newblock \emph{\bibinfo{journal}{ArXiv e-prints} \eprint{1203.6563}}.

\bibitem{LEP1Cal}
\bibinfo{author}{{A{\ss}mann}, R.} \emph{et~al.}
\newblock \bibinfo{title}{{Calibration of centre-of-mass energies at
    LEP1 for precise measurements of Z properties}}
\newblock \emph{\bibinfo{journal}{Eur.Phys.J.}} \textbf{\bibinfo{volume}{C6}},
  \bibinfo{pages}{187--223} (\bibinfo{year}{1999}).

\bibitem{ement_of_the_W_boson_mass_2005}
\bibinfo{author}{A{\ss}mann, R.} \bibinfo{author}{{\it et al.}}
  \bibinfo{author}{{(The LEP Energy Working Group)}}.
\newblock \bibinfo{title}{{Calibration of centre-of-mass energies at LEP2 for a
  precise measurement of the W boson mass}}.
\newblock \emph{\bibinfo{journal}{The European Physical Journal C}}
  \textbf{\bibinfo{volume}{39}}, \bibinfo{pages}{253--292}
  (\bibinfo{year}{2005}).
\newblock \urlprefix\url{http://dx.doi.org/10.1140/epjc/s2004-02108-8}.

\bibitem{altay_Band_Barklow_et_al__2000}
\bibinfo{author}{Abe, K.} \emph{et~al.}
\newblock \bibinfo{title}{{High-Precision Measurement of the Left-Right Z Boson
  Cross-Section Asymmetry}}.
\newblock \emph{\bibinfo{journal}{Physical Review Letters}}
  \textbf{\bibinfo{volume}{84}}, \bibinfo{pages}{5945--5949}
  (\bibinfo{year}{2000}).
\newblock \urlprefix\url{http://dx.doi.org/10.1103/PhysRevLett.84.5945}.

\bibitem{1206.2913}
\bibinfo{author}{Fernandez, J. L.~A.} \bibinfo{author}{{\it et al}}
  \bibinfo{author}{(The LHeC Working~Group)}.
\newblock \bibinfo{title}{{A Large Hadron Electron Collider at CERN: Report on
  the Physics and Design Concepts for Machine and Detector}}
\newblock \emph{\bibinfo{journal}{ArXiv e-prints} \eprint{1206.2913}}  (\bibinfo{year}{2012}).
\newblock \urlprefix\url{http://arxiv.org/abs/1206.2913}.

\bibitem{cite:Blondel-Jowett-LEP606}
\bibinfo{author}{Blondel, A.} \& \bibinfo{author}{Jowett, J.~M.}
\newblock \bibinfo{title}{{Dedicated wigglers for polarization}}.
\newblock \bibinfo{type}{Tech. Rep.} \bibinfo{number}{CERN-LEP-Note-606.
  LEP-Note-606}, \bibinfo{institution}{CERN}, \bibinfo{address}{Geneva}
  (\bibinfo{year}{1988}).
\newblock \urlprefix\url{https://cds.cern.ch/record/442913}.

\bibitem{cite:LEP-beam-beam-pol}
\bibinfo{author}{A{\ss}mann, R.} \bibinfo{author}{{\it et al.}}
\newblock \bibinfo{title}{{Experiments on beam-beam depolarization at LEP,
  Proceedings of the 1995 Particle Accelerator Conference, May 1{\textendash}5
  1995, Dallas, Texas}} (\bibinfo{year}{1995}).
\newblock
  \urlprefix\url{http://accelconf.web.cern.ch/AccelConf/p95/ARTICLES/RAA/RAA19.PDF}.

\bibitem{cite:HERA-beams}
\bibinfo{author}{Hoffstaetter, G.}, \bibinfo{author}{Vogt, M.} \&
  \bibinfo{author}{Willeke, F.}
\newblock \bibinfo{title}{{Experience with HERA beams}} (\bibinfo{year}{2003}).
\newblock
  \urlprefix\url{http://icfa-usa.jlab.org/archive/newsletter/icfa_bd_nl_30.pdf}.

\bibitem{Blondel:1987wr}
\bibinfo{author}{Blondel, A.}
\newblock \bibinfo{title}{{A scheme to measure the polarization asymmetry at
  the Z pole in LEP}}.
\newblock \emph{\bibinfo{journal}{Phys.Lett.}} \textbf{\bibinfo{volume}{B202}},
  \bibinfo{pages}{145} (\bibinfo{year}{1988}).

\bibitem{cite:Wienans-TLEP4}
\bibinfo{author}{Wienans, U.}
\newblock \bibinfo{title}{{Is Polarization possible in TLEP?, 4th TLEP
  workshop}} (\bibinfo{year}{2013}).
\newblock
  \urlprefix\url{https://indico.cern.ch/contributionDisplay.py?contribId=26&confId=240814}.

\bibitem{on_Blondel_Assmann_Dehning_1992}
\bibinfo{author}{Arnaudon, L.} \emph{et~al.}
\newblock \bibinfo{title}{{Measurement of LEP beam energy by resonant spin
  depolarization}}.
\newblock \emph{\bibinfo{journal}{Physics Letters B}}
  \textbf{\bibinfo{volume}{284}}, \bibinfo{pages}{431--439}
  (\bibinfo{year}{1992}).
\newblock \urlprefix\url{http://dx.doi.org/10.1016/0370-2693(92)90457-F}.

\bibitem{cite:0506115}
\bibinfo{author}{{Hinze}, A.} \& \bibinfo{author}{{Moenig}, K.}
\newblock \bibinfo{title}{{Measuring the Beam Energy with Radiative Return
  Events}} 
\newblock \emph{\bibinfo{journal}{ArXiv e-prints} \eprint{physics/0506115}}  (\bibinfo{year}{2005}).
\newblock \urlprefix\url{http://arxiv.org/abs/physics/0506115}.

\bibitem{ILC:Detectors}
\bibinfo{author}{{Behnke}, T.} \emph{et~al.}
\newblock \bibinfo{title}{{The International Linear Collider Technical Design
  Report - Volume 4: Detectors}}.
\newblock \emph{\bibinfo{journal}{ArXiv e-prints} \eprint{1306.6329}}  (\bibinfo{year}{2013}).
\newblock \urlprefix\url{http://arxiv.org/abs/arXiv:1306.6329}.

\bibitem{cite:1208.1662}
\bibinfo{author}{{Azzi}, P.} \emph{et~al.}
\newblock \bibinfo{title}{{Prospective Studies for LEP3 with the CMS Detector}}
\newblock \emph{\bibinfo{journal}{ArXiv e-prints} \eprint{1208.1662}}  (\bibinfo{year}{2012}).
\newblock \urlprefix\url{http://arxiv.org/abs/1208.1662}.

\bibitem{Bediaga}
\bibinfo{author}{{Bediaga}, I.} \bibinfo{author}{{\it et al.}}
  \bibinfo{author}{{(LHCb Collaboration)}}.
\newblock \bibinfo{title}{{Framework TDR for the LHCb Upgrade: Technical Design
  Report}}.
\newblock \bibinfo{type}{Tech. Rep.} \bibinfo{number}{CERN-LHCC-2012-007.
  LHCb-TDR-12}, \bibinfo{institution}{CERN}, \bibinfo{address}{Geneva}
  (\bibinfo{year}{2012}).
\newblock \urlprefix\url{http://cds.cern.ch/record/1443882}.

\bibitem{OsbornePrivate}
\bibinfo{author}{Osborne, J.}
\newblock \bibinfo{title}{{Private communnication}}.
\newblock \bibinfo{type}{} (\bibinfo{year}{2013}).

\bibitem{cite:HZHA}
\bibinfo{author}{Janot, P.} \& \bibinfo{author}{Ganis, G.}
\newblock \bibinfo{title}{{The HZHA Generator, in Physics at LEP2, Eds, G.
  Altarelli, T. Sj{\o}strand and F. Zwirner}}.
\newblock \bibinfo{type}{CERN Report} \bibinfo{number}{96/01 (Vol.2) 309}
  (\bibinfo{year}{1996}).

\bibitem{ILC:Summary}
\bibinfo{author}{{Behnke}, T.} \emph{et~al.}
\newblock \bibinfo{title}{{The International Linear Collider Technical Design
  Report - Volume 1: Executive Summary}}.
\newblock \emph{\bibinfo{journal}{ArXiv e-prints} \eprint{1306.6327}}  (\bibinfo{year}{2013}).
\newblock \urlprefix\url{http://arxiv.org/abs/1306.6327}.

\bibitem{1305.5251}
\bibinfo{author}{{Craig}, N.}, \bibinfo{author}{{Englert}, C.} \&
  \bibinfo{author}{{McCullough}, M.}
\newblock \bibinfo{title}{{A New Probe of Naturalness}}.
\newblock \emph{\bibinfo{journal}{ArXiv e-prints} \eprint{1305.5251}}  (\bibinfo{year}{2013}).
\newblock \urlprefix\url{http://arxiv.org/abs/1305.5251}.

\bibitem{cite:durig}
\bibinfo{author}{D{\"{u}}rig, C.~F.}
\newblock \bibinfo{title}{{Determination of the Higgs Decay Width at ILC
  (Masterarbeit in Physik)}}.
\newblock \bibinfo{type}{Tech. Rep.} (\bibinfo{year}{2012}).
\newblock
  \urlprefix\url{http://lhc-ilc.physik.uni-bonn.de/thesis/Masterarbeitduerig.pdf}.

\bibitem{cite:1207.2516}
\bibinfo{author}{{Peskin}, M.~E.}
\newblock \bibinfo{title}{{Comparison of LHC and ILC Capabilities for Higgs
  Boson Coupling Measurements}} 
\newblock \emph{\bibinfo{journal}{ArXiv e-prints} \eprint{1207.2516}}  (\bibinfo{year}{2012}).
\newblock \urlprefix\url{http://arxiv.org/abs/1207.2516}.

\bibitem{1208.1533}
\bibinfo{author}{{Martin}, S.~P.}
\newblock \bibinfo{title}{{Shift in the LHC Higgs diphoton mass peak from
  interference with background}}.
\newblock \emph{\bibinfo{journal}{Physical Review D}}
  \textbf{\bibinfo{volume}{86}}, \bibinfo{pages}{073016}
  (\bibinfo{year}{2012}).
\newblock \emph{\bibinfo{journal}{ArXiv e-prints} \eprint{1208.1533}}.

\bibitem{1305.3854}
\bibinfo{author}{{Dixon}, L.~J.} \& \bibinfo{author}{{Li}, Y.}
\newblock \bibinfo{title}{{Bounding the Higgs Boson Width Through
  Interferometry}}.
\newblock \emph{\bibinfo{journal}{ArXiv e-prints} \eprint{1305.3854}}  (\bibinfo{year}{2013}).
\newblock \urlprefix\url{http://arxiv.org/abs/arXiv:1305.3854}.

\bibitem{1306.5770B}
\bibinfo{author}{{Bodwin}, G.~T.}, \bibinfo{author}{{Petriello}, F.},
  \bibinfo{author}{{Stoynev}, S.} \& \bibinfo{author}{{Velasco}, M.}
\newblock \bibinfo{title}{{Higgs boson decays to quarkonia and the Hccbar
  coupling}}.
\newblock \emph{\bibinfo{journal}{ArXiv e-prints} \eprint{1306.5770}}  (\bibinfo{year}{2013}).
\newblock \urlprefix\url{http://arxiv.org/abs/1306.5770}.

\bibitem{Barger_Ishida_Keung_2012}
\bibinfo{author}{Barger, V.}, \bibinfo{author}{Ishida, M.} \&
  \bibinfo{author}{Keung, W.-Y.}
\newblock \bibinfo{title}{{Total Width of 125 GeV Higgs Boson}}.
\newblock \emph{\bibinfo{journal}{Physical Review Letters}}
  \textbf{\bibinfo{volume}{108}} (\bibinfo{year}{2012}).
\newblock \urlprefix\url{http://dx.doi.org/10.1103/PhysRevLett.108.261801}.

\bibitem{1307.7135}
\bibinfo{author}{CMS Collaboration}.
\newblock \bibinfo{title}{{Projected Performance of an Upgraded CMS Detector at
  the LHC and HL-LHC: Contribution to the Snowmass Process}}
\newblock \emph{\bibinfo{journal}{ArXiv e-prints} \eprint{1307.7135}}  (\bibinfo{year}{2013}).
\newblock \urlprefix\url{http://arxiv.org/abs/1307.7135}.

\bibitem{1307.7292}
\bibinfo{author}{ATLAS Collaboration}.
\newblock \bibinfo{title}{{Physics at a High-Luminosity LHC with ATLAS}}
\newblock \emph{\bibinfo{journal}{ArXiv e-prints} \eprint{1307.7292}}  (\bibinfo{year}{2013}).
\newblock \urlprefix\url{http://arxiv.org/abs/1307.7292}.

\bibitem{Isidori_Marrouche_et_al__2012}
\bibinfo{author}{Buchmueller, O.} \emph{et~al.}
\newblock \bibinfo{title}{{The CMSSM and NUHM1 in light of 7 TeV LHC, $B_s \to
  \mu^+\mu^-$ and XENON100 data}}.
\newblock \emph{\bibinfo{journal}{The European Physical Journal C}}
  \textbf{\bibinfo{volume}{72}} (\bibinfo{year}{2012}).
\newblock \urlprefix\url{http://dx.doi.org/10.1140/epjc/s10052-012-2243-3}.

\bibitem{1307.1347}
\bibinfo{author}{Heinemeyer, S.}, \bibinfo{author}{Mariotti, C.},
  \bibinfo{author}{Passarino, G.}, \bibinfo{author}{Tanaka, R.}
  \bibinfo{author}{{\it et al.}}
\newblock \bibinfo{title}{{Handbook of LHC Higgs Cross Sections: 3. Higgs
  Properties}} 
\newblock \emph{\bibinfo{journal}{ArXiv e-prints} \eprint{1307.1347}}  (\bibinfo{year}{2013}).
\newblock \urlprefix\url{http://arxiv.org/abs/1307.1347}.

\bibitem{cite:Pietrzyk}
\bibinfo{author}{{Pietrzyk}, B.}
\newblock \bibinfo{title}{{LEP asymmetries and fits of the Standard Model}}
\newblock \emph{\bibinfo{journal}{ArXiv e-prints} \eprint{hep-ex/9406001}}  (\bibinfo{year}{1994}).
\newblock \urlprefix\url{http://arxiv.org/abs/hep-ex/9406001}.

\bibitem{_Lys_Murayama_Wohl_et_al__2012}
\bibinfo{author}{Beringer, J.} \emph{et~al.}
\newblock \bibinfo{title}{{Review of Particle Physics}}.
\newblock \emph{\bibinfo{journal}{Physical Review D}}
  \textbf{\bibinfo{volume}{86}} (\bibinfo{year}{2012}).
\newblock \urlprefix\url{http://dx.doi.org/10.1103/PhysRevD.86.010001}.

\bibitem{ements_on_the_Z_resonance_2006}
\bibinfo{author}{ALEPH}, \bibinfo{author}{DELPHI}, \bibinfo{author}{L3},
  \bibinfo{author}{OPAL} \& \bibinfo{author}{{The LEP Electroweak Working
  Group}}.
\newblock \bibinfo{title}{{Precision Electroweak measurements on the Z
  resonance}}.
\newblock \emph{\bibinfo{journal}{Physics Reports}}
  \textbf{\bibinfo{volume}{427}}, \bibinfo{pages}{257--454}
  (\bibinfo{year}{2006}).
\newblock \urlprefix\url{http://dx.doi.org/10.1016/j.physrep.2005.12.006}.

\bibitem{1302.3415}
\bibinfo{author}{ALEPH}, \bibinfo{author}{DELPHI}, \bibinfo{author}{L3},
  \bibinfo{author}{OPAL} \& \bibinfo{author}{{The LEP Electroweak Working
  Group}}.
\newblock \bibinfo{title}{{Electroweak Measurements in Electron-Positron
  Collisions at W-Boson-Pair Energies at LEP}} 
\newblock \emph{\bibinfo{journal}{ArXiv e-prints} \eprint{1302.3415}}  (\bibinfo{year}{2013}).
\newblock \urlprefix\url{http://arxiv.org/abs/1302.3415}.


\bibitem{Ellis:1990}
\bibinfo{author}{Ellis, J.~R.}, \bibinfo{author}{Kelley, S.} \&
  \bibinfo{author}{Nanopoulos, D.~V.}
\newblock \bibinfo{title}{{Probing the desert using gauge coupling
  unification}}.
\newblock \emph{\bibinfo{journal}{Phys.Lett.}} \textbf{\bibinfo{volume}{B260}},
  \bibinfo{pages}{131--137} (\bibinfo{year}{1991}).

\bibitem{Amaldi:1991}
\bibinfo{author}{Amaldi, U.}, \bibinfo{author}{de~Boer, W.} \&
  \bibinfo{author}{Furstenau, H.}
\newblock \bibinfo{title}{{Comparison of grand unified theories with
  electroweak and strong coupling constants measured at LEP}}.
\newblock \emph{\bibinfo{journal}{Phys.Lett.}} \textbf{\bibinfo{volume}{B260}},
  \bibinfo{pages}{447--455} (\bibinfo{year}{1991}).

\bibitem{Langacker:1991}
\bibinfo{author}{Langacker, P.} \& \bibinfo{author}{Luo, M.}
\newblock \bibinfo{title}{{Implications of precision electroweak experiments
  for $M_t$, $\rho_{0}$, $\sin^2\theta_W$ and grand unification}}.
\newblock \emph{\bibinfo{journal}{Phys.Rev.}} \textbf{\bibinfo{volume}{D44}},
  \bibinfo{pages}{817--822} (\bibinfo{year}{1991}).

\bibitem{Giunti:1991}
\bibinfo{author}{Giunti, C.}, \bibinfo{author}{Kim, C.} \&
  \bibinfo{author}{Lee, U.}
\newblock \bibinfo{title}{{Running coupling constants and grand unification
  models}}.
\newblock \emph{\bibinfo{journal}{Mod.Phys.Lett.}}
  \textbf{\bibinfo{volume}{A6}}, \bibinfo{pages}{1745--1755}
  (\bibinfo{year}{1991}).

\bibitem{1207.7355}
\bibinfo{author}{{Espinosa}, J.~R.}, \bibinfo{author}{{Grojean}, C.},
  \bibinfo{author}{{Sanz}, V.} \& \bibinfo{author}{{Trott}, M.}
\newblock \bibinfo{title}{{NSUSY fits}}.
\newblock \emph{\bibinfo{journal}{Journal of High Energy Physics}}
  \textbf{\bibinfo{volume}{12}}, \bibinfo{pages}{77} (\bibinfo{year}{2012}).
\newblock \emph{\bibinfo{journal}{ArXiv e-prints} \eprint{1207.7355}}.

\bibitem{Jarlskog_1990}
\bibinfo{author}{Jarlskog, C.}
\newblock \bibinfo{title}{{Neutrino counting at the Z-peak and right-handed
  neutrinos}}.
\newblock \emph{\bibinfo{journal}{Physics Letters B}}
  \textbf{\bibinfo{volume}{241}}, \bibinfo{pages}{579--583}
  (\bibinfo{year}{1990}).
\newblock \urlprefix\url{http://dx.doi.org/10.1016/0370-2693(90)91873-A}.

\bibitem{Barbiellini}
\bibinfo{author}{Barbiellini, G.} \emph{et~al.}
\newblock \bibinfo{title}{{Neutrino Counting}}  (\bibinfo{year}{1989}).
\newblock \urlprefix\url{http://cds.cern.ch/search?sysno=000112318CER}.

\bibitem{cite:Abbiendi2000hh}
\bibinfo{author}{Abbiendi, G.} \bibinfo{author}{{\it et al.}}
\newblock \bibinfo{title}{{Photonic events with missing energy in $e^+ e^-$
  collisions at $\sqrt{s} = 189$ GeV}}.
\newblock \emph{\bibinfo{journal}{Eur.Phys.J.}} \textbf{\bibinfo{volume}{C18}},
  \bibinfo{pages}{253--272} (\bibinfo{year}{2000}).
\newblock \emph{\bibinfo{journal}{ArXiv e-prints} \eprint{hep-ex/0005002}}.

\bibitem{cite:Heister2002ut}
\bibinfo{author}{Heister, A.} \bibinfo{author}{{\it et al.}}
\newblock \bibinfo{title}{{Single photon and multiphoton production in
  $e^{+}e^{-}$ collisions at $\sqrt{s}$ up to 209-GeV}}.
\newblock \emph{\bibinfo{journal}{Eur.Phys.J.}} \textbf{\bibinfo{volume}{C28}},
  \bibinfo{pages}{1--13} (\bibinfo{year}{2003}).
\newblock
  \urlprefix\url{http://link.springer.com/article/10.1140%2Fepjc%2Fs2002-01129-7}.

\bibitem{cite:Achard2003tx}
\bibinfo{author}{Achard, P.} \bibinfo{author}{{\it et al.}}
\newblock \bibinfo{title}{{Single photon and multiphoton events with missing
  energy in $e^{+} e^{-}$ collisions at LEP}}.
\newblock \emph{\bibinfo{journal}{Phys.Lett.}} \textbf{\bibinfo{volume}{B587}},
  \bibinfo{pages}{16--32} (\bibinfo{year}{2004}).
\newblock \emph{\bibinfo{journal}{ArXiv e-prints} \eprint{hep-ex/0402002}}.

\bibitem{cite:Abdallah2003np}
\bibinfo{author}{Abdallah, J.} \bibinfo{author}{{\it et al.}}
\newblock \bibinfo{title}{{Photon events with missing energy in $e^+ e^-$
  collisions at $\sqrt{s} = 130$ to $209$ GeV}}.
\newblock \emph{\bibinfo{journal}{Eur.Phys.J.}} \textbf{\bibinfo{volume}{C38}},
  \bibinfo{pages}{395--411} (\bibinfo{year}{2005}).
\newblock \emph{\bibinfo{journal}{ArXiv e-prints} \eprint{hep-ex/0406019}}.

\bibitem{Baikov_Chetyrkin_Kuhn_2008}
\bibinfo{author}{Baikov, P.}, \bibinfo{author}{Chetyrkin, K.} \&
  \bibinfo{author}{K{\"{u}}hn, J.}
\newblock \bibinfo{title}{{Order $\alpha_s^4$ QCD Corrections to Z and $\tau$
  Decays}}.
\newblock \emph{\bibinfo{journal}{Physical Review Letters}}
  \textbf{\bibinfo{volume}{101}} (\bibinfo{year}{2008}).
\newblock \urlprefix\url{http://dx.doi.org/10.1103/PhysRevLett.101.012002}.

\bibitem{Chetyrkin_Kuhn_Rittinger_2012}
\bibinfo{author}{Baikov, P.~A.}, \bibinfo{author}{Chetyrkin, K.~G.},
  \bibinfo{author}{K{\"{u}}hn, J.~H.} \& \bibinfo{author}{Rittinger, J.}
\newblock \bibinfo{title}{{Complete O($\alpha_s^4$) QCD Corrections to Hadronic
  Z Decays}}.
\newblock \emph{\bibinfo{journal}{Physical Review Letters}}
  \textbf{\bibinfo{volume}{108}} (\bibinfo{year}{2012}).
\newblock \urlprefix\url{http://dx.doi.org/10.1103/PhysRevLett.108.222003}.

\bibitem{Bethke_2004}
\bibinfo{author}{Bethke, S.}
\newblock \bibinfo{title}{{$\alpha_s$ at Zinnowitz 2004}}.
\newblock \emph{\bibinfo{journal}{Nuclear Physics B - Proceedings Supplements}}
  \textbf{\bibinfo{volume}{135}}, \bibinfo{pages}{345--352}
  (\bibinfo{year}{2004}).
\newblock \urlprefix\url{http://dx.doi.org/10.1016/j.nuclphysbps.2004.09.020}.

\bibitem{g_Manohar_Stewart_Teubner_2001}
\bibinfo{author}{Hoang, A.}, \bibinfo{author}{Manohar, A.},
  \bibinfo{author}{Stewart, I.} \& \bibinfo{author}{Teubner, T.}
\newblock \bibinfo{title}{{Threshold $t\bar t$ cross section at
  next-to-next-to-leading logarithmic order}}.
\newblock \emph{\bibinfo{journal}{Physical Review D}}
  \textbf{\bibinfo{volume}{65}} (\bibinfo{year}{2001}).
\newblock \urlprefix\url{http://dx.doi.org/10.1103/PhysRevD.65.014014}.

\bibitem{cite:1303.3758}
\bibinfo{author}{{Seidel}, K.}, \bibinfo{author}{{Simon}, F.},
  \bibinfo{author}{{Tesar}, M.} \& \bibinfo{author}{{Poss}, S.}
\newblock \bibinfo{title}{{Top quark mass measurements at and above threshold
  at CLIC}} 
\newblock \emph{\bibinfo{journal}{ArXiv e-prints} \eprint{1303.3758}}  (\bibinfo{year}{2013}).
\newblock \urlprefix\url{http://arxiv.org/abs/1303.3758}.

\bibitem{Martinez_Miquel_2003}
\bibinfo{author}{Martinez, M.} \& \bibinfo{author}{Miquel, R.}
\newblock \bibinfo{title}{{Multi-parameter fits to the $t\bar{t}$ threshold
  observables at a future $e^+e^-$ linear collider}}.
\newblock \emph{\bibinfo{journal}{The European Physical Journal C}}
  \textbf{\bibinfo{volume}{27}}, \bibinfo{pages}{49--55}
  (\bibinfo{year}{2003}).
\newblock \urlprefix\url{http://dx.doi.org/10.1140/epjc/s2002-01094-1}.

\bibitem{Ferroglia_Sirlin_2013}
\bibinfo{author}{Ferroglia, A.} \& \bibinfo{author}{Sirlin, A.}
\newblock \bibinfo{title}{{Comparison of the standard theory predictions of
  $m_W$ and $\sin^2\theta^{lept}_{eff}$ with their experimental values}}.
\newblock \emph{\bibinfo{journal}{Physical Review D}}
  \textbf{\bibinfo{volume}{87}} (\bibinfo{year}{2013}).
\newblock \urlprefix\url{http://dx.doi.org/10.1103/PhysRevD.87.037501}.

\bibitem{1007.5232}
\bibinfo{author}{{Heinemeyer}, S.} \& \bibinfo{author}{{Weiglein}, G.}
\newblock \bibinfo{title}{{Top, GigaZ, MegaW}}.
\newblock \emph{\bibinfo{journal}{ArXiv e-prints} \eprint{1007.5232}}  (\bibinfo{year}{2010}).
\newblock \urlprefix\url{http://arxiv.org/abs/arXiv:1007.5232}.

\bibitem{aQEDD}
\bibinfo{author}{{Jegerlehner}, F.}
\newblock \bibinfo{title}{{The running fine structure constant {$\alpha$}(E)
  via the Adler function}}.
\newblock \emph{\bibinfo{journal}{Nuclear Physics B Proceedings Supplements}}
  \textbf{\bibinfo{volume}{181}}, \bibinfo{pages}{135--140}
  (\bibinfo{year}{2008}).
\newblock \eprint{0807.4206}.

\bibitem{aQEDC}
\bibinfo{author}{{Actis}, S.} \emph{et~al.}
\newblock \bibinfo{title}{{Quest for precision in hadronic cross sections at
  low energy: Monte Carlo tools vs. experimental data}}.
\newblock \emph{\bibinfo{journal}{European Physical Journal C}}
  (\bibinfo{year}{2010}).
\newblock \emph{\bibinfo{journal}{ArXiv e-prints} \eprint{0912.0749}}.

\bibitem{aQEDB}
\bibinfo{author}{{Babusci}, D.} \emph{et~al.}
\newblock \bibinfo{title}{{Proposal for taking data with the KLOE-2 detector at
  the DA\$$\backslash$Phi\$NE collider upgraded in energy}}.
\newblock \emph{\bibinfo{journal}{ArXiv e-prints} \eprint{1007.5219}}  (\bibinfo{year}{2010}).
\newblock \urlprefix\url{http://arxiv.org/abs/1007.5219}.

\bibitem{aQEDA}
\bibinfo{author}{{Alesini}, D.} \emph{et~al.}
\newblock \bibinfo{title}{{IRIDE White Book, An Interdisciplinary Research
  Infrastructure based on Dual Electron linacs{\&}lasers}}.
\newblock \emph{\bibinfo{journal}{ArXiv e-prints} \eprint{1307.7967}}  (\bibinfo{year}{2013}).
\newblock \urlprefix\url{http://arxiv.org/abs/1307.7967}.

\bibitem{gler_Monig_Schott_Stelzer_2012}
\bibinfo{author}{Baak, M.} \emph{et~al.}
\newblock \bibinfo{title}{{The Electroweak fit of the Standard Model after the
  discovery of a new boson at the LHC}}.
\newblock \emph{\bibinfo{journal}{The European Physical Journal C}}
  \textbf{\bibinfo{volume}{72}} (\bibinfo{year}{2012}).
\newblock
\urlprefix\url{http://dx.doi.org/10.1140/epjc/s10052-012-2205-9}.

\bibitem{petruc}
\bibinfo{author}{{Petrucciani, G.}}
\newblock \bibinfo{title}{{The ttH coupling measurement at the HL-LHC,
  talk given on behalf of the CMS collaboration at the sixth TLEP
  Workshop}}
\newblock \urlprefix\url{https://indico.cern.ch/conferenceDisplay.py?ovw=True&confId=257713}






\bibitem{cite:ATLAS}
\bibinfo{author}{ATLAS Collaboration}.
\newblock \bibinfo{title}{{Physics at a High-Luminosity LHC with ATLAS}}.
\newblock \bibinfo{type}{Tech. Rep.} \bibinfo{number}{ATL-PHYS-PUB-2012-004}
  (\bibinfo{year}{2012}).
\newblock
  \urlprefix\url{http://cds.cern.ch/record/1484890/files/ATL-PHYS-PUB-2012-004.pdf}.

\bibitem{cite:CMS}
\bibinfo{author}{CMS Collaboration}.
\newblock \bibinfo{title}{{CMS at the High-Energy Frontier}}.
\newblock \bibinfo{type}{Tech. Rep.} \bibinfo{number}{CMS-2012-006}
  (\bibinfo{year}{2012}).
\newblock
  \urlprefix\url{https://cds.cern.ch/record/1494600/files/NOTE2012_006.pdf}.

\bibitem{Mangano_Rojo_2012}
\bibinfo{author}{Mangano, M.~L.} \& \bibinfo{author}{Rojo, J.}
\newblock \bibinfo{title}{{Cross section ratios between different CM energies
  at the LHC: opportunities for precision measurements and BSM sensitivity}}.
\newblock \emph{\bibinfo{journal}{Journal of High Energy Physics}}
  \textbf{\bibinfo{volume}{2012}} (\bibinfo{year}{2012}).
\newblock \urlprefix\url{http://dx.doi.org/10.1007/JHEP08(2012)010}.

\bibitem{Snowmass}
\bibinfo{author}{Dawson, S.} \emph{et~al.}
\newblock \bibinfo{title}{{The Snowmass Higgs Working group report (in
  preparation)}}.
\newblock \bibinfo{type}{Tech. Rep.} (\bibinfo{year}{2013}).
\newblock
  \urlprefix\url{http://www.snowmass2013.org/tiki-index.php?page=The+Higgs+Boson}.

\bibitem{cite:1305.6397}
\bibinfo{author}{{Gupta}, R.~S.}, \bibinfo{author}{{Rzehak}, H.} \&
  \bibinfo{author}{{Wells}, J.~D.}
\newblock \bibinfo{title}{{How well do we need to measure the Higgs boson mass
  and self-coupling?}} 
\newblock \emph{\bibinfo{journal}{ArXiv e-prints} \eprint{1305.6397}}  (\bibinfo{year}{2013}).
\newblock \urlprefix\url{http://arxiv.org/abs/1305.6397}.

\bibitem{QuarticHiggs}
\bibinfo{author}{{Plehn}, T.} \& \bibinfo{author}{{Rauch}, M.}
\newblock \bibinfo{title}{{Quartic Higgs coupling at hadron colliders}}.
\newblock \emph{\bibinfo{journal}{Physical Review D}}
  \textbf{\bibinfo{volume}{72}}, \bibinfo{pages}{053008}
\newblock \emph{\bibinfo{journal}{ArXiv e-prints} \eprint{hep-ph/0507321}}  (\bibinfo{year}{2005}).
\newblock \urlprefix\url{http://link.aps.org/doi/10.1103/PhysRevD.72.053008}.

\end{thebibliography}
